\documentclass[11pt]{article}
\usepackage{graphicx}
\usepackage{amssymb}
\usepackage{amsmath}
\usepackage{graphicx}
\usepackage{amstext}
\usepackage{leftidx}
\usepackage{esint}
\usepackage[utf8]{inputenc}
\usepackage{color}
\usepackage{cite}
\usepackage{hyperref}
\usepackage{multido}

\makeatletter
\ams@newcommand{\vardot}[2]{%
  {\mathop{#2\kern0pt}\limits^{\vbox to-1.4\ex@{\kern-\tw@\ex@
   \hbox{\normalfont\multido{}{#1}{.}}\vss}}}}
\makeatother
\newcommand{\rL}{\rho_\Lambda}

\newcommand{\CC}{\Lambda}

\newcommand{\rv}{\rho_{\rm vac}}
\newcommand{\Pv}{P_{\rm vac}}
\newcommand{\rvo}{\rho^0_{\rm vac}}

\newcommand{\nueff}{\nu_{\rm eff}}

\newcommand{\bk}{{\bf k}}
\newcommand{\mpl}{m_{\rm Pl}}

\newcommand{\be}{\begin{equation}}
\newcommand{\ee}{\end{equation}}

\newcommand{\jtext}[1]{{\textcolor{black}{#1}}}

\newcommand{\cH}{\mathcal{H}}

\newcommand{\astar}{a_{*}}

\newcommand{\rI}{\rho_I}

\newcommand{\wv}{w_{\rm vac}}

\begin{document}

\hyphenation{theo-re-ti-cal gra-vi-ta-tio-nal theo-re-ti-cally mo-dels cos-mo-lo-gi-cal cor-res-pon-ding}

\begin{center}
%{\bf \LARGE Running Vacuum from QFT in curved spacetime}\\
\vskip 2mm
{\bf \LARGE  Running vacuum in QFT in  FLRW spacetime:} \vskip 2mm
{\bf \Large The dynamics of $\rv(H)$ from the quantized matter fields}\footnote{Dedicated to Harald Fritzsch}

 \vskip 8mm

\textbf{Cristian Moreno-Pulido$^{a,b}$, Joan Sol\`a Peracaula$^a$ and Samira Cheraghchi$^c$}

\vskip 0.4cm

$^a$Departament de F\'isica Qu\`antica i Astrof\'isica, \\
and   Institute of Cosmos Sciences,\\ Universitat de Barcelona,
Av. Diagonal 647, E-08028 Barcelona, Catalonia, Spain

\vskip 0.4cm

$^b$Universitat Carlemany,\\
 Av. Verge de Canòlich, 47, AD600, Sant Julià de Lòria, Andorra \\
 
\vskip 0.4cm

$^c$Faculty of Mathematics and Computer Science,\\ 
Transilvania University, Iuliu Maniu Str. 50, 500091 Brasov, Romania

\vskip 0.5cm

\vskip0.4cm

E-mails:  cristian.moreno@fqa.ub.edu, sola@fqa.ub.edu, samira.cheraghchi@unitbv.ro

 \vskip2mm

\end{center}
\vskip 15mm

\begin{quotation}
\noindent {\large\it \underline{Abstract}}.
Phenomenological work in the last few years has provided  significant support to the idea that the vacuum energy density (VED) is a running quantity with the cosmological evolution and that this running helps to alleviate the cosmological tensions afflicting the $\Lambda$CDM. On the theoretical side,  recent devoted studies have shown that  the properly renormalized   $\rho_{\rm vac}$ in QFT in  FLRW spacetime  adopts the `running vacuum model' (RVM) form.  While in three previous studies by two of us (CMP and JSP) such computations focused solely on scalar fields non-minimally coupled to gravity, in the present work we compute the spin-$1/2$ fermionic contributions and combine them both.
The calculation is performed using a new version of the adiabatic renormalization procedure based on subtracting the UV divergences at an off-shell renormalization point $M$. The quantum scaling of $\rho_{\rm vac}$ with $M$ turns into  cosmic evolution with the Hubble rate, $H$.  As a result the  `cosmological constant'  $\Lambda$  appears in our framework as the nearly sustained value of  $8\pi G(H)\rho_{\rm vac}(H)$  around (any)  given epoch $H$, where $G(H)$ is the  gravitational coupling, which is also running, although very mildly (logarithmically).
 We find that the VED evolution at present reads  $\delta \rho_{\rm vac}(H)\sim \nu_{\rm eff}\, m_{\rm Pl}^2 \left(H^2-H_0^2 \right)\  (|\nu_{\rm eff}|\ll 1)$. The coefficient $\nu_{\rm eff}$ receives contributions from all the quantized fields, bosons and fermions, which we compute here for an arbitrary number of matter fields.
Remarkably, there also exist  higher powers  ${\cal O}(H^{6})$   which can trigger inflation  in the early universe. Finally, the equation of state (EoS) of the vacuum receives also quantum corrections from bosons and fermion fields, shifting its value from -1. The striking consequence is that the EoS of the quantum vacuum may nowadays effectively appears as quintessence.

\end{quotation}

\newpage

\tableofcontents

\newpage
\section{Introduction}
Despite having coexisted for many decades,  a completely successful theory of gravity that combines Quantum Field Theory (QFT) and General Relativity (GR) does not exist yet, unfortunately. However, a variety of different approaches and techniques are available in the literature which allow one to study the subject of quantum fields in the gravitational context, and more specifically the physics of the expanding  universe and its current speeding up. Our aim is to understand such an acceleration on fundamental grounds.  To be precise, in this work we will concentrate on the well-known semiclassical approach which goes under the name of QFT in curved spacetime\,\cite{Birrell:1982ix,Parker:2009uva,Fulling:1989nb}. This means that gravity is still a classical external (background) field, whereas the matter fields are quantum field operators obeying suitable commutation or anticommutation relations\,\cite{DeWitt:1975ys,Parker:1978gh}.   A further step in the path of understanding gravity in the QFT context is quantum gravity (QG), in which spacetime itself (the metric) is quantized and hence  functional integration over metrics is mandatory, see e.g. \cite{Christensen:1979iy,Christensen:1978md,Christensen:1978gi}, and the review \cite{Woodard:2009ns} and references therein.
At the same time  a lot of exciting QG phenomenology is being investigated in the current  multi-messenger era, characterized by an outburst  of  experimental data that are being obtained from
the detection of the various cosmic messengers (photons, neutrinos, cosmic rays and gravitational waves) from numerous origins\,\cite{Addazi:2021xuf}. On the theoretical side, effective field theory methods and the possibility of quantum gravitational effects leading to quantum hair may  provide useful hints of QG which have been explored recently\,\cite{Calmet:2022bpo,Calmet:2022hvn,Calmet:2021cip}.  However, while the QG option has, of course, to be kept in mind since it can be very important when QG can be (hopefully) formulated in a fully consistent way\cite{Woodard:2009ns}, QFT in curved spacetime may still be of great help to further describe the role of quantum fields in a gravitational context. In this work, we will continue dwelling upon these lines and shall focus exclusively on the, more modest, but effective, semiclassical approach. It goes without saying that the latter has had also its own problems and successes over the years, and still has\,\cite{Ford:1997hb}. However, new perspectives have recently been explored in this context concerning the vacuum energy and the cosmological constant (CC)\cite{Moreno-Pulido:2020anb, Moreno-Pulido:2022phq,Moreno-Pulido:2022upl} which may be of significance, and for this reason we wish to further pursue this line of approach here.

The agent responsible for the accelerated cosmic expansion is generically called {\it Dark Energy} (DE), an entity which constitutes a key piece in the cosmological puzzle, but whose fundamental nature is still undisclosed \cite{Amendola:2015ksp}. Although it might be caused by deviations from GR connected with extended theories of gravity\,\cite{Capozziello:2007ec,Sotiriou:2008rp,Capozziello:2011et,CapozzielloFaraoni2011,Cai:2015emx,Horndeski:1974wa},  the canonical possibility is still that the DE is related to the CC in Einstein's equations, $\CC$,  as done routinely in the standard (or `concordance') model of cosmology, aka $\Lambda$CDM\,\cite{peebles1993principles,Krauss:1995yb,Ostriker:1995su}.   The model  has been a rather successful  paradigm for the phenomenological description of the universe for about three decades, but it became consolidated only in the late nineties\cite{Turner:2022gvw,Turner:2018bcg} and especially after the consistent measurements of $\CC$ made in the last twenty years using independent cosmological sources, in particular including distant type Ia supernovae (SnIa), baryon-acoustic oscillations (BAO), the data on large-scale structure formation and, of course,  the anisotropies of the cosmic microwave background (CMB). All in all they have put the very experimental basis for the concordance  $\CC$CDM model of cosmology\,\cite{SupernovaSearchTeam:1998fmf,SupernovaCosmologyProject:1998vns,SDSS:2005xqv,Pan-STARRS1:2017jku,Planck:2015fie,Planck:2018vyg}. The situation is far from being satisfactory, though.  The problems with the $\CC$CDM are both of theoretical and observational nature.  As for the theoretical problems, recall that the value of $\CC$  is traditionally associated to a parameter called the vacuum energy density (VED) in the universe, which in the context of the $\CC$CDM  is nothing but a name for the following quantity with dimensions of energy density:  $\rv=\CC/(8\pi G_N)$ ($G_N$ being Newton's constant).  Its theoretical  significance  is not explained at all in the context of the standard cosmological model. If, however, we take quantum theory seriously,  \jtext{the most universal contribution to vacuum energy density} is the zero-point energy (ZPE) of the massive quantum fields in the standard model of particle physics, and in fact also in any realistic QFT model.  However, it is well-known that a naive calculation of this quantity leads to very large contributions proportional to the quartic power of the mass of the particles\,\cite{Zeldovich1967a},
$\rho_{ \rm ZPE}\sim m^4$,  which is in blatant discordance with the order of magnitude obtained for this quantity from cosmological observations:
$\rho_{\rm vac}^{\rm obs}\sim 10^{-47} \textrm{GeV}^4$
(expressed in natural units, with $\hbar  = c = 1$). Even taking, for instance,  the electron field  one finds a mismatch of $34$ orders of magnitude:  $\rho_{\rm vac}^{\rm obs}/\rho_{\rm ZPE}\sim  10^{-34}$.  The huge discrepancy between a typical standard model  contribution to the ZPE and the measured value of VED constitutes the so-called Cosmological Constant Problem (CCP) \cite{Weinberg:1988cp,Peebles:2002gy,Padmanabhan:2002ji}.  See also \cite{Sola:2013gha,Peracaula:2022vpx} \jtext{for more recent accounts}. Despite the enormous discrepancies between usual theory predictions and factual measurements, estimates on the value of $\Lambda$ within the right order of magnitude have been attempted under certain assumptions in the context of QG in different approaches, see e.g.\,\cite{Adler:2021arz,Padmanabhan:2016eld,Padmanabhan:2017naf,Ward:2022eps,Ward:2010qs}.

While the aforementioned measurements of $\rho_{\rm vac}$  indicate that the vacuum can gravitate within an energy density order of magnitude of $\sim 10^{-47} \textrm{GeV}^4$, what is difficult to understand theoretically is why the vacuum can  only gravitate in that tiny range, given the fact that any typical quantum effect rockets its contribution to much larger values. This is of course a rephrasing of the same puzzle associated to the CCP, expressed in the QFT context.  However, new avenues for a possible solution have been suggested recently.  The renormalization approach presented in the present work and in the preceding studies\,\cite{Moreno-Pulido:2020anb, Moreno-Pulido:2022phq,Moreno-Pulido:2022upl} offers some hope  to eschew part of these difficulties. First and foremost, the renormalized quantum effects found here endow the VED with a mild dynamical nature. The latter thus appears as a slowly varying function of the cosmic expansion, specifically of the Hubble rate $H$,  see below.  Second, the renormalized VED  as reported here proves well behaved and can perfectly accommodate the measured value of $\CC$ from observational cosmology without fine-tuning. Technically, this is because the ``running'' of $\CC$ is proportional to the tiny values of the $\beta$-function coefficients for bosons and fermions, which are  responsible for the renormalization group evolution of the VED.  As a result, at any given epoch of the late universe  $\CC$ appears essentially as constant, but it is not strictly so.  Finally, a third crucial ingredient of our approach is that,  in the very early universe, the  VED becomes, in contrast, very large and fast evolving. There it can take the capital role of bringing about inflation, as we shall see.

As previously mentioned, in addition to the traditional theoretical problems,  other issues of more practical and mundane nature have been perturbing  cosmologists in the last few years, which put the concordance $\CC$CDM against the wall. The practical problems are the presently  irreconcilable  observational differences between the concordance model predictions and  a number of different kinds of cosmological  observations. For example, those involving structure formation data (the so-called $\sigma_8$ tension), remain at a moderate  level of $ 2-3\sigma$ (where $\sigma_8$ is the root mean square of fluctuations in matter density perturbations within spheres of radius  $8h^{-1}$ Mpc); and,  above all, the notorious conflict  between the local value of the Hubble parameter,  $H_0$, obtained from the traditional distant ladder techniques,  and the value extracted from the early universe using CMB data.  \jtext{The latter is arguably the most puzzling open question within the current cosmological paradigm as it points to a severe discrepancy of $\sim 5 \sigma$ c.l. or more.  In the constant pursue for possible solutions to these tensions, it has been argued that within the class of models where the DE is dealt with as a fluid with EoS $w(z)$, solving the $H_0$ tension demands the phantom condition $w(z)<-1$ at some $z$, whilst solving both the $H_0$ and $\sigma_8$ tensions may require $w(z)$ to cross the phantom divide and/or other sorts of transitions, see e.g. \cite{Heisenberg:2022gqk,Marra:2021fvf,Alestas:2020zol,Perivolaropoulos:2021bds,Alestas:2021luu,Perivolaropoulos:2022khd}. As we shall see, among the possible signatures implied by the framework presented in  the current work (cf. Sec.\,\hyperref[sec:signatures]{\ref{sec:signatures}})  one may obtain transitions of this sort. Recently it has also been put forward the idea  that supposing $H=H(z)$ as a function of redshift and considering the look-back time as a cosmological  probe it could help fixing the $H_0$ tension from data with intriguing quantitative success, see\,\cite{Capozziello:2023ewq}.}  We refer the reader to \cite{DiValentino:2020zio,DiValentino:2020vvd} for a summary of these problems,  and \cite{Perivolaropoulos:2021jda,Abdalla:2022yfr,Dainotti:2023yrk}  for comprehensive reviews and a long list of related references. For some people the disagreement  in the $H_0$ case is sufficiently severe  (and persistent over time) as to still be pretending that it can be attributed to a fluke, thus increasing the odds that its origin may come \jtext{from physics  beyond the $\CC$CDM\,\cite{Riess:2019cxk,Riess:2021jrx}}. So, even if  we remain agnostic, we cannot exclude that the prevailing model of cosmology might well be facing  a crisis. Science, however, thrives on crisis since new ideas are then stimulated  which could help to overcome it and maybe  refine some aspects of the  paradigm in force, \jtext{or even originate a new one capable of subsuming the old paradigm	}.   Many proposals have indeed been made to alleviate these tensions, which include different forms of DE as well, despite that many of them are essentially \textit{ad hoc}.  As indicated above, clues to eventually substantiate the nature  of the DE may come from a variety of cosmic and even astrophysical messengers\,\cite{Addazi:2021xuf}.  For instance, the possibility of measuring the bending of light in the Solar System scale has been proposed\,\cite{Zhang:2021ygh}. But whatever  the nature of the DE might be, we must provide an explanation for the role played by the vacuum energy in QFT.  Indeed, in the absence of a correct understanding of the VED from first principles many DE proposals may look as an escape forward rather than a real alternative. This work, in contrast,  intends to dwell further on the  methods of  QFT in curved spacetime so as to shed some useful light on these difficult problems. Recall that in QFT we treat the  $\CC$ term in the gravitational action  as a formal quantity  from which together with the ZPE (a computable quantity in curved spacetime) one can determine the  renormalized VED, $\rv$,  a fundamental concept in QFT.  The physical cosmological term can then be defined from the VED as follows: $\CC_{\rm phys}=8\pi G\rv$. This is nevertheless not just the usual naive relation since $G$ and $\rv$ are now properly renormalized quantities in QFT in curved spacetime.  In fact, in this work we report on progress made along the lines of the preceding comprehensive works\,\cite{Moreno-Pulido:2020anb, Moreno-Pulido:2022phq,Moreno-Pulido:2022upl}, where a detailed account was made of the virtual contribution to the VED from quantized scalar matter fields.  We found that these effects, when appropriately renormalized,  translate into a (mild) dynamical evolution of the VED and $G$ with the cosmic expansion, $\rv=\rv (H)$ and $G=G (\ln H)$.  This is not excluded by the cosmological principle, as it permits a homogeneous and isotropic dependence in time of physical quantities. More specifically, it was shown by explicit calculation and adequate renormalization that the VED behaves in the characteristic  manner of the running vacuum model (RVM), see \cite{Peracaula:2022vpx} for a recent comprehensive review (for a shorter summary, see e.g.\cite{SolaPeracaula:2022mlg}).  Let us also note that a similar RVM interpretation of the vacuum energy density can be achieved in the string context, what has been called `stringy-RVM'\,\cite{Mavromatos:2020kzj,Mavromatos:2021urx,Mavromatos:2021sew,Basilakos:2020qmu,Basilakos:2019acj,Mavromatos:2022gtr}. \jtext{In  the recent work \cite{Gomez-Valent:2023hov}  a confrontation of the  stringy variant of RVM with the data is performed}.  The fact that a  QFT calculation and an effective string theory approach can lead to the same kind of RVM solution seems to indicate that the VED as a rigid concept is not very natural and that a dynamical evolution of the vacuum energy (density) should be more plausible.  In effect, it has been recently shown that this  can help significantly to improve the description of the overall cosmological  data and in particular opens a viable solution to the well-known tensions afflicting the $\CC$CDM, \jtext{see particularly the last phenomenological analyses\,\cite{SolaPeracaula:2021gxi,SolaPeracaula:2023swx}  which define the state-of-the-art results on the RVM}. They were preceded  by several other akin works, such as e.g. \cite{Gomez-Valent:2014rxa,Sola:2015wwa,Sola:2016jky,SolaPeracaula:2016qlq,Sola:2017znb,SolaPeracaula:2017esw,Gomez-Valent:2018nib,Gomez-Valent:2017idt}.  It is also interesting to remark that the RVM structure of the vacuum energy has been successfully tested against competing models (e.g. ghost models and holographic models of the DE) using cosmographical methods, which are essentially  model-independent --  see e.g.\cite{Rezaei:2022bkb,Rezaei:2021qwd,Rezaei:2019xwo} for details. The model has indeed passed a battery of different tests\,\cite{GomezValent:2017kzh,Perez:2021kdh} and the outcome is that the quality fit provided by the overall cosmological data  is comparable, actually better,  than that of the $\CC$CDM, if we attend to the verdict of the standard information criteria\,\cite{SolaPeracaula:2021gxi,SolaPeracaula:2023swx}. \jtext{It is also remarkable that well-known alternative theories to GR, such as Brans \& Dicke (BD) theories of gravitation\,\cite{Brans:1961sx}, turn out to mimic the RVM framework under appropriate conditions \cite{SolaPeracaula:2018dsw,deCruzPerez:2018cjx}. For example, adding a rigid cosmological term $\CC$ in the original BD theory one can show that it behaves as a running vacuum model when viewed from the GR perspective and one finds that the $H_0$ tension (as well as the $\sigma_8$ tension)  become significantly relieved, see \,\cite{SolaPeracaula:2019zsl,SolaPeracaula:2020vpg}.  The improvement is even more pronounced if the RVM structure is explicitly encoded in the BD framework\,\cite{deCruzPerez:2023wzd}.}

In this paper we continue the  task of computing the dynamics of $\rv(H)$  induced  by the quantum effects of the  quantized matter fields in Friedmann-Lemaître-Robertson-Walker (FLRW) spacetime, which two of us (CMP and JSP)  initiated in  previous works  \cite{Moreno-Pulido:2020anb,Moreno-Pulido:2022phq,Moreno-Pulido:2022upl}, see also \cite{Peracaula:2022vpx} for a comprehensive review.  The result of the present, more complete, calculation (involving for the first time the spin-1/2 fermionic contributions)  reconfirms that the combined dynamics of  the vacuum adopts the RVM form indicated below.  In these works, a new (off-shell) implementation  of the adiabatic regularization prescription (ARP) was used to compute the renormalized  $\rv$  for   a non-minimally coupled real scalar field. The method is based on a series expansion in the number of derivatives of the scale factor which introduces a hierarchy in some physical quantities evolving in a dynamical background\,\cite{Birrell:1982ix,Parker:2009uva,Fulling:1989nb}. Not only it was shown that the running of the VED was free from dangerous large contributions proportional to $m^4$ (quartic powers of the mass of the particles), but $\rv$ was shown to be mildly evolving with the Hubble rate and hence with the cosmic expansion. As a result,  if $t_1$ and $t_2$ are two particular values of the cosmic time, both close to the present, the corresponding values of $\rho_{\rm vac}$ are connected through the approximate relation
\begin{equation}\label{eq:RVMcanonicalLowEnerg}
\rho_{\rm vac}(H_2)-\rho_{\rm vac}(H_1)\approx \nu_{\rm eff}\,\mpl^2\left(H^2_2-H^2_1\right)\,,
\end{equation}
where $H_1\equiv H(t_1)$ and $H_2\equiv H(t_2)$ are the values of the Hubble function at times $t_1$ and $t_2$, respectively and $|\nueff|\ll1$ is a small parameter.  While the above relation is relevant for the (very mild) evolution of the VED in the current universe, the corresponding analysis of the early universe leads to a new mechanism of inflation called `RVM-inflation', which relies on the existence of quantum effects of  $6th$ adiabatic order, {\it i.e.} up to terms $\sim \mathcal{O}(H^6)$ which have been first accounted for in the case of scalar fields in  \cite{Moreno-Pulido:2022phq}.  Despite of the fact that the  family of Running Vacuum Models (RVM) has been in the literature for quite some time (cf. \cite{Sola:2013gha,Peracaula:2022vpx} and  \cite{Sola:2011qr,Sola:2015rra,Sola:2014tta}, and references therein), a full-fledged account based on QFT principles is much more recent\,\cite{Moreno-Pulido:2020anb,Moreno-Pulido:2022phq,Moreno-Pulido:2022upl}.

This work is tightly related to the preceding studies, in which the adiabatic regularization was applied to the  `simple' case of one real scalar field. It was  natural to perform the next step and check if spin-1/2 fermions do preserve the main conclusions derived from scalars, above all to verify if the corresponding vacuum fluctuations induce  also a running of the VED independent of the quartic powers of their masses and hence  remain also free from the traditional fine-tuning illness.  So the main goal of this paper is to extend the computations done for the scalar field, by considering the quantization of  spin-$1/2$ fermions in the FLRW background.  The extension proves rather non-trivial since we are in curved spacetime and the computations with fermions are no less involved  owing to the Fermi-Dirac statistics and the  peculiarities  inherent to the spinor calculus.  It is however reassuring to find that despite the many new technicalities involved in the computation they do not alter the main conclusions derived from the calculation with scalars.  The RVM form \eqref{eq:RVMcanonicalLowEnerg} at low energies  is once more attained, but the contribution to the coefficient $\nueff$ is, of course, different and involves non-trivial computational details.  Similarly, we compute the fermionic contribution to the  $\sim \mathcal{O}(H^6)$ terms which are involved in the RVM mechanism of inflation occurring in the very early universe. The final results concerning the renormalized VED can be obtained by considering  the contributions from an arbitrary number of quantized scalar  and fermion fields. We combine the two types of effects and present a final formula which we refer to as the bosonic and fermionic contributions to the VED, with the understanding that additional effects from gauge fields and their interactions with matter would be necessary in our calculation in a more complete approach. It is nevertheless  not necessary in our study since our fields are free, except for the non-minimal coupling of the quantized scalar fields with the external (non-quantized) gravitational field and the necessary spinorial affine connection of the fermion fields.  All that said, the computation of the free field contribution from bosons and fermions in curved spacetime is already a rather formidable task. So,  for the sake of a stepwise and clearer presentation,  we will address the fermionic contributions here on equal footing to the presentation of the scalar part performed in the preceding studies\,\,\cite{Moreno-Pulido:2020anb,Moreno-Pulido:2022phq}.

\vspace{0.25cm}
This work is structured as follows: In \hyperref[ZPEScalar]{Sec.\,\ref{ZPEScalar}}, we consider a quantized scalar field $\phi$ non-minimally coupled to curvature and  review  the computation of its  energy-momentum tensor  (EMT) and corresponding vacuum expectation value (VEV) induced by the vacuum fluctuations of that field. The original study of this part was made in\,\cite{Moreno-Pulido:2020anb,Moreno-Pulido:2022phq}. The  $00th$ component of the  VEV of the EMT constitutes the ZPE  of $\phi$.  We define also its associated  vacuum energy density (VED),  $\rv$,  and vacuum pressure, $P_{\rm vac}$.  All these quantities are unrenormalized at this point  and hence UV-divergent.  In the same section we review the off-shell adiabatic renormalization of the VED introduced  in the previous references, which involves as a distinctive feature  the subtraction of the on-shell EMT at an off-shell renormalization point $M$ up to $4th$ adiabatic order (in $4$ spacetime dimensions).   The main subject of this work is to address the corresponding calculation for spin-1/2 fermions and combine it with the scalar field case.  In \hyperref[sec:QuantizFermions]{Sec.\,\ref{sec:QuantizFermions}}, we review the quantization of a Dirac fermion in a curved background, the corresponding Dirac equation and its spinor solutions obtained from adiabatic expansion of the field modes.  The computation of the EMT and of its VEV for the case of a free quantized fermionic field in a spatially flat FLRW background is  performed in \hyperref[sec:ZPEFermions]{Sec.\,\ref{sec:ZPEFermions}}.  The off-shell adiabatic renormalization of the  EMT for spin-1/2 fermions is addressed and we extract  the renormalized  ZPE,  VED ($\rv$) and  vacuum pressure ($\Pv$)  in this context.  Additionally, some remarks on the trace anomaly and its role in our approach are discussed.  \hyperref[sec:CombinedBandF]{Sec.\,\ref{sec:CombinedBandF}} contains the combined results from all the quantized matter fields. Specifically,  we compute the renormalized VED for a system made of  an arbitrary number of quantized scalar fields non-minimally coupled to curvature (with different masses and non-minimal couplings) and an arbitrary number of quantized  spin-$1/2$ free  fermion fields.  In the same section we report on the corresponding running of the gravitational coupling $G=G(H)$, which goes hand in hand with the running of  $\rv(H)$ in order to preserve the Bianchi identity.  We also discuss the mechanism of `RVM-inflation' with the combined contribution from all these fields, and derive the equation of state (EoS) of the quantum vacuum for that system of quantized bosons and fermions fluctuating in it.   Remarkably, the vacuum EoS is \textit{no} longer equal to $\wv=-1$,  the reason being that  the vacuum pressure and the VED are \textit{not} exactly related in the usual manner (viz. $P_{\rm vac}=-\rv$)  since $P_{\rm vac}$ and $\rv$  are independent functions of the Hubble rate $H$ and its time derivatives owing to the quantum effects.   In the current universe, there is still some remnant of these quantum effects which induce a small (but potentially measurable) departure  making the quantum vacuum mimic quintessence or phantom.  \jtext{In Sec. \ref {sec:signatures} we discuss some phenomenological signatures  of the RVM, including a possible alleviation of the current tensions}. The conclusions are delivered in \hyperref[sec:Conclusions]{Sec.\,\ref{sec:Conclusions}} together with an additional discussion.
Finally, three appendices are included. In \hyperref[sec:appendixA]{Appendix\,\ref{sec:appendixA}}, we define our conventions and some useful formulas. The last two appendices,  \hyperref[sec:appendixB]{\ref{sec:appendixB}} and \hyperref[sec:appendixC]{\ref{sec:appendixC}}, are rather bulky since they collect a number of  cumbersome expressions related to the adiabatic expansion of the VEV of the EMT and the Fourier modes of the fermionic field (computed up to 6{\it th} order for the first time in the literature).

\section{Vacuum energy density of a non-minimally coupled scalar field}\label{ZPEScalar}
In this section, we summarize the results for the vacuum energy density and pressure  associated  with a quantized scalar field in FLRW spacetime as obtained in\,\cite{Moreno-Pulido:2020anb,Moreno-Pulido:2022phq}. In passing we introduce some notation which will be useful also for  the  fermionic calculation that will be subsequently reported.  The Einstein-Hilbert (EH) action for gravity plus matter reads
\begin{equation}\label{eq:EH}
S_{\rm EH+m}=  S_{\rm EH}+S_{\rm m}=\int d^4 x \sqrt{-g}\,\left(\frac{1}{16\pi G}\, R  - \rL\right) + S_{\rm m}\,.
\end{equation}
The  term $\rL$ has dimension of energy density and sometimes is called the vacuum energy density, but this is inaccurate in the formal QFT context since renormalization is necessary and  the physical vacuum energy density, $\rv$, is not just that term.  In fact,  $\rL$ is at this point just a bare parameter of the action, as the gravitational coupling  $G$ itself.  Varying the action with respect to the metric provides Einstein's equations
\begin{equation} \label{EinsteinEqs}
\frac{1}{8\pi G}G_{\mu \nu}=-\rho_\Lambda g_{\mu \nu}+T_{\mu \nu}^{\rm m}\,,
\end{equation}
with  $G_{\mu\nu}=R_{\mu\nu}-(1/2) g_{\mu\nu} R$  the usual Einstein tensor and  $ T_{\mu \nu}^{\rm m}$   the EMT  of matter\,\footnote{Our geometric conventions and other formulas  of interest for this calculation are collected in the \hyperref[sec:appendixA]{Appendix\,\ref{sec:appendixA}}.}:
\begin{equation}\label{eq:deltaTmunu2}
  T_{\mu \nu}^{\rm m}=-\frac{2}{\sqrt{-g}}\frac{\delta S_{\rm m}}{\delta g^{\mu\nu}}\,.
\end{equation}
The matter action $S_{\rm m}$ may contain a variety of contributions, including those from  incoherent matter, but it will be enough to focus on fundamental effects from quantized scalar and fermion fields.  Here we shall compute  the fermionic contribution to the VED.  But let us summarize first the situation with the scalar field  part. The latter  was dealt with in great detail in the two previous studies\,\cite{Moreno-Pulido:2020anb, Moreno-Pulido:2022phq}, in which the VED calculation was addressed under the assumption that no effective potential was present.  However, we admitted a non-minimal coupling of the scalar field to gravity. That calculation  in curved spacetime is already sufficiently demanding and in addition it furnishes  the universal part of the VED through  the zero-point energy (ZPE) effects in the curved background, see next section.    The classical action for a non-minimally coupled real scalar field is the following:
\begin{equation}\label{PhiAction}
S_\phi=-\int d^4x \sqrt{-g}\left(\frac{1}{2}g^{\mu\nu}\partial_\mu \phi \partial_\nu \phi+\frac{1}{2}\left(m_\phi^2+\xi R\right)\phi^2\right),
\end{equation}
where $\xi$ is the non-minimal coupling with gravity. It is well known that this action enjoys (local) conformal symmetry in the massless case with $\xi=1/6$.  However, the value of $\xi$ is not fixed in our computation and in general we do not assume the presence of such a symmetry.  Varying the above action with respect to the scalar field leads to the  Klein-Gordon (KG) equation with non-minimal coupling:
\begin{equation}\label{KG}
\left(\Box-m_\phi^2-\xi R\right)\phi^2=0,
\end{equation}
where $\Box\phi=g^{\mu\nu}\nabla_\mu\nabla_\nu\phi=(-g)^{-1/2}\partial_\mu\left(\sqrt{-g}\, g^{\mu\nu}\partial_\nu\phi\right)$. The corresponding EMT of  $\phi$  follows from the metric variation of the action \eqref{PhiAction} according to the  recipe \eqref{eq:deltaTmunu2}, and yields
\begin{equation}
\begin{split}\label{EMTScalarField}
T_{\mu\nu}(\phi ) &= -\frac{2}{\sqrt{-g}}\frac{\delta S_\phi}{\delta g^{\mu\nu}}=\left(1-2\xi\right) \partial_\mu \phi \partial_\nu \phi +\left(2\xi-\frac{1}{2}\right)g_{\mu \nu}\partial^\sigma \phi \partial_\sigma \phi\\
&-2\xi\phi \nabla_\mu \nabla_\nu\phi + 2\xi g_{\mu\nu}\phi \box \phi+\xi G_{\mu\nu}\phi^2-\frac{1}{2}m_\phi^2 g_{\mu\nu}\phi^2\,.
\end{split}
\end{equation}
As indicated, we perform the calculation in cosmological (FLRW) spacetime with flat three-dimensional metric. For convenience we use the conformal frame  $ds^2=a^2(\tau)\eta_{\mu\nu}dx^\mu dx^\nu$, with  $\eta_{\mu\nu}={\rm diag} (-1, +1, +1, +1)$  the Minkowski  metric in our conventions (cf.  \hyperref[sec:appendixA]{Appendix\,\ref{sec:appendixA}}),  $a(\tau)$ is the scale factor and  $\tau$  the conformal time. Differentiation with respect to $\tau$ will be denoted with a prime, so for example $\mathcal{H}\equiv a^\prime/a$  is the corresponding Hubble function in conformal time. We will perform the explicit calculations using the conformal metric  but our final results will eventually  be rendered in terms of the usual Hubble function $H(t)=\dot{a}/a$ in cosmic time $t$ (where a dot denotes differentiation with respect to $t$).  Recall that $d\tau=dt/a$ and hence  $\mathcal{H}=a H$.

If we switch on the quantum fluctuations of the $\phi$  field it is natural to consider the following decomposition:
 \begin{equation}
 \phi\left(\vec{x},\tau \right) = \phi_b (\tau)+\delta\phi \left(\vec{x},\tau \right),
 \end{equation}
in which  the background $\phi_b$ and the fluctuating part $\delta\phi$  are understood to be independent. In
particular, this is also the case for the corresponding Fourier decomposition in frequency modes. The decomposition of the fluctuating part reads
\begin{equation}\label{FourierDecomposition}
\delta \phi(\tau,{\bf x})=\frac{1}{(2\pi)^{3/2}a}\int d^3{k} \left[ A_\bk e^{i{\bf k\cdot x}} h_k(\tau)+A_\bk^\dagger e^{-i{\bf k\cdot x}} h_k^*(\tau) \right]\,,
\end{equation}
with the usual commutation relations for the creation and annihilation operators, $A_k$ and $A_k^\dagger$:
\begin{equation}
[A_\bk, A_\bk'^\dagger]=\delta({\bf k}-{\bf k'}), \qquad [A_\bk,A_ \bk']=0\,. \label{CommutationRelation}
\end{equation}
By using these relations, the KG-equation \eqref{KG} in terms of the  frequency modes can be put as
\begin{equation} \label{KGmodes}
h^{\prime\prime}_{k}+\Omega_k^2(\tau)h_k=0\,,
\end{equation}
where $\Omega_k^2\equiv k^2+a^2 m_\phi^2+\left(\xi-1/6\right)R$ and we recall that $\left(\right)^\prime \equiv d / d\tau \left(\right) $. The above differential equation does not possess a close analytic solution for the entire cosmological evolution. However, it can be solved by means of what is called an adiabatic series expansion, which is essentially a WKB-type solution.  First of all, it is necessary to introduce the following ansatz for the mode functions:
\begin{equation}\label{eq:phaseintegral}
h_k (\tau)=\frac{1}{\sqrt{2W_k (\tau)}}\exp \left(-i\int^\tau W_k (\tilde{\tau})d\tilde{\tau}\right).
\end{equation}
Notice that the modes are normalized through the Wronskian condition
\begin{equation}
 h_k^{} h_k^{*\prime}-h_k^\prime h_k^* = i\,, \label{WrosnkianCondition}
\end{equation}
which is essential to preserve the canonical commutation relations for the quantized field $\phi$.
By introducing the above ansatz into \eqref{KGmodes}, the function $W_k$ (effective frequency) is the solution of the (WKB-type)  non-linear differential equation
\begin{equation}\label{Non-LinDiffEq}
W_k^2=\Omega_k^2 -\frac{1}{2}\frac{W_k^{\prime \prime}}{W_k}+\frac{3}{4}\left( \frac{W_k^\prime}{W_k}\right)^2\,.
\end{equation}

\subsection{Zero-point energy and adiabatic expansion}\label{ZPEAdiabatic}

For a slowly varying effective frequency $\Omega_k (\tau)$ one can proceed to solve this equation perturbatively with the help of an asymptotic series which can be  organized through adiabatic orders. This approach constitutes the basis for the aforementioned ARP\,\cite{Parker:1969au,Parker:1971pt,Parker:1974qw, Fulling:1974zr, Fulling:1974pu,Christensen:1978yd, Birrell1978,Bunch:1979uk,Bunch:1980vc}, see also   \cite{Agullo:2009zi,Agullo:2011qg,BarberoG:2018oqi,Landete:2013axa,Landete:2013lpa,delRio:2014cha}
 and \cite{Kohri:2016lsj,Kohri:2017iyl,Ferreiro:2018oxx,Ferreiro:2021pec,Ferreiro:2022hik} for more recent applications and extensions, and  the textbooks \,\cite{Birrell:1982ix,Parker:2009uva,Fulling:1989nb} for a more systematic presentation.

The quantities $k^2$ and $a$ are taken to be of adiabatic order 0. Of adiabatic order 1 are: $a^\prime$ and $\mathcal{H}$. The quantities $a^{\prime \prime},a^{\prime 2},\mathcal{H}^\prime$ and $\mathcal{H}^2$ and linear combinations are taken of adiabatic order 2. It can be summarized by saying that each time derivative increases one unit the adiabatic order.
So, the ansatz solution for $W_k$ can be written as an adiabatic series expansion:
\begin{equation}
W_k=W_k^{(0)}+W_k^{(2)}+W_k^{(4)}+W_k^{(6)}+\cdots,
\end{equation}
in which the superscript indicates the adiabatic order. We note that only even orders are allowed, which is justified from the general covariance of the result  since only tensors of even adiabatic order can be present in the effective action and the field equations. Explicit calculation indeed corroborates the absence of the odd adiabatic orders.
The ``seed'' to initiate the adiabatic series (i.e. the zeroth order contribution to $W_k$)  is
\begin{equation}
W_k^{(0)}\equiv \omega_k = \sqrt{k^2+a^2M^2},
\end{equation}
where $M$ is an arbitrary off-shell scale. Nothing enforces us to take the mass of the particle at this point, we only need to preserve the adiabaticity of the expansion. The floating quantity $M$ will play the role of renormalization scale, as it will be seen.  In fact, as it was shown in \cite{Moreno-Pulido:2022phq}, this parameter can also be used as the renormalization scale in the DeWitt-Schwinger expansion\cite{DeWitt:1975ys} of the vacuum effective action $W_{\rm eff}$\,\cite{Birrell:1982ix}, to wit: the effective action obtained from integrating out the vacuum fluctuations of the quantized matter fields.  From the explicit expression of $W_{\rm eff}$ one can also derive the VEV of the  EMT -- denoted $\left\langle T_{\mu\nu}^{\delta \phi} \right\rangle$ -- by computing its metric functional derivative as follows:
\begin{equation}\label{eq:DefWeff}
\langle T_{\mu\nu}^{\delta \phi}\rangle=-\frac{2}{\sqrt{-g}} \,\frac{\delta W_{\rm eff}}{\delta g^{\mu\nu}}\,.
\end{equation}
This formula is of course similar  to Eq.\,\eqref{eq:deltaTmunu2}, but for the vacuum effective action. This alternative method provides exactly the same result as the WKB expansion of the field modes, as  outlined below,  and it was illustrated in great detail in  \cite{Moreno-Pulido:2022phq} for the case of the quantized  scalar fields.  Such a cross-check in the determination of the VEV of the EMT involves a significant amount of calculations and provides a nontrivial validation of the entire renormalization procedure. The same holds good for the case of fermions but we shall not present the details of the DeWitt-Schwinger method for fermions here.

Next we summarize the mode expansion for scalar fields. By introducing $W_k^{(0)}$ given above in the \textit{r.h.s.} of \eqref{Non-LinDiffEq},  the terms of adiabatic order 2 can be collected to find the next term in the series $W_k^{(2)}$, with the result
\begin{equation}
W_k^{(2)}=\frac{a^2 \Delta^2}{2\omega_k}+\frac{a^2 R}{2\omega_k}\left(\xi-\frac{1}{6}\right)-\frac{\omega_k^{\prime\prime}}{4\omega_k^2}+\frac{3\left(\omega_k^{\prime}\right)^2}{8\omega_k^3}.
\end{equation}
Here $\Delta^2 \equiv m^2-M^2$ is the difference between the quadratic mass of the field and that of  the off-shell scale, and is of adiabatic order $2$ because it is necessary for renormalization. Loosely speaking,  since $M^2$ and $\Delta^2$ appear together in the expansion they need to be of different adiabatic order so as  to obtain a  consistent adiabatic expansion exploring the off-shell regime.   Since $M^2$ is of adiabatic order $0$,  the next-to-leading order  for $\Delta ^2$ to be compatible with general covariance is precisely order $2$. This fact is reconfirmed on using  the aforementioned DeWitt-Schwinger expansion\,\cite{Moreno-Pulido:2022phq}.

Introducing  $W_k^{(2)}$ on  the \textit{r.h.s.} of \eqref{Non-LinDiffEq} we can  obtain $W_k^{(4)},\dots$ and so on.  The higher order  adiabatic terms  become progressively more and more cumbersome since the number of terms involved in the expansion becomes larger and larger.  In our case we reach up to order $6$, i.e. we compute the series up to $W_k^{(6)}$.  This was done for the first time in the literature in \cite{Moreno-Pulido:2022phq} for scalar fields,  and we will also be done  here to order $W_k^{(6)}$ for the first time for fermions, see \hyperref[sec:ZPEFermions]{Sec.\,\ref{sec:ZPEFermions}}.  Notice that attaining the order $W_k^{(6)}$ is indispensable in order to study RVM-inflation in the early universe\cite{Moreno-Pulido:2022phq}. Extensive use of Mathematica\,\cite{Mathematica} has been made to handle these bulky calculations.

Once obtained the expansion of $W_k$  we can  compute the mode functions $h_k$ and other physical quantities such as the EMT to the suitable order, in particular  the EMT trace, which can be used to compute the vacuum pressure (see below). The technical details for the scalars are not going to be repeated here\footnote{They are  provided in \cite{Moreno-Pulido:2020anb, Moreno-Pulido:2022phq}. Analogous computations will be reported in \hyperref[sec:QuantizFermions]{Sec.\,\ref{sec:QuantizFermions}} and \hyperref[sec:ZPEFermions]{Sec.\,\ref{sec:ZPEFermions}} for the case of the fermion field.}, but it is important to remark that these quantities present divergent terms up to $4th$ adiabatic order (in $4$-dimensional spacetime). The approach we adopt here is the same as that which was proposed and  amply tested in \cite{Moreno-Pulido:2020anb, Moreno-Pulido:2022phq}, namely we define the renormalized vacuum expectation value (VEV) of the EMT by taking the on-shell value (at an arbitrary adiabatic order) and subtracting from it the divergent orders at an arbitrary scale, which we denote $M$:
\begin{equation}\label{RenormalizedEMTScalar}
\left\langle T_{\mu\nu}^{\delta\phi} \right\rangle_{\rm ren} (M)=\left\langle T_{\mu\nu}^{\delta\phi} \right\rangle (m_\phi)-\left\langle T_{\mu\nu}^{\delta\phi}\right\rangle^{(0-4)}(M).
\end{equation}
Here the superscript $(0-4)$  refers to the UV-divergent subtracted orders, i.e. from $0th$ up to $4th$ adiabatic order, all of them being UV-divergent (the higher  adiabatic orders being all finite in $n=4$ spacetime).   Notice that for $M=m_\phi$  the above definition provides  the natural generalization of the subtraction of divergent constants performed to obtain finite results on trivial backgrounds (such as Minkowski spacetime). However, in curved backgrounds the mode by mode  subtraction implied in the above prescription is not just a constant term; and moreover for arbitrary $M$ the corresponding renormalized result allows us to test the evolution of the VED with the scale $M$.  As previously noted, this feature obviously offers a floating scale which is characteristic of the renormalization group (RG)  analysis in cosmology\,\cite{Sola:2013gha,Peracaula:2022vpx}. Let us clarify, however,  that we distinguish $M$  from 't Hooft's mass unit $\mu$ in dimensional regularization (DR), which will not be used in this work at any point, although it can  be invoked as an intermediate regularization procedure (not at all for renormalization though) if one likes\cite{Moreno-Pulido:2020anb, Moreno-Pulido:2022phq}.  The parameter $\mu$ is unphysical and is used in the minimal subtraction scheme (MS) with DR to define the renormalization point\,\cite{Collins1984}.  We should emphasize that we do not use MS at all in the present work, although one could make  (optional) use of DR in intermediate steps.   In these cases, the quantity $\mu$  always cancels  out and the final renormalized expressions depend on $M$ only, as it is the case e.g. of the  effective action of vacuum.  But the full effective action (which involves the classical and quantum parts)  is of course independent of $M$ as well, as the running of the couplings exactly  compensates for the explicit $M$-dependence of the quantum effects. This is, of course, the standard lore of the RG program, see \cite{Moreno-Pulido:2022phq} for detailed considerations along these lines and making use  of the effective action.

We refrain from writing out the unrenormalized expression for the EMT in the case of scalar fields,  see \cite{Moreno-Pulido:2020anb, Moreno-Pulido:2022phq} for full details.  Let us however quote the renormalized result emerging from the ARP prescription \eqref{RenormalizedEMTScalar}. Expressing the final result in terms of the cosmic time and the corresponding Hubble function $H=\dot{a}/a$, we find for the ZPE the following result:
\begin{equation}\label{eq:T00Integrated}
\begin{split}
&\left\langle T_{00}^{\delta \phi}\right\rangle_{\rm ren}(M)
=\frac{a^2}{128\pi^2}\left(-M^4+4m_\phi^2M^2-3m_\phi^4+2m_\phi^4\ln \frac{m_\phi^2}{M^2}\right)\\
&- \left(\xi-\frac{1}{6}\right)\frac{3 a^2 H^2}{16\pi^2}\left(m_\phi^2-M^2-m_\phi^2\ln \frac{m_\phi^2}{M^2}\right)\\
&+\left(\xi-\frac{1}{6}\right)^2\frac{9a^2}{16\pi^2}\left(6H^2\dot{H}+2H\ddot{H}-\dot{H}^2\right)\ln \frac{m_\phi^2}{M^2}+\mathcal{O}\left(\frac{H^6}{m_\phi^2}\right),
\end{split}
\end{equation}
and similarly for the VEV of its trace
\begin{equation}%\label{eq:TraceIntegrated}
\begin{split}
&\left\langle T^{\delta \phi}\right\rangle_{\rm ren}(M)
=\frac{1}{32\pi^2}\left(3m_\phi^4-4m_\phi^2M^2+M^4-2m_\phi^2\ln \frac{m_\phi^2}{M^2}\right)\\
&+\left(\xi-\frac{1}{6}\right)\frac{3}{8\pi^2} \left(2H^2+\dot{H}\right) \left(m_\phi^2-M^2-m_\phi^2\ln \frac{m_\phi^2}{M^2}\right)\\
&-\left(\xi-\frac{1}{6}\right)^2\frac{9}{8\pi^2}\left(12H^2 \dot{H}+4\dot{H}^2+7H\ddot{H}+\vardot{3}{H}\right)\ln \frac{m_\phi^2}{M^2}+\mathcal{O}\left(\frac{H^6}{m_\phi^2}\right)\,.
\end{split}
\end{equation}
We used the notation $\mathcal{O}(H^6 / m_\phi^2)$ to collectively refer to the terms of adiabatic order 6 (consisting of 6 time derivatives of the scale factor). It may include terms such as  $H^6/m_\phi^2$, but also many other combinations such as  $\dot{H}^3/m_\phi^2$, $H^2\vardot{3}{H}/m_\phi^2,\dots$   There are actually many terms of this sort and they have been reported explicitly  in \cite{Moreno-Pulido:2022phq}.  We refrain from writing them down again here and invite the reader to check the aforementioned paper for more details.  We will report explicitly on the $6th$-order adiabatic terms only in the case of fermions (cf. \hyperref[sec:ZPEFermions]{Sec.\,\ref{sec:ZPEFermions}})  since these are computed for the first time in this work.

\subsection{Renormalized vacuum energy  and vacuum pressure}\label{RenZPEscalar}

We are now ready to compute the vacuum EMT, $\left\langle T_{\mu\nu}^{\rm vac} \right\rangle$,  which will lead us to the VED, $\rv$, and vacuum's pressure, $\Pv$.  As in \cite{Moreno-Pulido:2022phq}, we write the vacuum EMT as the sum of the renormalized parameter $\rL$ and  the renormalized VEV of the EMT, which embodies the finite form of the adiabatically renormalized vacuum fluctuations:
\begin{equation}
\label{VacuumEMT}
\left\langle T_{\mu\nu}^{\rm vac} \right\rangle=-\rho_{\Lambda} (M) g_{\mu\nu}+\left\langle T_{\mu\nu}^{\rm \delta\phi} \right\rangle_{\rm ren}(M)\,.
\end{equation}
Since the vacuum is expected to be a most symmetric state free of any  new parameter, this expression must take on the form of a perfect fluid:
$\langle T_{\mu\nu}^{{\rm vac}}\rangle=\Pv g_{\mu \nu}+\left(\rv+\Pv\right)u_\mu u_\nu$,
where $u^\mu$ is the $4$-velocity ($u^\mu u_\mu=-1$). In conformal coordinates, in  the comoving (FLRW)  frame, $u^\mu=(1/a,0,0,0)$ and $u_\mu=(-a,0,0,0)$. Taking the $00 th$ and $ii th$-component  (any $i=1,2,3$  is good owing to isotropy, so we take $i=1$)  one finds the precise form of the vacuum  energy density and pressure\,\cite{Moreno-Pulido:2022phq}:
\begin{equation}\label{ScalarFieldVED}
\rho_{\rm vac}(M)=\frac{\left\langle T_{00}^{\rm vac}\right\rangle}{a^2}=\rho_{\Lambda}(M)+\frac{\left\langle T_{00}^{\rm \delta\phi} \right\rangle^{\rm ren}(M)}{a^2}\,,
\end{equation}
\begin{equation}\label{ScalarFieldPressure}
\begin{split}
P_{\rm vac}(M)&=\frac{\left\langle T_{11}^{\rm vac}\right\rangle}{a^2}=-\rho_{\Lambda}(M)+\frac{\left\langle T_{11}^{\rm \delta\phi} \right\rangle_{\rm ren}(M)}{a^2}=-\rho_{\Lambda}(M)+\frac{1}{3}\left(\left\langle T^{\rm \delta\phi} \right\rangle_{\rm ren}(M)+\frac{\left\langle T_{00}^{\rm \delta\phi} \right\rangle_{\rm ren}(M)}{a^2}\right)\\
&=-\rho_{\rm vac}(M)+\frac{1}{3}\left(\left\langle T^{\rm \delta\phi} \right\rangle_{\rm ren}(M)+4\frac{\left\langle T_{00}^{\rm \delta\phi} \right\rangle_{\rm ren}(M)}{a^2}\right)\,,
\end{split}
\end{equation}
where  isotropy allows to express the result in terms of  the trace $T^{\rm \delta\phi} $ of the fluctuating part, if desired.
Notice that  $\rho_\Lambda (M)$ in the above expressions is the renormalized form of the corresponding bare parameter appearing in the EH action \eqref{eq:EH} and it has units of energy density. The VED, however,  is not just this renormalized parameter but the renormalized sum \eqref{ScalarFieldVED}.  Although is tantalizing to call  $\rho_\Lambda (M)$  ``the CC density'', and in fact this has been common in the literature (especially when the discussion is strictly classical without considering quantum effects),  this is not strictly correct since  the physical CC is not simply  $8\pi G\rho_\Lambda$ but $8\pi G\rv$, that is, the physical vacuum energy density is connected with the physical $\CC$ through  $\rho_{\rm vac}=\Lambda/(8\pi G)$.  The parameter $\CC$ which is measured in the observations is indeed defined  through this expression, which  is precisely computable in QFT from Eq.\,\eqref{ScalarFieldVED}.  We shall show once more for fermions (as we did for scalar fields in  the previous works \,\cite{Moreno-Pulido:2020anb, Moreno-Pulido:2022phq}), that the adiabatically renormalized form of the running VED is free from the huge $\sim m^4$ contributions that are usually attributed to the VED in other (inappropriate) renormalization schemes, and therefore the renormalized expression that we will obtain can be perfectly consistent with the measured $\CC$.  To be sure,  it is not our aim to predict this value  but rather to show that the theoretical formula points naturally to a value as small (in natural units) as measured by the observations.

A simple way to condense these ideas is to say that the VED is related with the ZPE and $\rho_\Lambda$ as follows: ``${\rm VED}=\rho_\Lambda+{\rm ZPE}$'', i.e. Eq.\,\eqref{ScalarFieldVED}.   Parameter $\rho_\Lambda$ is initially just a bare coupling in the effective action and it has no direct phenomenological interpretation, not even after renormalization.  On the other hand, the ZPE embodies the quantum fluctuations of the massive fields and calls also for renormalization since it is originally UV-divergent. The physical VED in this context is then the renormalized sum of these two contributions, and it can not be split apart since the separate terms make no sense in an isolated way.  Observations are sensitive only to the sum.  Furthermore, as we shall see explicitly for the fermionic case, there is a crucial cancellation of the quartic mass terms when we consider the evolution of the sum  $\rho_\Lambda+{\rm ZPE}$ as a function of the renormalization point, which does not occur if  the two terms are dealt with separately.  This was already pinpointed for the case of scalar fields in\,\cite{Moreno-Pulido:2020anb, Moreno-Pulido:2022phq,Moreno-Pulido:2022upl} .

With these provisos, the expression  for the  VED of the scalar field can be obtained. Notwithstanding, the final renormalized result still requires a  physical interpretation since it depends on the renormalization scale $M$.  In fact, recall that the result depends on both the values of $M$ and $H$ (and corresponding time derivatives), which are  independent arguments.  The scale $M$ can not be left arbitrary at this point since we wish to provide an estimate of the VED at a given expansion epoch.  As previously indicated, the vacuum effective action $W_{\rm eff}$ is explicitly dependent on $M$ despite the full effective action is of course RG-independent. Thus, following ,\cite{Moreno-Pulido:2020anb, Moreno-Pulido:2022phq}  an adequate choice of the renormalization point  $M$ is to select it equal to the value of $H$ at the epoch under consideration. This  corresponds to choose the RG scale around the characteristic energy scale of  FLRW spacetime at any given moment, and hence it should have physical significance.  In actual fact this is in analogy with the standard practice in ordinary gauge theories, where the choice of  the renormalization group scale  is  made near the typical energy of the process.  In what follows we derive the `low energy'  form  of the VED  along these lines. This is actually the form that applies for the current universe.  Subsequently we will focus on the running gravitational coupling $G(M)$ and its relation with the running $\rv(M)$.

We should also point out that the reach of our considerations concerns the calculation of the  evolution (or `running') of the VED only, rather than predicting its current value.  Given that value, however,  we can predict how it evolves with $H$ around our epoch, or any other epoch.  Now in the absence of an observational input at some expansion epoch $H(t)$  we cannot  compute $\rho_{\rm vac}(H)$ at other values  of $H$ (i.e. at other epochs of the cosmic evolution). To compute the value of the VED at present is out of the scope of the renormalization program since the latter is based on the RG  flow, which requires a boundary condition. This is exactly the same situation as in any renormalization calculation, we need the input values of the parameters at one scale to predict some observable  (e.g. a cross-section) at another scale.  The truly relevant feature of our calculational approach, as it should be clear at this point, is that the RG-flow is completely smooth. It  only depends on the evolution of $H$ and is completely free from spurious effects associated with the large contributions from the quartic masses of the fields. These quartic terms are the traditional  kind of undesirable effects which spoil the  physical interpretation of the renormalization program concerning the CC and the VED. They  are typically found in calculations of the VED whose renormalization is based on, say, the  MS scheme. Most existing approaches to the CC problem in the literature exhibit this unwanted feature, which is already at the basis of the Minkowskian calculation and  is of course unacceptable for a realistic description of the VED in curved spacetime\,\cite{Peracaula:2022vpx}.   A similar situation is found in Schwarzschild and in de Sitter backgrounds, see e.g.\,\cite{Firouzjahi:2022vij,Firouzjahi:2022xxb}.

Bearing in mind  the above considerations, the final result for the running  VED at low energies (specifically the part triggered by the quantized scalar fields), can  be best written in terms of the evolution between two expansion history times. It is natural that we choose the current epoch (characterized by the value $H_0$ of the Hubble parameter) and relate it with the value of the VED at a nearby epoch $H$ of the cosmic evolution\footnote{In this context, a nearby epoch does not necessarily mean that it is very close to the current epoch,  it rather refers to any cosmic span for which the VED running is still driven by the $\sim H^2$ terms and not by higher powers.  The higher powers are only relevant for the very early universe, namely during the inflationary time, and hence the low-energy formula applies virtually to any (post-inflationary) epoch. For more details, see \,\cite{Moreno-Pulido:2022phq}.}.  The approximate  final result can be rendered in the following  very compact form\,\cite{Moreno-Pulido:2022phq}:
\begin{equation}\label{VacuumEnergyDensityScalarField}
\rho_{\rm vac}(H)=\rho_{\rm vac}^0+\frac{3\nu_{\rm eff}}{8\pi}\left(H^2-H_0^2\right)\mpl^2,
\end{equation}
where
\begin{equation}\label{eq:nueffBososns}
\nu_{\rm eff}\equiv \frac{1}{2\pi}\left(\xi-\frac{1}{6}\right)\frac{m_\phi^2}{\mpl^2}\ln \frac{m_\phi^2}{H_0^2}\,,
\end{equation}
 $\rho_{\rm vac}^0 \equiv \rho_{\rm vac} (H_0)$ being the current value of the VED (accessible by observations) and  $H_0$   today's value of the Hubble function. It is necessary to remark that $\nu_{\rm eff}$ is an effective parameter expected to be small due to its proportionality to $m_\phi^2 / \mpl^2$.  Remarkably, the above dynamical form of the VED turns out to adopt the canonical RVM form, see \,\cite{Sola:2013gha,Peracaula:2022vpx} and references therein.  Phenomenological studies based on fitting the above RVM formula to the overall cosmological data  indeed provide an estimate for $\nueff$ at the level of  $\nueff \sim 10^{-5}-10^{-3}$ depending on the model\,\cite{SolaPeracaula:2021gxi,SolaPeracaula:2023swx}, see also \cite{Gomez-Valent:2014rxa,Sola:2015wwa,Sola:2016jky,SolaPeracaula:2016qlq,Sola:2017znb,SolaPeracaula:2017esw}. This phenomenological determination turns out to lie in the ballpark estimate of the theoretical expectations\,\cite{Sola:2007sv}. The order of magnitude is reasonable if we take into account that the masses involved here pertain of course to the scale of a typical Grand Unified Theory (GUT) where, in addition, a large factor must be included to account for the  large multiplicity of heavy particles.  In \hyperref[sec:CombinedBandF]{Sec.\,\ref{sec:CombinedBandF}} we provide a more general formula where an arbitrary number of species of bosons and fermion fields are included. It is worth noticing that the order of magnitude of $\nueff$ picked out in the mentioned study is perfectly compatible with the result recently obtained from the Big Bang nucleosynthesis (BBN)  bound in \cite{Asimakis:2021yct}, although in the latter case the bound was not sensitive to the sign of $\nueff$.

On the other hand, the computation of the pressure in an analogous way (we refrain from providing more details on  the scalar field contribution, see once more \cite{Moreno-Pulido:2022phq} and \cite{Moreno-Pulido:2022upl}  for a full-fledged account)  enables us to write an explicit expression for the equation of state (EoS) of the vacuum\,\cite{Moreno-Pulido:2022upl}. The leading expression for the current universe is the following:
\begin{equation}
w_{\rm vac}=\frac{P_{\rm vac}(H)}{\rho_{\rm vac}(H)} \approx -1-\nu_{\rm eff}\frac{\dot{H}\mpl^2}{4\pi \rho^0_{\rm vac}}\,.
\end{equation}
For very low redshift  $z$ and in terms of the current cosmological parameters $\Omega^0_i=\rho^0_i/\rho^0_c=8\pi G_N\rho^0_i/(3H_0^2)$ the above expression reduces to
\begin{equation}\label{EqStateScalar}
w_{\rm vac}(z)\approx -1+\nu_{\rm eff}\frac{\Omega_{\rm vac}^0}{\Omega_{\rm m}^0}(1+z)^3.
\end{equation}
This result is especially worth emphasizing since it predicts a small departure  from -1 which could perhaps be measured around the present time. Recall that the traditional value of the EoS of the  Cosmological Constant is just $-1$.  The above result implies that the quantum vacuum receives small quantum effects which trigger a departure of its EoS from $-1$.   For  instance, if we adopt the positive sign for $\nueff$, as  obtained in most cases from the latest fitting analysis to a large set of different kinds of observational data\,\cite{SolaPeracaula:2021gxi,SolaPeracaula:2023swx},  then  Eq.\,\eqref{EqStateScalar} predicts that the vacuum energy behaves as quintessence around the current time.

As noted, the EoS formula \eqref{EqStateScalar} is valid only for small values of the redshift $z$, but one can show that the departure is even bigger in the past, adopting a kind of chameleonic behavior by which the EoS of the quantum vacuum tracks the EoS of matter at high redshifts, see\,\cite{Moreno-Pulido:2022upl} and  \hyperref[sec:EoS-QVacuum]{Sec.\,\ref{sec:EoS-QVacuum}} for more details. All in all, these  results have been predicted from first principles, namely from explicit QFT calculations in the FLRW background. In particular, the fact that the quantum vacuum  may currently mimic quintessence is remarkable since the result does not rely on ad-hoc fields or on any other phenomenological ansatz.

\section{Quantization of a spin-1/2 fermion field in curved spacetime}\label{sec:QuantizFermions}
As pointed out  in the introduction, the main goal of this work is to extend the QFT results for the VED obtained for quantized scalar fields, which we have summarized in the previous section, to the case of  quantized spin-$1/2$ Dirac fermion fields and then combine the two types of contributions in closed form.  The calculation of the renormalized VED  for free spin-1/2 fermions is also nontrivial and rather cumbersome, and requires a devoted study, which we present here  (see also the appendices provided at the end for bulky complementary details).  While  the QFT treatment is analogous to the case of scalars,  the specific technicalities are  quite different and no less intricate, but fortunately the final result proves to be in consonance with the one previously derived for the scalars,  so it is perfectly possible to furnish a close form for the combined contribution to the VED involving an arbitrary number of non-interacting scalar and spin-1/2 fermion fields, cf. \hyperref[sec:CombinedBandF]{Sec.\,\ref{sec:CombinedBandF}}.

The study of the solutions of the Dirac equation in curved spacetime goes back to the works from many decades ago by Fock, Tetrode, Schr\"{o}dinger, McVittie, Bargmann,   Wheeler and others: see e.g. \cite{Parker:1980kw,Barut:1987mk,Villalba:1990bd,Finster:2009ah,Pollock:2010zz}, where the relevant historical references are given  and different aspects of spin-1/2 fermions in curved spacetime are studied, including  a detailed account  for the solutions in  FLRW spacetime -- see also the review \cite{Collas:2018jfx}, with a rather complete list of references.  On the other hand, the  subject of adiabatic regularization for fermions has been previously treated in the literature in different applications,  see e.g.  \cite{Christensen:1978yd} as well  as the more recent papers\,\cite{Landete:2013axa, Landete:2013lpa, delRio:2014cha, BarberoG:2018oqi} where  emphasis is made on exact solutions e.g. in de Sitter spacetime. The calculation of the renormalized VED in FLRW spacetime is,  however,  more complicated for it does not admit an exact solution.  Our strategy to circumvent this problem   is based on using an off-shell variant of the ARP framework\,\cite{Moreno-Pulido:2020anb, Moreno-Pulido:2022phq} which leads to the RVM  behavior of the vacuum energy\,\cite{Sola:2013gha,Peracaula:2022vpx}. The RVM framework has proven rather successful in mitigating the cosmological tensions\,\cite{Perivolaropoulos:2021jda,Abdalla:2022yfr,Dainotti:2023yrk}, as shown in  different phenomenological analyses, such as\cite{SolaPeracaula:2021gxi,SolaPeracaula:2023swx} and previous works such as \cite{Gomez-Valent:2018nib,Gomez-Valent:2017idt}.   On the theoretical side, attempts at computing the VED with other procedures has led to the traditional calamity with the quartic powers of the masses.  Here we will show that using the off-shell ARP to tackle the VED contribution from fermions generates a result which is free from these difficulties and fully along the lines of what has been obtained for the scalar fields in the previous sections and originally in\,\cite{Moreno-Pulido:2020anb, Moreno-Pulido:2022phq}.  Therefore, the combined contribution from fermions and scalar fields to the VED is compatible with a smooth running of the cosmological vacuum energy  and is consistent with the aforementioned phenomenological analysis of the RVM as a possible solution to the cosmological tensions.

Since it will be necessary a considerable amount of formalism to treat  fermions within the adiabatic approach, it is  convenient  to summarize first the necessary aspects of that formalism  before we can put forward our main results concerning their contribution to the vacuum energy density. It  will be useful to fix some notation as well.  Once more we perform the calculations  in FLRW spacetime with flat three-dimensional metric.
Consider a free Dirac spin-$1/2$ field, described by the  four-component spinor $\psi$. In our conventions,  the Dirac action in curved spacetime is given by
\begin{equation}\label{FermionAction}
S_\psi(x)=-\int d^4x \sqrt{-g}\left[\frac{1}{2}i\left( \Bar{\psi}\underline{\gamma}^\mu \nabla_\mu \psi -\left(\nabla_\mu \Bar{\psi}\right)\underline{\gamma}^\mu \psi \right)+m_{\psi} \Bar{\psi} \psi\right]\,.
\end{equation}
In the above expression,  $m_{\psi}$ denotes the mass of the Dirac field and   $\Bar{\psi}\equiv \psi^\dagger \gamma^0$  the adjoint spinor.  Since we are in a curved background,  the partial derivative of a spinor $\partial_\mu\psi$  has been replaced with the corresponding covariant derivative $\nabla_\mu\psi$, which is defined below.  Moreover,  gamma matrices in curved spacetime are also needed, they are sometimes indicated (as above) with an underline to distinguish them from the Minkowski space gamma matrices. The former are  $\underline{\gamma}^\mu (x)$ (which are generally functions of the coordinates) whereas the latter are the constant matrices ${\gamma}^\alpha$ in flat spacetime. As it is well-known, to obtain a representation for the curved spacetime gamma matrices in terms of the Minkowskian gamma matrices we need  to introduce the local tetrad or  vierbein field (in $4$-dimensional spacetime)  $e^{\,\mu}_\alpha$. It is defined in each tangent space of the spacetime manifold and relates the curved spacetime metric with the Minkowskian one in the usual way:  $g^{\mu\nu}(x)=e^{\mu}_{\,\alpha}(x) e^{\nu}_{\beta}(x) \eta^{\alpha \beta}$,  where $\eta_{\alpha\beta}$ is the Lorentz metric in the local inertial frame specified by the normal coordinates at the given spacetime point.  The general relation between the  two sorts of gamma matrices is $\underline{\gamma}^\mu(x)=e^{\mu}_{\,\alpha}(x)\gamma^\alpha $.  Specifically, in a spatially flat FLRW spacetime the vierbein in conformal coordinates is  $e^{\mu}_{\alpha}={\rm diag}\left(1/a(\tau), 1/a(\tau), 1/a(\tau), 1/a(\tau)\right)$ where $a(\tau)$ is the scale factor as a function of the conformal time.  Whence the  gamma matrices in this background are  time-dependent and  related to the constant flat spacetime ones as follows:  $\underline{\gamma}^\mu(\tau)=\gamma^\mu /a(\tau)$.
This relation insures that they satisfy the following anti-commutation relations:
\begin{equation}\label{dirac matrices relation}
\left\{ \underline{\gamma}^\mu, \underline{\gamma}^\nu \right\} =-2g^{\mu\nu}\mathbb{I}_4\,,
\end{equation}
provided,  of course,  the (constant) flat space gamma matrices satisfy $\left\{ {\gamma}^\alpha, {\gamma}^\beta \right\} =-2\eta^{\alpha\beta}\mathbb{I}_4$.
In order to obtain the equation of motion, i.e. the covariant Dirac equation in curved spacetime, one has to vary the covariant action \eqref{FermionAction} with respect to the spinor field, giving
\begin{equation}\label{CovDiracEq}
i \underline{\gamma}^\mu \nabla_\mu \psi+m_\psi \psi=i {e}^\mu_\alpha \gamma^\alpha \nabla_\mu \psi+m_\psi \psi=i \frac{1}{a}\left( \gamma^\alpha\partial_\alpha-  \gamma^\alpha\Gamma_\alpha\right) \psi+m_\psi \psi=0\,.
\end{equation}
The covariant derivative is defined through the spin connection, $\nabla_\mu \equiv \partial_\mu-\Gamma_\mu$.  The spinorial affine connection $\Gamma_\mu$  satisfies the equation\,\cite{Barut:1987mk}
\begin{equation}\label{eq:spinconnection}
\left[\Gamma_\nu, \underline{\gamma}^\mu(x)\right]=\frac{\partial \underline{\gamma}^\mu(x)}{\partial x^\nu}+\Gamma^\mu_{\nu\rho} \underline{\gamma}^\rho(x)\,,
\end{equation}
where $\Gamma^\mu_{\nu\rho}$ are the  Christoffel symbols.   The above equation is tantamount to require the vanishing of the covariant derivative of the curved space gamma matrices:  $\nabla_\nu \underline{\gamma}^\mu(x)=0$\,\cite{Parker:2009uva}, i.e. the curved-space gamma matrices are defined to be covariantly constant over the spacetime manifold.
Using the Christoffel symbols  in the conformally flat FLRW metric as given in \hyperref[sec:appendixA]{Appendix\,\ref{sec:appendixA}}, the explicit solution of Eq.\,\eqref{eq:spinconnection}  can be found, with the following result:  $\Gamma_0=0, \Gamma_j=-\left(\mathcal{H}/2\right)\gamma_j\gamma_0=-\left(a'/2a\right)\gamma_j\gamma_0$. Therefore,   $\gamma^\alpha\Gamma_\alpha=3(a'/2a)\gamma_0=-3(a'/2a)\gamma^0$. This expression can then  be inserted in Eq.\,\eqref{CovDiracEq}.

In this way we have obtained an explicit form for the Dirac equation in  FLRW spacetime with spatially flat metric. We are now in position to attempt a solution by  expanding the quantized fermion field  in mode functions:
\begin{align}
	\psi(x)=\int d^3k\sum_{\lambda=\pm 1}(B_{\vec{k},\lambda}u_{\vec{k},\lambda}\left(\tau,\vec{x}\right)+D^\dagger_{\vec{k},\lambda} v_{\vec{k},\lambda}\left(\tau, \vec{x}\right)).
\end{align}
Here $B_{\vec{k},\lambda}$ and $D^\dagger_{\vec{k},\lambda}$ are creation and annihilation operators which satisfy the standard anticommutation relations,
\begin{align}\label{eq:anticommutators}
\begin{split}
&\left\{D_{\vec{k},\lambda},D^\dagger_{\vec{q},\lambda^\prime}\right\}=\left\{B_{\vec{k},\lambda},B^\dagger_{\vec{q},\lambda^\prime}\right\}=\delta_{\lambda,\lambda^\prime}\delta^{(3)}\big(\vec{k}-\vec{q}\big)\,,\\
&\left\{D_{\vec{k},\lambda},D_{\vec{q},\lambda^\prime}\right\}=\left\{D^\dagger_{\vec{k},\lambda},D^\dagger_{\vec{q},\lambda^\prime}\right\}=\left\{B_{\vec{k},\lambda},B_{\vec{q},\lambda^\prime}\right\}=\left\{B^\dagger_{\vec{k},\lambda},B^\dagger_{\vec{q},\lambda^\prime}\right\}=0\,.
\end{split}
\end{align}
The momentum expansion of the mode functions $u_{\vec{k},\lambda}$ and their charge conjugates $v_{\vec{k},\lambda}$ can be conveniently  written in terms of two $2$-component spinors $\xi_\lambda(\vec{k})$ and corresponding spinor modes $h^{\rm I}_k$ and $h^{\rm II}_k$:
\be\label{2-2component spinors}
\begin{split}
&u_{\vec{k},\lambda}(\tau, \vec{x})=	\frac{e^{i\vec{k}\cdot\vec{x}}}{\sqrt{(2\pi a)^3}}\begin{pmatrix}
	h^{\rm I}_k( \tau )\xi_\lambda(\vec{k})\\
	h^{\rm II}_k(\tau)\frac{\vec\sigma.\vec{k}}{k}\xi_\lambda(\vec{k})
\end{pmatrix},\\
&v_{\vec{k},\lambda}(\tau , \vec{x})=	\frac{e^{-i\vec{k}\cdot\vec{x}}}{\sqrt{(2\pi a)^3}}\begin{pmatrix}
	-h^{\rm II*}_k ( \tau )\frac{\vec\sigma.\vec{k}}{k}\xi_{-\lambda}(\vec{k})\\
	-h^{\rm I*}_k (\tau) \xi_{-\lambda}(\vec{k})
\end{pmatrix}\,,
\end{split}
\ee
with
\begin{align}
\frac{\vec\sigma\cdot\vec{k}}{k}\xi_\lambda(\vec{k})=\lambda\xi_\lambda(\vec{k}), \hspace{0.3in}\lambda=\pm 1\,,\ \ \ \ \ \ \  \xi_\lambda^\dagger (\vec{k})\xi_\lambda(\vec{k})=1\,.
\end{align}
Using this representation,  Eq. (\ref{CovDiracEq}) splits into a system of two coupled first order equations for each of the two types of spinor modes $h^{\rm I}_k$ and $h^{\rm II}_k$:
\be\label{h1 and h2 eq}
h^{\rm I}_k=\frac{i a}{k}(\frac{1}{a}\partial_\tau+i m_\psi)h^{\rm II}_k(\tau),\hspace{0.5 in}h^{\rm II}_k=\frac{i a}{k}(\frac{1}{a}\partial_\tau-
i m_\psi)h^{\rm I}_k(\tau).
\ee
After straightforward calculation, these  equations can be rewritten as two second order decoupled equations:
\be\label{dc eq of spinors}
\begin{split}
	&\bigg(\partial_\tau^2-i m_{\psi} a'+a^2 m_\psi^2+k^2
\bigg)h^{\rm I}_k(\tau)=0\rightarrow \big(\partial_\tau^2+\Omega_k^2(\tau)\big)h^{\rm I}_k(\tau)=0,\\
	&\bigg(\partial_\tau^2+i m_\psi a'+a^2 m_\psi^2+k^2
	\bigg)h^{\rm II}_k(\tau)=0\rightarrow \big(\partial_\tau^2+(\Omega_k^2(\tau))^*\big)h^{\rm II}_k(\tau)=0,
\end{split}	
\ee
where
\begin{equation}\label{eq:Omegak}
\Omega_k^2 \equiv\omega_k^2+a^2\Delta^2-i\sigma(\tau)\,,
\end{equation}
with
\be\label{adiabatic order}
\begin{split}
&\omega_k (M) \equiv \sqrt{k^2+M^2 a^2},\\
&\sigma \equiv m_\psi a'=\sqrt{M^2+\Delta^2} \ a^\prime.
\end{split}
\ee
The fact that \eqref{dc eq of spinors} only depends on the modulus of the momentum, $k$, justifies the notation used for the modes $h^{\rm I}_k, h^{\rm II}_k$, with no arrows.
Following the same prescription as in the case of scalar fields (cf. \hyperref[ZPEScalar]{Sec.\,\ref{ZPEScalar}}),  we have  introduced an off-shell scale $M$, which again will take the role of  renormalization scale. Correspondingly, we have defined  $\Delta^2\equiv m_\psi^2-M^2$  and once more assigned adiabaticity order $2$ to it.  We did not change the notation $\Delta$ as compared to the scalar case since the final formulas do not depend on $\Delta$ but on $M$ and the respective physical masses.  The argument of $\omega_k$ will be omitted  from now on, unless it takes a different value from $M$.  The normalization conditions for the mode functions involved in  $\psi$ are implemented through the Dirac scalar product:
\be\label{eq:NormPrescription}
(u_{\vec{k},\lambda},u_{\vec{k}^\prime,\lambda^\prime})=\int d^3 x\, a^3 u_{\vec{k},\lambda}^\dagger u_{\vec{k}^\prime,\lambda^\prime} =\delta_{\lambda\lambda^\prime}\delta^3(\vec{k}-\vec{k}^\prime)
\ee
and similarly for $ (v_{\vec{k},\lambda},v_{\vec{k}^\prime,\lambda^\prime})=\delta_{\lambda\lambda^\prime}\delta^3(\vec{k}-\vec{k}^\prime)$. It follows that
\be\label{normalization relation}
|h_{k}^{\rm I}|^2+|h_{k}^{\rm II}|^2=1.
\ee
As mentioned in the previous section, the number of time derivatives of the cosmological scale factor $a(\tau)$ that appear in a term of the expansion is called adiabatic order of the term.

In order to solve the differential equations \eqref{dc eq of spinors} we may follow a recursive process which preserves the adiabatic hierarchy, just as we did with the scalar fields.
Let us first  redefine $h^{\rm I}_k$ and the time variable as follow
\be\label{redefine field and time}
h^{\rm I}_{k,1} \equiv\sqrt{\Omega_k}h^{\rm I}_k\hspace{0.5in}d\tau_1=\Omega_k d\tau.
\ee
Substituting these relations into the equation for $h^{\rm I}_k$ in \eqref{dc eq of spinors} we find
\be\label{1st order of Eom}
\frac{d^2}{d\tau^2_1}h^{\rm I}_{k,1}+\Omega_{k,1}^2 h^{\rm I}_{k,1}=0,\hspace{0.5in} \Omega_{k,1}^2 \equiv 1+\epsilon_2,\hspace{0.5 in}\epsilon_2\equiv-\Omega_{k}^{-1/2}\frac{d^2}{d\tau_1^2}\Omega_{k}^{1/2}.
\ee
Since $\epsilon_2$ includes two derivatives, it contains terms of second and higher adiabatic order. We can ignore it to find the leading order solution
\be
h_{k,1}^{\rm I} \approx e^{-i\tau_1},
\ee
so that we get a first approximation
\be\label{FirstApprox}
h_{k}^{\rm I}\approx \frac{e^{-i\int^\tau \Omega_{k}d\tilde{\tau}}}{\sqrt{\Omega_{k}}}.
\ee
Notice that $h_{k,1}^{\rm I}$ formally satisfies a differential equation with the same form as \eqref{dc eq of spinors} for $h_{k}^{\rm I}$. So that, we can repeat the process:
\be
h_{k,2}^{\rm I} \equiv \sqrt{\Omega_{k,1}} h_{k,1}^{\rm I}, \hspace{0.5 in} d\tau_2 \equiv \Omega_{k,1}d\tau_1.
\ee
The corresponding differential equation for $h_{k,2}^{\rm I}$ is
 \be\label{Epsilon4}
\left(\frac{\partial^2}{\partial \tau_2^2}+\Omega_{k,2}^2\right)h_{k,2}^{\rm I}=0,\hspace{0.5in} \Omega_{k,2}^2 \equiv 1+\epsilon_4,\hspace{0.5 in}\epsilon_4\equiv-\Omega_{k,1}^{-1/2}\frac{d^2}{d\tau_2^2}\Omega_{k,1}^{1/2}.
\ee
Once again, $\epsilon_4$ consists of terms of adiabatic order 4 and higher. We can approximate a solution of \eqref{Epsilon4} by neglecting $\epsilon_4$:
\be
h_{k,2}^{\rm I} \approx e^{-i\tau_2}\,,
\ee
whereby the approximation to  $h_{k}^{\rm I}$ can be further improved:
\be\label{2ndIterationh_kI}
h_{k}^{\rm I} \approx \frac{e^{-i \int^\tau \Omega_{k}\Omega_{k,1}\,d\tilde{\tau}} }{\sqrt{\Omega_{k}\Omega_{k,1}}}.
\ee
By  iterating the procedure, we can obtain a better and better approximation to $h_{k}^{\rm I}$, and after $\ell>1$ steps we find
\be\label{GeneralSolution}
h_{k}^{\rm I} \approx \frac{e^{-i\int^\tau \Omega_{k}\cdots \Omega_{k,\ell-1}\,d\tilde{\tau}}}{\sqrt{\Omega_{k}\cdots \Omega_{k,\ell-1}}},
\ee
where, for $\ell \geq 1$,
\be
\Omega_{k,\ell}^2\equiv 1+\epsilon_{2\ell}, \hspace{0.5 in} d\tau_\ell \equiv \Omega_{k,\ell-1} d\tau_{\ell-1}, \hspace{0.5 in} \epsilon_{2\ell}\equiv -\Omega_{k,\ell-1}^{-1/2} \frac{d^2}{d\tau_\ell^2} \Omega_{k,\ell-1}^{1/2}.
\ee
Now that the general method has been set up, let's find the $0th$ order solution for $h_{k}^{\rm I}$. From \eqref{FirstApprox}, the most generic solution for $h_{k}^{\rm I}$ is
\be\label{h_I(0)}
h^{\rm I}_k(\tau)\approx\frac{f_k^{(0)}}{\sqrt{\Omega_k}}e^{-i\int^\tau {\Omega_k\, d\tilde{\tau}}}=\frac{f_k^{(0)}}{(\omega^2_k+a^2\Delta^2-i\sigma )^{1/4}}e^{-i\int^\tau \sqrt{\omega_k^2+a^2\Delta^2-i\sigma}\,d\tilde{\tau}},
\ee
where the time independent function $f_k^{(0)}$ (of adiabatic order $0$) accounts for the integration `constant' ({strictly speaking, a function of the momentum but not of conformal time) in the exponential.
As for $h_{k}^{\rm II}$, by comparing both lines of \eqref{dc eq of spinors} it is clear that it is possible to proceed in an analogous manner. So we  obtain
\be\label{h_II(0)}
h^{\rm II}_k(\tau)\approx\frac{g_k^{(0)}}{\sqrt{\Omega_k^*}}e^{-i\int^\tau {\Omega_k^*\, d\tilde{\tau}}}=\frac{g_k^{(0)}}{(\omega^2_k+a^2\Delta^2+i\sigma )^{1/4}}e^{-i\int^\tau \sqrt{\omega_k^2+a^2\Delta^2+i\sigma}\,d\tilde{\tau}}\,,
\ee
where $g_k^{(0)}$ has the same paper as $f_k^{(0)}$.
To find the zeroth adiabatic order it is just enough to expand this solution and keep zero order terms. However, some extra caution is needed when dealing with the integrand in the exponential of \eqref{h_I(0)},  which may be expanded up to $1st$ order as
\be
\Omega_k^{(0-1)}=\omega_k+\omega_k^{(1)}\,,
\ee
where
\be
\omega_k^{(1)}\equiv -\frac{i a M }{2\omega_k}\frac{a^\prime}{a}\,.
\ee
The reason is that the integration of the second term in the exponential factor is:
\be
e^{-i\int{\omega_k^{(1)}d\tau}}= \left( \frac{ \omega_k+aM }{k}\right)^{-1/2}=\left( \frac{ \omega_k-aM }{\omega_k+aM}\right)^{1/4},
\ee
so it yields a real term of adiabatic order zero, meaning that the expansion of $\Omega_k$ up to $1st$ order in the integral was mandatory. We have not included an explicit multiplicative factor related with the constant of integration\footnote{The same situation happens with indefinite integrals of higher order terms in the imaginary exponential of Eq.\,\eqref{GeneralSolution}. They are written in an appropriate manner, contributing at bigger adiabatic orders. The final results, though, just depend on $f_k^{(0)}$ and not on the other higher order integrations constants, as dictated by the normalization condition \eqref{normalization relation}. See  \hyperref[sec:appendixB]{Appendix\,\ref{sec:appendixB}} for more details.} since it is already represented by $f_k^{(0)}$. We choose $f^{(0)}_k$ such that the above solution can be compatible with mode functions in Minkowskian spacetime, so we can write
\be\label{Zero order solution}
\begin{split}
&f_k^{(0)}=\sqrt{\frac{k}{2}},\hspace{0.5in}h^{\rm I(0)}_k(\tau)=\sqrt{\frac{\omega_k-aM}{2\omega_k}}e^{-i\int^\tau \omega_k \,d\tilde{\tau}},\\
&g_k^{(0)}=\sqrt{\frac{k}{2}},\hspace{0.5in}h^{\rm II(0)}_k(\tau)=\sqrt{\frac{\omega_k+aM}{2\omega_k}}e^{-i\int^\tau \omega_k \,d\tilde{\tau}}.
\end{split}
\ee
Next we move on to the solution at $1st$ adiabatic order. As we have mentioned, the quantity $\epsilon_2$ defined in \eqref{1st order of Eom}, contains terms of adiabatic order two and higher, so it is not necessary to find the first order solution. It is enough to find the first order term from the denominator of (\ref{h_I(0)}).  So,
\be
\begin{split}\label{h_I(1)}
h^{\rm I(0-1)}_k&\approx \left[\frac{1}{\sqrt{\omega_k}}\left(f^{(0)}_k+f^{(1)}_k \right)\left(1+\frac{iM a^\prime}{4\omega_k^2 }\right)e^{-\int^\tau \frac{M a^\prime}{2\omega_k}\,d\tilde{\tau}}\right] e^{-i\int^\tau \omega_k\, d\tilde{\tau}}\\
&=\sqrt{\frac{\omega_k-aM}{2\omega_k}}\left(1+i\frac{M a^\prime}{4\omega_k^2}+\sqrt{\frac{2}{k}}f^{(1)}_k\right)e^{-i\int^\tau {\omega_k\,d\tilde{\tau}}}.
\end{split}
\ee
Similarly for the second spinor mode $h^{\rm II}_k$:
\be
\begin{split}\label{h_II(1)}
h^{\rm II(0-1)}_k&\approx \left[\frac{1}{\sqrt{\omega_k}}\left(g^{(0)}_k+g^{(1)}_k \right)\left(1-\frac{iM a^\prime}{4\omega_k^2 }\right)e^{\int^\tau \frac{M a^\prime}{2\omega_k}\,d\tilde{\tau}}\right] e^{-i\int^\tau \omega_k\, d\tilde{\tau}}\\
&=\sqrt{\frac{\omega_k+aM}{2\omega_k}}\left(1-i\frac{M a^\prime}{4\omega_k^2}+\sqrt{\frac{2}{k}}g^{(1)}_k\right)e^{-i\int^\tau {\omega_k\, d\tilde{\tau}}}\,,
\end{split}
\ee
where $f^{(1)}_k$ and $g^{(1)}_k$ come from integration constants, as mentioned in the footnote of the previous page. By imposing the normalization condition \eqref{normalization relation}, which has to be satisfied at each adiabatic order, it is possible to see that these constants are purely imaginary, that is
\be \label{ConditionsI}
\operatorname{\mathbb{R}e}f_k^{(1)} =\operatorname{\mathbb{R}e} g_k^{(1)} = 0.
\ee
To continue, we deal with the $2nd$ adiabatic order of the mode functions, i.e. $h_k^{\rm  I,II (2)}$. At this time, we have to include $\Omega^2_{k,1}=1+\epsilon_2$ in our considerations (this term contains $2nd$ order adiabatic terms and beyond). Starting from Eq. \eqref{2ndIterationh_kI}, we have
\begin{align}\label{SecondOrderMode}
	h^{\rm I}_k\approx\frac{f_k^{(0)}+f_k^{(1)}+f_k^{(2)}}{\sqrt{\Omega_k(1+\epsilon_2)^{1/2}}}e^{-i\int^\tau \Omega_k(1+\epsilon_2)^{1/2}\,d\tilde{\tau}}\,,
\end{align}
where $\epsilon_2$ can be computed to be
\be
\begin{split}\label{Epsilon2}
&\epsilon_2 = \frac{5}{16\Omega_k^6}\left(2a a^\prime m_\psi^2-i m_\psi a^{\prime\prime}\right)^2-\frac{1}{4\Omega_k^4}\left(2{a^\prime}^2 m_\psi^2+2a a^{\prime \prime} m_\psi^2-i m_\psi a^{\prime\prime\prime} \right)\,.
\end{split}
\ee
With this result, it is immediate to obtain an approximation for $\Omega_{k,1}$ valid up to the third adiabatic order:
\begin{equation}
\begin{split}
\Omega_{k,1}=\left(1+\epsilon_2\right)^{1/2}&=1-\frac{a^2 M^2}{4\omega_k^4}\frac{a^{\prime\prime}}{a}-\frac{a^2 M^2}{4\omega_k^4}\left(\frac{a^\prime}{a}\right)^2+\frac{5}{8}\frac{a^4 M^4}{\omega_k^6}\left(\frac{a^\prime}{a}\right)^2+\frac{i a M}{8\omega_k^4}\frac{a^{\prime\prime\prime}}{a}-\frac{ia^3 M^3}{2\omega_k^6}\left(\frac{a^\prime}{a}\right)^3\\
&+\frac{15i a^5 M^5}{8\omega_k^8}\left(\frac{a^\prime}{a}\right)^3-\frac{9ia^3 M^3}{8\omega_k^6}\frac{a^\prime}{a}\frac{a^{\prime\prime}}{a}+\dots
\end{split}
\end{equation}
On the other hand, an expansion of the product $\Omega_k\Omega_{k,1}$ is necessary to improve the approximation of $h_k^{\rm I,II}$, as one can see from equation\,\eqref{2ndIterationh_kI}. As earlier, if we wish to present a second order approximation of the modes we have to expand that product up to $3rd$ adiabatic order in the exponential. The expansion can be presented as follows:
\be\label{ExpansionofOmegakOmegak1}
\Omega_k \Omega_{k,1}=\Omega_k  \left(1+\epsilon_2\right)^{1/2}=\omega_k+\omega_k^{(1)}+\omega_k^{(2)}+\omega_k^{(3)}+\dots
\ee
where the dots represent the contributions of adiabatic order bigger than 3, and the indicated terms in the expansion read
\be\label{termsExpansionofOmegakOmegak1}
\begin{split}
\omega_k^{(1)}&\equiv -\frac{i a M}{2\omega_k}\frac{a^\prime}{a}\,,\\
\omega_k^{(2)}&\equiv -\frac{a^2 M^2}{8\omega_k^3}\left(\frac{a^\prime}{a}\right)^2-\frac{a^2 M^2}{4\omega_k^3}\frac{a^{\prime\prime}}{a}+\frac{5a^4 M^4}{8\omega_k^5}\left(\frac{a^\prime}{a}\right)^2+\frac{a^2 \Delta^2}{2\omega_k}\,,\\
\omega_k^{(3)}&\equiv -\frac{5 i  a^3 M^3}{16\omega_k^5}\left(\frac{a^\prime}{a}\right)^3 -\frac{ i a^3 M^3}{\omega_k^5}\frac{a^\prime}{a}\frac{a^{\prime\prime}}{a}+\frac{i aM}{8\omega_k^3}\frac{a^{\prime\prime\prime}}{a}-\frac{i a\Delta^2}{4 M\omega_k}\frac{a^\prime}{a}+\frac{25 i a^5 M^5}{16\omega_k^7}\left(\frac{a^\prime}{a}\right)^3+\frac{i a^3 M\Delta^2}{4\omega_k^3}\frac{a^\prime}{a}\,.
\end{split}
\ee
As noted before, $\omega_k^{(1)}$ and $\omega_k^{(3)}$ are purely imaginary, while $\omega_k$ and $\omega_k^{(2)}$ are real. Again, when integrated inside the exponential of equation \eqref{2ndIterationh_kI} the former two give a real contribution, whereas the latter two become part of the phase of the mode and play the role of an effective frequency:
\be
\begin{split}\label{IntegralsExp}
&\exp\left(-i\int^\tau \Omega_k \left(1+\epsilon_2\right)^{1/2}d\tilde{\tau}\right) \approx \exp\left(-i\int^\tau \left(\omega_k^{(1)}+\omega_k^{(3)}\right)d\tilde{\tau}\right)\exp\left(-i\int^\tau \left(\omega_k+\omega_k^{(2)}\right)d\tilde{\tau}\right)\\
&=\left(\frac{\omega_k-a M}{\omega_k+a M}\right)^{1/4}\exp\left(-\frac{5 a^3 M^3}{16 \omega_k^5}\left(\frac{a^\prime}{a}\right)^2+\frac{a M}{8\omega_k^3}\frac{a^{\prime\prime}}{a}-\frac{a\Delta^2}{4M\omega_k}\right)\exp\left(-i\int^\tau \left(\omega_k+\omega_k^{(2)}\right)d\tilde{\tau}\right)\\
&\approx\left(\frac{\omega_k-a M}{\omega_k+a M}\right)^{1/4}\left(1-\frac{5 a^3 M^3}{16 \omega_k^5}\left(\frac{a^\prime}{a}\right)^2+\frac{a M}{8\omega_k^3}\frac{a^{\prime\prime}}{a}-\frac{a\Delta^2}{4M\omega_k}\right)\exp\left(-i\int^\tau \left(\omega_k+\omega_k^{(2)}\right)d\tilde{\tau}\right)\,.
\end{split}
\ee
The last result holds good up to an arbitrary function of momentum (constant in conformal time) multiplying the whole result. We account for this arbitrary constant by introducing the functions $f_k^{(0)},f_k^{(1)},f_k^{(2)},\dots$ at each order.

An efficient strategy to compute the integrals involved in the above calculation (and many other ones of a similar sort, see \hyperref[sec:appendixB]{Appendix\,\ref{sec:appendixB}} for a  sample of them) is to set up an ansatz which respects the adiabaticity order of the calculation. The ansatz consists of a finite number of terms (in fact, a linear combination of them) taken each at the given adiabatic order and with coefficients (or `form factors') which must be determined. The terms of the ansatz are constructed out  of the derivatives of the scale factor and the parameter $\Delta^2$ (which we recall is of second adiabatic order). For instance, in order to compute the integral of $\omega_k^{(3)}$ in Eq.\,\eqref{termsExpansionofOmegakOmegak1}, we know that the result must be of second adiabatic order. Hence as a suitable ansatz we use a linear combination of second order adiabatic terms:
\be
-i\int^\tau \omega_k^{(3)}d\tilde{\tau}=Q_1\left(a,\omega_k\right)\left(\frac{a^\prime}{a}\right)^2+Q_2 \left(a,\omega_k\right)\frac{a^{\prime\prime}}{a}+Q_3\left(a,\omega_k\right)\Delta^2+\textrm{const.}
\ee
where again  the term `const.' at the end means that it does not depend on the integration variable, $\tilde{\tau}$.
By taking derivatives with respect to (conformal) time of the last expression and comparing with $\omega_k^{(3)}$ one can identify the form factors $Q_1=-\frac{5a^3 M^3}{16\omega_k^5}$, $Q_2=\frac{aM}{8\omega_k^3}$ and $Q_3=-\frac{a}{4M\omega_k}$.

Using \eqref{IntegralsExp} together with \eqref{ExpansionofOmegakOmegak1} and \eqref{SecondOrderMode}, the expansion of $h_{k}^{\rm I}$ up to $2nd$ order is
\be\label{hI(2)}
\begin{split}
h_k^{\rm I (0-2)}=\left(\frac{\omega_k-a M}{2\omega_k}\right)^{1/2}\Bigg(&1+\frac{i a^\prime M}{4\omega_k^2}+\sqrt{\frac{2}{k}}f_k^{(1)}\left(1+i\frac{a^\prime M}{4\omega_k^2}\right)+\sqrt{\frac{2}{k}}f_k^{(2)}+\frac{M a^{\prime \prime}}{8\omega_k^3}-\frac{5M^3{a^\prime}^2 a}{16\omega_k^5}\\
&-\frac{5a^2 {a^\prime}^2M^4}{16\omega_k^6}-\frac{{a^\prime}^2 M^2}{32\omega_k^4}+\frac{a a^{\prime\prime}M^2}{8\omega_k^4}-\frac{a\Delta^2}{4M\omega_k}-\frac{a^2 \Delta^2}{4\omega_k^2}\Bigg) e^{-i\int^\tau \left(\omega_k +\omega_k^{(2)}\right)d\tilde{\tau}}\,.
\end{split}
\ee
In a similar way,
\be\label{hII(2)}
\begin{split}
h_k^{\rm II (0-2)}=\left(\frac{\omega_k+a M}{2\omega_k}\right)^{1/2}\Bigg(&1-\frac{i a^\prime M}{4\omega_k^2}+\sqrt{\frac{2}{k}}g_k^{(1)}\left(1-i\frac{a^\prime M}{4\omega_k^2}\right)+\sqrt{\frac{2}{k}}g_k^{(2)}-\frac{M a^{\prime \prime}}{8\omega_k^3}+\frac{5M^3{a^\prime}^2 a}{16\omega_k^5}\\
&-\frac{5a^2 {a^\prime}^2M^4}{16\omega_k^6}-\frac{{a^\prime}^2 M^2}{32\omega_k^4}+\frac{a a^{\prime\prime}M^2}{8\omega_k^4}+\frac{a\Delta^2}{4M\omega_k}-\frac{a^2 \Delta^2}{4\omega_k^2}\Bigg) e^{-i\int^\tau \left(\omega_k +\omega_k^{(2)}\right)d\tilde{\tau}}\,.
\end{split}
\ee
The normalization condition fixes the following relations:
\be
\begin{split} \label{ConditionsII}
&\left|f_k^{(1)}\right|^2=-\sqrt{2 k} \operatorname{\mathbb{R}e}f_k^{(2)},\\
&\left|g_k^{(1)} \right|^2=-\sqrt{2 k} \operatorname{\mathbb{R}e}g_k^{(2)}.
\end{split}
\ee
So far, the expansion for the modes $h^{\rm I}_k$ and $h^{\rm II}_k$ up to $2nd$ order has been presented. One can continue with the procedure formerly described to reach higher orders, although of course the calculation becomes more and more involved.  We should keep in mind, though, that the adiabatic expansion is an asymptotic expansion. While for renormalization purposes it is enough to stop the expansion at $4th$ adiabatic order (in $4$-dimensional spacetime), it is nonetheless necessary  to reach  up to $6th$ order to meet the finite terms $\sim H^6$ that are dominant in the early universe and capable of triggering inflation in this framework (cf. \hyperref[sec:RVM-inflation]{Sect.\,\ref{sec:RVM-inflation}})\footnote{As explained in \cite{Moreno-Pulido:2020anb},  owing to the renormalization prescription of the EMT -- see e.g. Eq. \eqref{RenormalizedEMTScalar} for the scalar case and its fermionic counterpart,  Eq.\,\eqref{RenormalizedEMTFermion} below -- the explicit  $4th$ order powers $H^4$ just cancel out. As a result,  the $6th$ order is the first non-vanishing contribution on-shell.}. We shall refrain from presenting these cumbersome formulas in the main text, see  \hyperref[sec:appendixB]{Appendix\,\ref{sec:appendixB}}.

It is worth noticing that there is some residual freedom in the previous calculations since, we can not determine entirely the set of integration constants that appear during the calculations $f_k^{(1)},g_k^{(1)},f_k^{(2)},g_k^{(2)},\dots$
Because of the normalization condition \eqref{normalization relation} of the mode functions, some restrictions such as \eqref{ConditionsI} and \eqref{ConditionsII} apply.
Fortunately, as commented in more detail in \hyperref[sec:appendixB]{Appendix\,\ref{sec:appendixB}}, the satisfaction of these restrictions is enough for the observables to be independent of this residual freedom. So that, is enough to set all of them to 0 to get, for instance, the desired values of the energy density and pressure.

\section{ZPE and VED for a spin-1/2 field in FLRW spacetime}\label{sec:ZPEFermions}
The computation of  the Fourier modes for a quantized fermion field through adiabatic expansion as explained  in the previous section is just the first step to compute the vacuum energy density (VED).  The next  step towards the VED  is to obtain the  ZPE  associated with Dirac fermions in curved spacetime. As it well known, traditional computations of ZPE suffer from the well-known headache of carrying highly unacceptable contributions proportional to the quartic powers of the masses, $\sim m^4$. This is so both for scalar and fermion fields, and it is already the case in flat, Minkowskian, spacetime, see e.g. \cite{Sola:2013gha,Peracaula:2022vpx} for a detailed discussion and more references. In curved spacetime we have the same situation, in principle, but  in addition we encounter subleading,  curvature dependent,  contributions which do not exist in the flat case, as we shall see in a moment. To handle this issue, an adequate renormalization prescription is called for.

The calculation of the ZPE performed here for spin-$1/2$ fermions is closely related with the one previously put forward for scalar fields in \cite{Moreno-Pulido:2020anb,Moreno-Pulido:2022phq} and summarized in \hyperref[ZPEScalar]{Sec.\,\ref{ZPEScalar}}. Once more the computation will be done through adiabatic expansion of the field modes and will be carried out up to $6th$ adiabatic order, since this is the first non-vanishing order on-shell, i.e. when fixing the renormalization scale $M$ to the value of the mass of the fermion $m_\psi$.  However, the off-shell computation at $4th$ order is already very useful as a means to determine the RG running of the VED as a function of the scale $M$.  This is actually one of the main new features of the off-shell ARP method proposed in \cite{Moreno-Pulido:2020anb,Moreno-Pulido:2022phq}, which leads to the cosmic evolution of the VED. Next we consider the actual calculation for spinor fields.

To find out the ZPE, we start from the definition of EMT in Eq.\,\eqref{eq:deltaTmunu2}. In this case we have to evaluate the functional derivative
\begin{align}
    T_{\mu\nu}^{\psi}=-\frac{2}{\sqrt{-g}}\frac{\delta S_\psi}{\delta g^{\mu\nu}},
\end{align}
applied to the fermion action \eqref{FermionAction}.  Upon  a straightforward calculation we arrive at the following symmetric expression:
\begin{align}\label{EMTFermion}
	T_{\mu\nu}^{\psi}=\frac{i}{4}\bar{\psi}\left(\underline{\gamma}_\mu \nabla_\nu+\underline{\gamma}_\nu \nabla_\mu\right)\psi-\frac{i}{4}\left(\left(\nabla_\mu \bar{\psi} \right) \underline{\gamma}_\nu+ \left( \nabla_\nu \bar{\psi}\right)\underline{\gamma}_\mu\right)\psi\,,
\end{align}
in which the equation of motion \eqref{CovDiracEq} and its hermitian conjugate have been used.  We treat this spinor field as a field operator and upon using its expansion  in Fourier modes and utilizing the anticommuting algebra of the creation and annihilation operators, Eq.\,\eqref{eq:anticommutators},  we can compute the  VEV of the various components, which reflect  the contribution from the vacuum fluctuations of the quantized fermion fields. The method is the same as for the scalar fields\,\cite{Moreno-Pulido:2020anb, Moreno-Pulido:2022phq}, but details are of course different and shall be omitted here.   After significant work, we find that the VEV of the $00 th $ component of the EMT can be cast as follows:
\begin{align}
	\left<T^{\delta\psi}_{00}\right>=\frac{1}{2\pi^2 a}\int{dk k^2 \rho_k  },
\end{align}
where $\rho_k$ is a function of the previously defined mode functions (which can be computed through adiabatic expansion):
\begin{align}\label{eq:producths}
\rho_k=\frac{i}{a}\left(h^{\rm I}_k h'^{\rm I*}_k+h^{\rm II}_k h'^{\rm II*}_k-h^{\rm I*}_k h'^{\rm I}_k-h^{\rm II*}_k h'^{\rm II}_k\right)\,.
\end{align}
The explicit form of the adiabatic expansion of $\rho_k$  is rather cumbersome; the  reader may find the final result for  $\langle T_{00}^{\delta\psi} \rangle$ in  the \hyperref[sec:appendixB]{Appendix\,\ref{sec:appendixB}}. Let us note that for off-shell renormalization at a point $M$  it suffices to adiabatically  expand the solution up to $4th$ order  (as it was prescribed in   Eq.\,\eqref{RenormalizedEMTScalar}), see  Eq.\,\eqref{RenormalizedEMTFermion} below. However,  we will provide the result up to $6th$ order so as to be sensitive to the on-shell result (occurring for  $M=m_\psi$)  and also because it is important for the inflationary mechanism in the early universe (cf. \hyperref[sec:RVM-inflation]{Sec.\,\ref{sec:RVM-inflation}}).  Renormalization of the above expressions is indeed necessary since the VEV of the EMT is formally divergent. The UV-divergent contributions  appear up to $4th$ adiabatic order  (in $n=4$ spacetime dimensions), so that one has to subtract terms up to this order  to obtain a finite result.

\subsection{Divergence balance between  bosons and fermions in vacuum}\label{sec:LeadingSubleading}
The unrenormalized  VEV  of the EMT can be split into two different parts, divergent (in the UV sense) and non-divergent.  Explicit calculation using the formulas of \hyperref[sec:appendixC]{Appendix\,\ref{sec:appendixC}}) shows that the divergent part  reads as follows:
\begin{align}\label{D part of energy density}
	\begin{split}
		\left<T^{\delta\psi}_{00}\right>=\frac{1}{2\pi^2 a^2}\int_0^\infty dk k^2\left(-2\omega_k-\frac{a^2 \Delta^2}{\omega_k}+\frac{a^4\Delta^4}{4\omega_k^3}\right)
		+\frac{1}{2\pi^2 }\int_0^\infty dk k^2\left(\frac{ M^2}{4\omega_k^3}+\frac{\Delta^2}{4\omega_k^3}\right){\cal H}^2\,.
	\end{split}
\end{align}
As it is easy to see, there are terms diverging quartically, quadratically and logarithmically.
The non-divergent part contains the remaining terms, all of them being finite.
The above ZPE is, as warned,  an unrenormalized result at this point. However, before we proceed to  renormalize that expression in the next section, it may be instructive  to check if there is a chance for a cancellation between UV-divergent terms between fermions and bosons in the supersymmetric (SUSY) limit, if only  for the leading divergences.  In the on-shell case ($M=m$ and hence $\Delta^2=0$) the above equation \eqref{D part of energy density}  simplifies to
\begin{equation}\label{eq:FermUVpartOnshell}
 \left.\langle T_{00}^{\delta \psi}\rangle\right|_{(M=m)}= -\frac{1}{\pi^2 a ^2}\int dk k^2 \omega_k(m)+\frac{1}{8\pi^2 }\int_0^\infty dk k^2 \frac{ m^2}{\omega^3_k(m)} {\cal H}^2\,.
\end{equation}
It coincides with the Minkowskian result for $a=1$ (since ${\cal H}=0$).
Now, in a SUSY theory, in which the number of boson and fermion degrees of freedom (d.o.f.)  is perfectly balanced, we should expect that the leading (quartic) divergences cancel among the fermionic and bosonic contributions in the vacuum state\,\cite{Zumino:1974bg,Wess:1974tw} since in such case the  scalar and fermionic fields have the same mass $m$. Thus the quartically divergent contribution from the first term of \eqref{eq:FermUVpartOnshell} should be minus four times the corresponding result for one real scalar field\footnote{ In the chiral supermultiplets of SUSY theories  the number of d.o.f. from  bosons and from fermions is balanced, and they have the same mass.}.  We can check it is indeed so using the above formulas,  for in the on-shell limit and projecting  the UV-divergent terms of the first two adiabatic orders only,  we find that  the contribution from one real scalar field in  FLRW spacetime with spatially flat metric  is \cite{Moreno-Pulido:2022phq}
  \begin{equation}\label{eq:T002}
  \left.\langle T_{00}^{\delta \phi}\rangle^{ (0-2)}\right|_{(M=m)}  =\frac{1}{4\pi^2 a^2}\int dk k^2  \omega_k(m)
-\frac{3\left(\xi-\frac{1}{6}\right)}{4\pi^2 a^2}\int dk k^2\left(\frac{\mathcal{H}^2}{\omega_k(m)}+\frac{ a^2 m^2\mathcal{H}^2}{\omega_k^3(m)}\right)\,.
\end{equation}
We confirm that  the first term (the quartically divergent one) of this expression  is of opposite sign to  the first one in \eqref{eq:FermUVpartOnshell} and is a factor of $4$ smaller, as noted. So, in a SUSY theory,  where we would have $4$ real scalar d.o.f.  for each Dirac fermion, there would be an exact cancellation of the  leading UV-divergent terms.
In addition, we can see at once  from \eqref{eq:T002}  that both the quadratic and logarithmic  divergences of  bosons  hinge on effects of the spacetime curvature since they are proportional to $\mathcal{H}^2$.  These terms, therefore,  vanish in Minkowski spacetime but are unavoidably present in the  FLRW background (except if $\xi=1/6$).
 On the other hand,  from the second term on the \textit{r.h.s.} of  Eq. \eqref{eq:FermUVpartOnshell} it is clear that for fermions we only have subleading divergences of  logarithmic type, which  also hinge on curvature effects since they are again proportional to $\mathcal{H}^2$ and  would also vanish in Minkowski space.    Hence there is no possible cancellation of these subleading divergences between bosonic and fermionic d.o.f., in FLRW spacetime, even in the exact SUSY limit.  Of course, our framework is not placed in the context of supersymmetry, but it serves as a consistency check of our calculations.  See also the discussion in \cite{Maggiore:2010wr,Bilic:2011zm}.

 Although it is possible to introduce a cutoff for a preliminary treatment of the subleading divergences (and maybe to speculate on its possible meaning)  it is not really necessary.  One simply has to implement appropriate renormalization since  renormalization is anyway necessary to deal meaningfully with the VED, as there is no way to cure the divergences from the combined contributions from bosons and fermions and it is not useful to be left with a ``physical'' cutoff.  Dealing with a cutoff is always ambiguous as it is generally not a covariant quantity.  Renormalization gets rid of cutoffs and one can preserve covariance, which is safer for a  physical interpretation of the final results. The adiabatic renormalization is ideal in this sense since the adiabatic expansion generates automatically a covariant result.

 It is well-known that the renormalization program in QFT  requires the presence of a renormalization point, as well as a renormalization prescription. The renormalization point is a floating scale characteristic of the RG. As in the ordinary adiabatic procedure, to implement the renormalization of the EMT in $4$ spacetime dimensions we perform a subtraction of the first four adiabatic  orders, which are the only ones that can be UV-divergent\,\cite{Birrell:1982ix,Parker:2009uva,Fulling:1989nb}. However, in contrast to the usual recipe, in which the subtraction is performed on the mass shell value $m$ of the quantum field,  we perform it at an arbitrary scale  $M$  since this enables  us to explore the RG evolution of the VED and ultimately connect it with its cosmic evolution. This is the specific feature of the  adiabatic  renormalization procedure (ARP)  for the VED that was proposed in \,\cite{Moreno-Pulido:2020anb, Moreno-Pulido:2022phq}  -- see also \cite{Peracaula:2022vpx} for additional details and a comparison with other renormalization schemes. The resulting  renormalized VED ensuing from this procedure  is free from the usual troubles  with the quartic powers of the  masses and their inherent fine tuning problems.

Finally, let us note that dealing with the CCP in Minkowski spacetime  using,  for instance,  the MS scheme and assigning some value to 't Hooft's mass unit $\mu$ in DR (as discussed so many times in the literature),  is entirely  meaningless. It is not only devoid of meaning in that a non-vanishing cosmological constant cannot be defined in Minkowski space without manifestly violating Einstein's equations; it is meaningless also on account of the fact that there is no sense in associating the scale $\mu$ with a cosmological variable, say $H$,  since, if  Einstein's equations are invoked,  the $\CC$ term as such in these equations cannot exist in Minkowski spacetime unless the VED is exactly $\rL+{\rm ZPE}=0$.   So there  is no cosmology whatsoever to do in  flat spacetime, despite some stubborn  attempts in the literature.  Persisting in this attitude leads to the nonsense of having to cope with $\sim m^4$ effects which must then be fine tuned among all the particles involved. This point has  been driven home repeatedly e.g.  in \cite{Sola:2013gha} and also recently in \cite{Peracaula:2022vpx},  see also  \cite{Mottola:2022tcn}.  A realistic approach to the VED  within QFT in curved spacetime  must get rid of Minkowski space pseudo-argumentations. The approach that we present here is fully formulated in curved spacetime and the vacuum energy density evolves with the participation of the  curvature effects (powers of $H$) rather than with only powers of the masses, i.e. we pursue the successful renormalization program of \,\cite{Moreno-Pulido:2022phq,Moreno-Pulido:2020anb}. Therefore, when the background curvature vanishes, we consistently predict that the non-trivial effects which are responsible for the value of the vacuum energy density and the cosmological constant disappear (and hence we are left with no $\CC$ nor VED in the universe).  Such is, of course, the situation in Minkowski space.  In practice, however,  we cannot reach that flat spacetime situation in our universe since there exists four-dimensional curvature at all times during the indefinite process of expansion. But by the same token such an impossibility evinces the fact that the VED and its dynamical nature is a direct consequence of the expansion process (and of the spacetime curvature inherent to it). The expected size of the VED and of $\CC$ in our framework is indeed provided  by the magnitude of the spacetime curvature, which is of the typical value of the measured $\CC$. It is therefore not caused by the quartic power of the masses of the fields (which is the very root of the CC problem in most approaches). These powers  do not affect the running  of the VED in our framework. To put it in a nutshell: the renormalized VED in our framework is like a small quantum  `ripple' imprinted on the existing (classical) background curvature owing to the vacuum fluctuations of the quantized matter fields. In the absence of the background curvature, the ripple would disappear too since it is proportional to it through the coefficient $\nueff$, which encodes the quantum effects from the quantized matter fields.

Following the same approach as for scalar fields, in the next section we compute the quantum effects contributing to the VED from the quantized spin-1/2 fields and express them in renormalized form using the same subtraction scheme devised in \cite{Moreno-Pulido:2022phq,Moreno-Pulido:2020anb}.

\subsection{Renormalized ZPE for fermions }\label{sec:RenZPEfermions}

Thus, following the same prescription \eqref{RenormalizedEMTScalar} as in the case of the scalar field, the renormalized form of the fermionic VEV of the EMT reads
\begin{align}\label{RenormalizedEMTFermion}
&\left<T_{\mu\nu}^{\delta\psi}\right>_{\rm ren}(M)\equiv\left<T^{\delta\psi}_{\mu\nu}\right>(m_\psi)-\left<T^{\delta\psi}_{\mu\nu}\right>^{(0-4)}(M)\,.
\end{align}
Since our aim is to study the ZPE we will focus into the 00{\it th}-component of the former equation for the moment. Alternatively, it is written as \footnote{The subscript `Div' refers to the part of the EMT calculation comprising divergent integrals. These  appear only  up to the $4th$ adiabatic order. The subscript  `Non-Div', on the other hand, refers, of course, to the part of the EMT calculation involving finite integrals only. }
\begin{align*}%\label{T00_ren0}
\begin{split}
\left<T_{00}^{\delta\psi}\right>_{\rm ren}(M)&=\left<T_{00}^{\delta\psi}\right>_{\rm Div}(m_\psi)-\left<T_{00}^{\delta\psi}\right>_{\rm Div}(M)+\left<T_{00}^{\delta\psi}\right>_{\rm Non-Div}^{(0-4)}(m_\psi)-\left<T_{00}^{\delta\psi}\right>_{\rm Non-Div}^{(0-4)}(M)\\
&+\left<T_{00}^{\delta\psi}\right>^{(6)}(m_\psi)+\dots
\end{split}
\end{align*}
\begin{align}\label{T00_ren}
\begin{split}
&=\frac{1}{2\pi^2 a}\int_0^\infty dk k^2 \left(-\frac{2\omega_k (m_\psi)}{a}+\frac{2\omega_k (M)}{a}+\frac{a\Delta^2}{\omega_k(M)}-\frac{a^3\Delta^4}{4\omega_k^3 (M)}\right)\\
&+\frac{1}{2\pi^2 a}\int_0^\infty dk k^2 \left(\frac{a m_\psi^2}{4\omega_k^3(m_\psi)}-\frac{aM^2}{4\omega_k^3 (M)}-\frac{a\Delta^2}{4\omega_k^3(M)}\right)\left(\frac{a^\prime}{a}\right)^2\\
&+\left<T_{00}^{\delta\psi}\right>_{\rm Non-Div}^{(0-4)}(m_\psi)-\left<T_{00}^{\delta\psi}\right>_{\rm Non-Div}^{(0-4)}(M)+\left<T_{00}^{\delta\psi}\right>^{(6)}(m_\psi)+\dots\\
\end{split}
\end{align}
where we have used the calculational results for the unrenormalized components of the VEV of the EMT recorded in \hyperref[sec:appendixC]{Appendix\,\ref{sec:appendixC}}  and we have introduced the notation $\omega_k(M) \equiv \sqrt{k^2+a^2 M^2}$ and  $\omega_k(m_\psi) \equiv \sqrt{k^2+a^2 m_\psi^2}$.
The last line of \eqref{T00_ren} contains all the non-divergent terms, which constitute a perfectly finite contribution and   is made of finite parts from the $4th$ order expansion and of the entire $6th$ order term, which is fully finite but rather cumbersome.
On the other hand, the first two lines in the last equality are a collection of terms that are individually divergent, but whose combination makes the integral convergent. In fact, by making use of simple algebraic manipulations at the level of the integrand one can show that explicitly. For instance, the rearrangement in the integrand
\begin{align}
dk k^2\left(\omega_k(m_\psi)-\omega_k(M)-\frac{a^2\Delta^2}{2\omega_k(M)}+\frac{a^4\Delta^4}{8\omega_k^3(M)}\right)= dk k^2a^6\Delta^6\frac{\omega_k(m_\psi)+3\omega_k(M)}{8\omega_k^3(M)(\omega_k(m_\psi)+\omega_k(M))^3}
\end{align}
shows that terms seemingly diverging  as  $\sim k^4$ organize themselves to eventually  converge as $\sim 1/k^2$.  Needless to say, this is the consequence of the subtraction that has been operated. Similarly with the second integral in \eqref{T00_ren}, whose individual terms are logarithmically divergent, but overall the integral  is once more convergent thanks to the involved subtraction.

The above renormalized result \eqref{T00_ren} would, of course,  vanish for $M=m_\psi$ if we were to stop the calculation at $4th$ adiabatic order, so in case that one wishes to obtain the renormalized on-shell result one has to either  compute the exact unrenormalized EMT on-shell  before subtracting the divergent adiabatic orders -- which is  possible but only in simpler cases such as in de Sitter space\,\cite{Landete:2013axa, Landete:2013lpa} -- or one has to face the calculation of the adiabatic expansion  up to $6th$-order at least. In the last case  the  term $\langle T_{00}^{\delta\psi}\rangle^{(6)}(m_\psi)$  indicated at the end of  Eq.\,\eqref{T00_ren} must be computed. This is what we have done here since an exact solution in the FLRW case is not possible. The necessary work to reach up to $6th$ adiabatic  order for fermions  is  again significant, as it was previously  for  the scalar case\,\cite{Moreno-Pulido:2022phq,Moreno-Pulido:2020anb}.  The  unrenormalized components of the EMT up to the desired order are explicitly collected  in \hyperref[sec:appendixC]{Appendix\,\ref{sec:appendixC}}.  To subsequently obtain  the renormalized EMT  one has to  implement the subtraction \eqref{RenormalizedEMTFermion} and compute all the involved integrals. Despite the considerable amount of work involved,  the  final result to  the desired order  can nevertheless  be presented through a rather compact formula, as follows\footnote{We refer the reader to  Appendix A.2 of \cite{Moreno-Pulido:2022phq} for the computation/regularization of the involved integrals (depending on whether they are convergent or divergent) with the help of the master DR formula quoted there. Use of DR can be convenient since in certain cases the needed rearrangement of terms in the integrand to verify that the overall integral is actually convergent can be complicated.  Let us emphasize, however,  that DR is only used as an auxiliary regularization tool  for intermediate steps.   The final result has no memory of this intermediate step, see e.g. Appendix B of \cite{Moreno-Pulido:2020anb} for an explicit nontrivial example.  To be sure,  no MS prescription is used for renormalization at any point of our calculation.  The crucial difference between the ARP and the  MS-like schemes is that the subtraction \eqref{RenormalizedEMTFermion}  involves not just the UV-divergences  but also the  finite parts. }:
\begin{align*}
	\begin{split}
\left<T^{\delta\psi}_{00}\right>_{\rm ren}^{(0-6)}(M,H)&=\frac{a^2}{32\pi^2}\left(3m_\psi^4-4m_\psi^2M^2+M^4-2m_\psi^4\ln \frac{m_\psi^2}{M^2}\right)\\
&+\frac{\mathcal{H}^2}{16\pi^2}\left(m_\psi^2-M^2-m_\psi^2\ln\frac{m_\psi^2}{M^2}\right)
\end{split}
\end{align*}
\begin{align}\label{EMT00}
	\begin{split}
&+\frac{1}{20160 \pi^2 a^4  m_\psi^2}\bigg( 204 \mathcal{H}^4 \mathcal{H}^\prime + 26 \left(\mathcal{H}^\prime\right)^3 - 30 \mathcal{H}^3\mathcal{H}^{\prime\prime} + 9 \left(\mathcal{H}^{\prime\prime}\right)^2 + 27 \mathcal{H}^2 \left( \mathcal{H}^{\prime}\right)^2 \\
&\phantom{aaaaaaaaaaaaaaa}- 72 \mathcal{H}^2 \mathcal{H}^{\prime\prime\prime}- 18 \mathcal{H}^\prime \mathcal{H}^{\prime\prime\prime} +
        \mathcal{H} (-78\mathcal{H}^\prime \mathcal{H}^{\prime\prime} + 18 \mathcal{H}^{\prime\prime\prime\prime})\bigg)\\
        &=\frac{a^2}{32\pi^2}\left(3m_\psi^4-4m_\psi^2M^2+M^4-2m_\psi^4\ln \frac{m_\psi^2}{M^2}\right)\\
        &+\frac{a^2 H^2}{16\pi^2}\left(m_\psi^2-M^2-m_\psi^2\ln\frac{m_\psi^2}{M^2}\right)\\
		&+\frac{a^2}{20160 \pi^2 m_\psi^2}\bigg( -31 H^6 - 108 H^4 \dot{H} - 46 \dot{H}^3 + 126 H^3 \ddot{H} + 9 \ddot{H}^2 -
         18 \dot{H} \vardot{3}{H}\\
        &\phantom{aaaaaaaaaaaaaa}+ 27 H^2 \left(7 \dot{H}^2 + 4 \vardot{3}{H} \right) + 6 H (23 \dot{H} \ddot{H} + 3 \vardot{4}{H})\bigg).
		\end{split}
\end{align}
The final equality corresponds to the expression in terms of the cosmic time ($d()/dt\equiv\dot{()})$ with $H\equiv \dot{a}/a$. We point out that there is an explicit dependency on the Hubble function (and its derivatives) coming from $G_{\mu\nu}$. This justifies the notation %$\rho^{\delta\psi}_{\rm vac}(M,H)$ and
 $\langle T_{00}^{\delta\psi} \rangle_{\rm ren}(M,H)$, with two arguments, where the dependence on the time derivatives of $H$ is omitted for simplicity.
We note that in the fermionic case there are no terms of ${\cal O}(H^4)$ in the evolution of the ZPE (and the VED, see next section), in stark contrast to the situation with scalars, see the last line of Eq.\,\eqref{eq:T00Integrated}, where we can recognize terms of the form $H^2\dot{H}, H\ddot{H}$ and $\dot{H}^2$ all of them of ${\cal O}(H^4)$.   We also remark what has been previously anticipated:  for $M=m_\psi$ (on-shell point) only the $6th$-order terms remain, which are the ones in the last two lines of Eq.\,\eqref{EMT00}.  These terms are relevant for the RVM mechanism of inflation in the  very early universe (cf. Sec.\,\ref{sec:RVM-inflation}).  However, for the study of the renormalized theory at the point $M$ (generally different from the on-shell mass point  $m_\psi$) it is enough to consider the terms  up to $4th$ adiabatic order.

So far,  we have been able to provide the desired formula for the renormalized ZPE at the energy scale $M$ up to $6th$ adiabatic order,  as expressed by Eq.\,\eqref{EMT00}. This is, however, not the end of the story, since a proper expression for the VED needs to take into account also the renormalized parameter $\rho_\Lambda$  in the Einstein-Hilbert action \eqref{eq:EH}, as this parameter is part of the unrenormalized vacuum action and after renormalization it also runs with the scale $M$, i.e. $\rL(M)$.  Both the renormalized ZPE and  $\rL(M)$ run with the scale and this will be crucial to study the properties of the renormalized VED. The running of the ZPE part between two different scales $M$ and $M_0$ can be illustrated by considering the difference of the respective ZPE values at these scales. From \eqref{EMT00} we find
\begin{align}\label{eq:TooTwoScales}
\begin{split}
\left<T^{\delta\psi}_{00}\right>_{\rm ren}(M,H)-\left<T^{\delta\psi}_{00}\right>_{\rm ren}(M_0,H)&=\frac{a^2}{32\pi^2}\left(M^4-M_0^4-4m_\psi^2 (M^2-M_0^2)+2m_\psi^4\ln \frac{M^2}{M_0^2}\right)\\
&+\frac{a^2 H^2}{16\pi^2}\left(-M^2+M_0^2+m_\psi^2\ln\frac{M^2}{M_0^2}\right)\,.
\end{split}
\end{align}
The finite parts, and in particular the $6th$ order terms cancel of course in the above difference, but the latter will be essential in the on-shell case since the result would be zero without these higher order effects\,\footnote{Let us remark that the difference \eqref{eq:TooTwoScales} is an exact result, in the sense that it does not depend on the adiabaticity order we are working.  This is obvious from the renormalization prescription \eqref{RenormalizedEMTFermion}, as all higher orders beyond the $4th$ one (not only the $6th$) cancel out in the subtraction,  the reason being that these  adiabatic orders  are independent of the renormalization point $M$.  The latter is involved  in the calculation of the EMT up to $4th$ order only (as these are the only adiabatic orders that are UV-divergent). }.
We should notice that, in contradistinction to the case with scalar fields, there are no contributions of ${\cal O}(H^4)$ such as $H^2\dot{H}$,  $H\ddot{H}$  or  $\dot{H}^2$ in the expression for the ZPE, as can be seen on comparing equations \eqref{eq:T00Integrated} and \eqref{EMT00}. For this reason it is unnecessary to use the higher derivative (HD) tensor  $\leftidx{^{(1)}}{\!H}_{\mu\nu}$ (cf. \hyperref[sec:appendixA]{Appendix\,\ref{sec:appendixA}})  as part of the renormalized Einstein's equations in the case of the fermion fields, again  in contrast to the situation with the scalar fields -- see \cite{Moreno-Pulido:2022phq} for details.
Therefore, for fermions the subtraction at the two scales  of the  renormalized form of Einstein's equations can be done using the ordinary form of Einstein equations \eqref{EinsteinEqs} without higher order curvature terms, and  we find
\begin{align}\label{EinsteinSubstractionScales}
\left<T^{\delta\psi}_{\mu\nu}\right>_{\rm ren}(M,H)-\left<T^{\delta\psi}_{\mu\nu}\right>_{\rm ren}(M_0,H)=\left( \rho_\Lambda (M)-\rho_\Lambda (M_0) \right) g_{\mu\nu}+\left(\frac{1}{8\pi G(M)}-\frac{1}{8\pi G(M_0)}\right)G_{\mu\nu}.
\end{align}
By comparison equations \eqref{eq:TooTwoScales} and \eqref{EinsteinSubstractionScales}, and taking into account the tensorial structure of \eqref{EinsteinSubstractionScales} and the explicit form of $G_{\mu\nu}$ in FLRW spacetime (cf. \hyperref[sec:appendixA]{Appendix\,\ref{sec:appendixA}}) , we can perform the following identifications:
\begin{align}\label{RunningOfRhoLambda}
\begin{split}
&\rho_\Lambda(M)-\rho_\Lambda(M_0)=-\frac{1}{32\pi^2}\left(M^4-M_0^4-4m_\psi^2(M^2-M_0^2)+2m_\psi^4\ln \frac{M^2}{M_0^2}\right),\\
&\frac{1}{8\pi G(M)}-\frac{1}{8\pi G(M_0)}=\frac{1}{48\pi^2}\left(-M^2+M_0^2+m_\psi^2\ln\frac{M^2}{M_0^2}\right).
\end{split}
\end{align}

\subsection{Renormalized  VED }\label{sec:RenVEDfermions}

Once the renormalized  ZPE has been obtained, the  same consideration as for the scalar field case (see \eqref{VacuumEMT} and \eqref{ScalarFieldVED} )  can be repeated  intact here, thus  leading to the expression for the renormalized VED  of the fermionic field:
\begin{align}\label{DensityDefinition}
\rho_{\rm vac}^{\delta\psi} (M,H)=\frac{\left\langle T_{00}^{\rm vac}\right\rangle (M,H)}{a^2}=\rho_\Lambda (M)+\frac{\left\langle T_{00}^{\delta\psi} \right\rangle_{\rm ren} (M,H)}{a^2}.
\end{align}
Now if the subtraction of scales is done,
\begin{align}
\begin{split}\label{DensitiesDifference}
\rho_{\rm vac}^{\delta\psi} (M,H)-\rho_{\rm vac}^{\delta\psi} (M_0,H)&=\frac{ \left\langle T_{00}^{\rm vac} \right\rangle (M,H)-\left\langle T_{00}^{\rm vac} \right\rangle (M_0,H)}{a^2}\\
&=\rho_\Lambda (M) - \rho_\Lambda (M_0)+\frac{\left\langle T_{00}^{\delta\psi} \right\rangle_{\rm ren} (M,H) -\left\langle T_{00}^{\delta\psi} \right\rangle_{\rm ren} (M_0,H)}{a^2}\\
&=\rho_\Lambda (M) - \rho_\Lambda (M_0)-\left( \rho_\Lambda (M) - \rho_\Lambda (M_0)\right)+\frac{3H^2}{8\pi}\left(\frac{1}{ G(M)}-\frac{1}{G(M_0)}\right)\\
&=\frac{H^2}{16\pi^2}\left(-M^2+M_0^2+m_\psi^2\ln\frac{M^2}{M_0^2}\right)\,,
\end{split}
\end{align}
where in the last equality \eqref{RunningOfRhoLambda} was used.  As expected, when written in terms of the ordinary Hubble function $H$ in cosmic time, the evolution of the VED  does not depend explicitly on the scale factor. For the sake of emphasizing the point, in the above equation we have explicitly indicated the cancellation of the terms carrying along the quartic powers of the masses, see the third equality in the above equation. As we can see, it is essential that the structure of the VED  is obtained from the sum  ``${\rm VED}=\rho_\Lambda+{\rm ZPE}$'', i.e. as in Eq.\,\,\eqref{ScalarFieldVED},  since the referred cancellation occurs between  the renormalized expressions of $\rho_\Lambda$  and ${\rm ZPE}$ upon being subtracted at the two arbitrary scales $M$ and $M_0$. This means that the two values of the VED at these scales are related in a very smooth manner: in fact, they differ only by a term proportional to $H^2$, as it is obvious from \eqref{DensitiesDifference}.

Even though Eq.\,\eqref{DensitiesDifference} is formally correct, our job is not finished in the physical arena  yet.  Despite of the fact that such an equation describes the precise mathematical evolution of the VED with the renormalization scale, $M$, it is necessary to associate the latter with a suitable physical scale in order to extract useful phenomenological information out of it,  exactly as in the companion studies of the VED for  scalar fields previously presented in \cite{Moreno-Pulido:2022phq,Moreno-Pulido:2020anb}.  As pointed out in these references and also in \hyperref[RenZPEscalar]{Sec.\,\ref{RenZPEscalar}} regarding the contribution from the scalar fields, the Hubble rate $H$  is a characteristic energy scale (in natural units) of the expanding universe in the FLRW metric,  and hence proves to be a natural candidate for a representative physical scale in this context.  Whereby by following the same prescription used in the aforementioned references,  we set the renormalization energy scale to $M=H(t)$ (at the end of our calculations) in order  to track the physical evolution of the VED.  In other words, this prescription should allow us to explore the VED at different expansion history  times $H(t)$ in a physically meaningful way. In this way we obtain a well behaved evolution of the VED, which means that, given its value at one scale all other values at nearby scales are very close to it.  The dynamics of the VED is  slow and can be encoded into an effective contribution to the $\nueff$ parameter, as we did for bosons in Eq.\,\eqref{eq:nueffBososns}.  The combined contribution from bosons and fermions to this parameter  will be given in \hyperref[sec:CombinedBandF]{Sec. \,\ref{sec:CombinedBandF}}. Let us finally clarify the  sense of this scale setting, Namely, the full effective action does not depend on $M$, of course, but the renormalized VED indeed does since the effective action of vacuum is only a part of the full effective action. The scale dependence on $M$ from the other terms of the action, for example the terms carrying the running couplings of the RG-improved classical action, compensates for the $M$-dependence of the vacuum action.  Put another way,  only the full effective action (involving the classical part plus the nontrivial quantum vacuum effects) is scale- (i.e. RG-) independent.  This is of course the standard lore of the renormalization group (RG), see also \cite{Peracaula:2022vpx} for an expanded discussion.  The choice of a particular scale helps of course in enhancing the physical significance of particular sectors of the full effective action. The procedure is of course akin to the usage of the RG in conventional gauge theories of strong and electroweak interactions, except that here one has to pick out an appropriate cosmological energy scale which is most adequate for the description of the universe's expansion. The distinguished  scale $H$  appears to be the natural choice if the universe where we live is indeed suitably described by the FLRW metric. In the next section we apply this approach to derive the  important RG equation of the VED itself.

\subsection{Renormalization group equation for the vacuum energy density}\label{sec:RGEforVED}

One can also compute the $\beta$ function of the running vacuum generated by the fermionic quantum fluctuations. Only the adiabatic terms below $4th$ order carry $M$-dependence by definition since the higher orders are finite and hence are not subtracted in the renormalization procedure. As it was noted before, in contrast to the scalar case the terms of $4th$ adiabatic order are not present for fermions. The computation follows the same strategy as for scalars\,\cite{Moreno-Pulido:2022phq}. In this case we make use of equations \eqref{eq:TooTwoScales} and \eqref{DensityDefinition}, and we find
\begin{equation}
\begin{split}\label{BetaFunctionFermion}
    \beta_{\rv}^{\delta\psi}=&M\frac{\partial\rv^{\delta\psi}(M)}{\partial M}= \beta^{\delta\psi}_{\rL}+\frac{1}{8\pi^2}\left(M^2-m_\psi^2\right)^2-\frac{1}{8\pi^2}{H}^2\left(M^2-m_\psi^2\right)=-\frac{1}{8\pi^2}{H}^2\left(M^2-m_\psi^2\right).
    \end{split}
\end{equation}
The second equality holds immediately after computing the $\beta$-function of the parameter $\rho_\Lambda$.  From the first equation \eqref{RunningOfRhoLambda}  we find that
\begin{equation}\label{BetaFunctionrLfermion}
    \beta_{\rL}^{\delta\psi}=M\frac{\partial\rL(M)}{\partial M}= -\frac{1}{8\pi^2}\left(M^2-m_\psi^2\right)^2
    \end{equation}
and hence contains a term proportional to the quartic power of the particle mass; what's more, there  is an exact cancellation between the terms of the ZPE containing quartic powers of $M$ and $m_\psi$  and the expression of $\beta_{\rL}$.   The result \eqref{BetaFunctionFermion} can also be consistently obtained directly from Eq.\,\eqref{DensitiesDifference}.  Notice that neither the parameter $\rL$ nor the ZPE have physical meaning in an isolated way, only the sum makes physical sense and defines the VED in the present context.  Let us compare the above results with those following from the contribution of one real scalar field $\phi$\,\cite{Moreno-Pulido:2022phq}:
\begin{equation}\label{eq:RGEVED1}
\beta_{\rv}^{\delta\phi}=\left(\xi-\frac{1}{6}\right)\frac{3 {H}^2 }{8 \pi^2}\left(M^2-m_\phi^2\right)+ {\cal O}(H^4)
\end{equation}
 and
 \begin{equation}\label{eq:BetaFunctionrL}
\beta_{\rL}^{\delta\phi} (M)=\frac{1}{2(4\pi)^2}(M^2-m_\phi^2)^2\,,
\end{equation}
where we omit the $ {\cal O}(H^4)$ terms in the scalar case (not present in the fermionic case) since it is enough to check the comparison at low energies.
 We can see that in both cases  the $\beta$-function of the VED  is proportional to $\beta_{\rv}\propto {H}^2 \left(M^2-m^2\right)$,  where $m=m_\phi$ or $m_\psi$, and therefore has a very smooth behavior thanks to the factor $H^2$.  In contrast, the $\beta$-function for the parameter $\rL$ in the gravitational action (which is often incorrectly identified as the VED in some  explicit QFT calculations of the vacuum energy in the literature) behaves in both cases as $\beta_{\rL}\propto \left(M^2-m^2\right)^2$ and hence  leads to undesired quartic contributions  $\sim m^4$  to the running.  These are the problematic terms leading to fine tuning problems, but as can be seen these terms exactly cancel in  $\beta_{\rv}$ for the vacuum energy both for fermions and bosons in our renormalization scheme.  Notice that there is a factor of $4$ between equations \eqref{BetaFunctionrLfermion} and \eqref{eq:BetaFunctionrL} and have opposite sign.   In a SUSY context,  Eq.\,\eqref{eq:BetaFunctionrL} should be multiplied by $4$ to equalize bosonic and fermionic d.o.f in a given matter supermultiplet, all of whose members possess the same mass. Then  $\beta_{\rL}^{\delta\phi}\rightarrow 4\beta_{\rL}^{\delta\phi}\equiv \beta_{\rL}^{\delta\phi\,( {\rm SUSY})} $, and  the sum of the two coefficients will indeed vanish in a supersymmetric context:
  \begin{equation}\label{eq:SUSYbetafunctions}
   \beta_{\rL}^{\delta\psi\,( {\rm SUSY})}+ \beta_{\rL}^{\delta\phi\,( {\rm SUSY})}=0\,.
  \end{equation}
   But this is, of course, not a cancellation of the $\beta$-function coefficients for the VED of bosons and fermions in the SUSY limit, but only the cancellation of the contributions to the $\beta$-function coefficient for the formal parameter $\rL$ in the EH action \eqref{eq:EH}.
  This property is obviously connected with the discussion in \hyperref[sec:LeadingSubleading]{Sec.\,\ref{sec:LeadingSubleading}} about the balance of UV-divergences between fermions and bosons.  In a SUSY theory the quartic divergences cancel prior to any renormalization process, as we have noticed, and the resulting $\beta$-function for the parameter $\rL$ is zero.  By the same token the running of the VED is freed from  $\sim m^4$ effects, which cancel among fermions and bosons in a SUSY context. The quartic powers  are independent of the curvature of spacetime.  However, the subleading divergences do depend on the background curvature and do not cancel at all, even in the exact SUSY limit \footnote{The SUSY considerations we have made here in passing only intend to clarify that in curved spacetime, irrespective of whether the quantized matter fields belong to a supersymmetric theory or not, the renormalization program is in any case mandatory to finally get rid of all the UV divergences. The calculations in this work, however,  do not presume any SUSY context at all, not even a SUSY-broken theory.  Our treatment of scalar and fermion fields is indeed completely general, in the sense that we are dealing with an arbitrary number of matter fields of both species without enforcing any balance between bosonic and fermionic d.o.f. -- see Sec.\,\ref{sec:CombinedBandF} for more details.}.    The ``residual'' (finite) parts left in the renormalization process do not cancel either; they are actually  proportional to the curvature of the FLRW background, $R\sim H^2$. This fact translates into a correction to the physical vacuum energy density of order  $\sim m ^2 H^2$  both for bosons and fermions, which is far smaller than $m^4$. So the finite,  curvature dependent,  terms that remain after ARP renormalization are de facto the most important ones for our purposes since they lead to the RVM form of the VED! The renormalization of the formal parameter $\rL$, in contrast, has no physical imprint in the final result for the VED, except that the unwanted $m^4$ terms cancel against those involved in the ZPE, thus rendering the renormalized $V{\rm ED}=\rL+{\rm ZPE}$ free from quartic mass dependencies.

 From the above RG equations we may write down the total contribution to the $\beta$-function of the VED from the matter fields.  Consider $N_{\rm f}$ species of fermion fields with  masses $m_{\psi,\ell}$ for each species  $\ell\in\{1,2,\dots , N_{\rm f}\}$, and similarly let $N_{\rm s}$ be the number of  scalar field species with  masses $m_{\phi, j}$, $j\in \{1,2,\dots, N_{\rm s}\}$. Some of these species may have the same mass, but this aspect is not relevant here, our formulas will include a summation over all contributions irrespective if some of them may be equal.
  The total $\beta$-function of the VED from an arbitrary number of  quantized matter fields can now be cast as follows:
  \begin{equation}\label{eq:Totalbetafunctions}
  \beta_{\rv}\equiv\sum_{j=1}^{N_{\rm s}} \beta_{\rv}^{\delta\phi_j}+ \sum_{\ell=1}^{N_{\rm f}}\beta_{\rv}^{\delta\psi_\ell}=\frac{3H^2}{8\pi^2}\left[\sum_{j=1}^{N_{\rm s}}\left(\xi_j-\frac{1}{6}\right) (M^2-m_{\phi_j}^2)-\frac{1}{3}\sum_{\ell=1}^{N_{\rm f}}(M^2-m_{\psi_\ell}^2)\right]+\mathcal{O}(H^4)\,.
  \end{equation}
 The net outcome, therefore,  is that the $\beta$-function of the vacuum energy density is free from undesirable contributions proportional to quartic mass powers of the quantized fields,  $\sim m^4$,  and hence these contributions do not appear in the renormalized theory.  This is of course an extremely welcome feature of our renormalization framework, which is, on inspection of the above equation,  fully shared by both  scalar and fermion fields.  Indeed, up to numerical factors fermions and scalar fields provide the same kind of leading contribution to the time evolution of the  cosmological vacuum energy.  Overall we find that  the running of $\rv$ depends  only on quadratic terms in the fermion mass, namely {$\sim m_{\psi_\ell}^2 H^2$, which are of the same type as in the case of bosons, namely $\sim m_{\phi_j}^2 H^2$, as discussed in \hyperref[RenZPEscalar]{Sec.\,\ref{RenZPEscalar}} and previously demonstrated in great detail in\,\cite{Moreno-Pulido:2020anb, Moreno-Pulido:2022phq}. These terms are actually very smooth owing to the presence of the $H^2$ factor.  Integrating the RG equation corresponding to the $\beta$-function \eqref{eq:Totalbetafunctions} one finds  the expression for the evolution of the VED as a function of the  renormalization scale $M$ in the presence of any number of matter fields, see \hyperref[sec:CombinedBandF]{Sec. \,\ref{sec:CombinedBandF}}. In particular, integrating \eqref{BetaFunctionFermion} for the case of one single fermion it is easy to verify that it leads exactly to \eqref{DensitiesDifference}.

The kind of  much tempered behavior of the VED evolution that we have found here within our ARP renormalization program  is of the sort that was expected on the basis of semi-qualitative RG arguments and constitutes the characteristic running law of the so-called Running Vacuum Models (RVM), see \cite{Sola:2013gha,Peracaula:2022vpx} and references therein. Thus, there is no need for fine-tuning in this scenario, since in such a renormalization procedure we have already gotten rid of the ugly contributions carried along by the quartic powers of the masses. In other words, the `problem' with the quartic powers of the masses does not appear in the physically renormalized theory. While the running of the formal parameter  $\rho_\Lambda$ with $M$ indeed carries $\sim m^4$ contributions, as it is obvious from the formulas above, this fact has no physical implication since $\rL$ is not itself a physical parameter (if taken in isolation) and  the unwanted terms carried by it exactly cancel out in the $\beta$-function for the  VED, as we have just proven.  As a result, the running of the VED is much softer, the `slope' is $\sim m^2 H^2$ rather than $\sim m^4$.   At variance with this result,  in the context of the MS renormalization approach, in which $\rho_\Lambda$ runs with the unphysical mass unit $\mu$ coming from dimensional regularization,  one is enforced to fine tune $\rho_\Lambda (\mu)$ against the large contribution proportional to $\sim m^4$ terms\,\cite{Peracaula:2022vpx}.

\subsection{Renormalization of the fermionic  vacuum pressure}
Taking into account the perfect fluid form of the EMT associated with the vacuum,  the corresponding pressure is defined through the $ii$th-components. Any of them can be utilized owing to the assumed  homogeneity and isotropy. So, it is just enough to compute the VEV of the  $11$th-component\footnote{One can either compute the VEV of the  $T_{11}$  component, as we do here,  or use the  formula \eqref{ScalarFieldPressure},  which allows to compute the vacuum pressure from the $00th$ component and the trace of the EMT. The result is the same, of course, owing to the isotropy of vacuum. In Ref. \cite{Moreno-Pulido:2022phq},  for instance, we presented the computation of the pressure for the scalar fields using this second method.}:
\begin{equation}\label{ScalarFieldPressure2}
P_{\rm vac}(M)=\frac{\left\langle T_{11}^{\rm vac}\right\rangle}{a^2}=-\rho_{\Lambda}(M)+\frac{\left\langle T_{11}^{\rm \delta\psi} \right\rangle^{\rm ren}(M)}{a^2}\,.
\end{equation}
From \eqref{EMTFermion} and using once more the expansion of the spin-1/2 fermion fields in Fourier modes  (cf. \hyperref[sec:appendixB]{Appendix\,\ref{sec:appendixB}} and \hyperref[sec:appendixC]{Appendix\,\ref{sec:appendixC}})  the result can be expressed  in the following way:
\begin{align}
    \begin{split}
        \left\langle T_{11}^{\delta\psi} \right\rangle=\frac{1}{2\pi^2a}\int_0^\infty dk k^2 P_k,
    \end{split}
\end{align}
with
\begin{align}
    P_k\equiv -\frac{2k}{3a}\left(h_{k}^{\rm I} h_k^{\rm II *}+h_k^{\rm I*}h_{k}^{\rm II}\right)
\end{align}
and where the explicit expressions (in WKB-expanded form) for the fermion modes $h_k^{\rm I}$ and $h_k^{\rm II }$ can be found in the appendices.
Notice that there is a relation between $\rho_k$ and $P_k$,
\begin{align}
    P_k=-\frac{\rho^\prime_k}{3\mathcal{H}}\,,
\end{align}
which follows from  \eqref{eq:producths} using the mode equations \eqref{h1 and h2 eq}.
This relation  can be used as an alternative way to calculate $\langle T_{11}^{\delta\psi} \rangle$ from $\langle T_{00}^{\delta\psi} \rangle$:
\begin{equation}\label{AlternativeExpression}
    \left\langle T_{11}^{\delta\psi} \right\rangle = -\frac{1}{3\mathcal{H}}\left(\left\langle T_{00}^{\delta\psi} \right\rangle^\prime+\mathcal{H}\left\langle T_{00}^{\delta\psi} \right\rangle\right)\,.
\end{equation}

For the sake of simplicity, the remaining discussions of this section will be restricted to the case of one single field. We shall retake the multifield case in \hyperref[sec:CombinedBandF]{Sec.\,\ref{sec:CombinedBandF}}.  Following the same steps and considerations made in the previous sections for the $00th$-component of the EMT, we reach the following expression for the renormalized value of the VEV of the $11th$-component of the EMT up to $6th$ adiabatic order:
\begin{align*}
\begin{split}
\left<T^{\delta\psi}_{11}\right>_{\rm ren}^{(0-6)}(M)&=-\frac{a^2}{32\pi^2}\big(M^4+3m_\psi^4-4m_\psi^2M^2-2m_\psi^4\ln\frac{m_\psi^2}{M^2}\big)\\
&+\frac{1}{48\pi^2}\left(M^2-m_\psi^2+m_\psi^2\ln\frac{m_\psi^2}{M^2}\right)\mathcal{H}^2+\frac{1}{24 \pi^2}\left(M^2-m_\psi^2+m_\psi^2\ln\frac{m_\psi^2}{M^2}\right) \mathcal{H}^\prime
\end{split}
\end{align*}
\begin{align}\label{ren ED sol}
\begin{split}
&+\frac{1}{20160 \pi^2 a^4 m_\psi^2}\left(-245\mathcal{H}^2\left(\mathcal{H}^\prime\right)^2+8\left(\mathcal{H}^\prime\right)^3-98\mathcal{H}^3\mathcal{H}^{ \prime\prime}+35\left(\mathcal{H}^{\prime\prime}\right)^2-62\mathcal{H}^2\mathcal{H}^{\prime\prime\prime}\right. \\
&\left.+204\mathcal{H}^4\mathcal{H}^\prime -66\mathcal{H}\mathcal{H}^\prime\mathcal{H}^{\prime\prime}+56\mathcal{H}^\prime\mathcal{H}^{\prime\prime\prime}+42\mathcal{H}\mathcal{H}^{\prime\prime\prime\prime}-6\mathcal{H}^{\prime\prime\prime\prime\prime}\right)\\
&=-\frac{a^2}{32\pi^2}\big(M^4+3m_\psi^4-4m_\psi^2M^2-2m_\psi^4\ln\frac{m_\psi^2}{M^2}\big)\\
&+\frac{a^2}{16\pi^2}\left(M^2-m_\psi^2+m_\psi^2\ln\frac{m_\psi^2}{M^2}\right)H^2+\frac{a^2}{24 \pi^2}\left(M^2-m_\psi^2+m_\psi^2\ln\frac{m_\psi^2}{M^2}\right)\dot{H}\\
&+\frac{a^2}{20160 \pi^2  m_\psi^2}\left(31H^6+170H^4\dot{H}-45H^2\dot{H}^2-80\dot{H}^3-90H^3\ddot{H}-55\ddot{H}^2\right. \\
&\left.-150H^2\vardot{3}{H}-100\dot{H}\vardot{3}{H}-6H(65\dot{H}\ddot{H}+9\vardot{4}{H})-6\vardot{5}{H}\right).
\end{split}
\end{align}	
We may now  proceed to compute the vacuum EoS for the fermion fields up to the sixth adiabatic order. The best strategy is to compute first the pressure  through  Eq.\,\eqref{ren ED sol}, which can be inserted  into the relation \eqref{ScalarFieldPressure2}.
Using next the VED expression\,\eqref{DensityDefinition} for fermions  --  with $\langle T^{\delta\psi}_{00}\rangle$  given by \eqref{EMT00} --  the vacuum pressure can be seen to be equal to minus the VED plus  some additional terms:
\begin{align}\label{eq:PressureEnergyVacuum}
\begin{split}
P_{\rm vac}(M)&=-\rho_{\rm vac}(M)+\frac{1}{24 \pi^2}\left(M^2-m_\psi^2+m_\psi^2\ln\frac{m_\psi^2}{M^2}\right)\dot{H}\\
    	&+\frac{1}{20160 \pi^2  m_\psi^2}\left(62H^4\dot{H}+144H^2\dot{H}^2-126\dot{H}^3+36H^3\ddot{H}-46\ddot{H}^2-42H^2\vardot{3}{H}\right. \\
		&\left.\phantom{xxxxxxxxxxxx}-118\dot{H}\vardot{3}{H}-6H(42\dot{H}\ddot{H}+6\vardot{4}{H})-6\vardot{5}{H}\right)+\dots
\end{split}
\end{align}
The additional terms represent a small (but worth noticing)  deviation from the classical vacuum EoS relation $P_{\rm vac}=-\rho_{\rm vac}$. The dominant vacuum  EoS is still the classical one up to a leading correction of ${\cal O}(\dot{H})$ (the second term on the \textit{r.h.s} of the above equation)  and several sorts of higher order corrections of  ${\cal O}(H^6)$ indicated in the last two lines.  The $\sim\dot{H}$  correction  in the first line of Eq.\,\eqref{eq:PressureEnergyVacuum} can obviously be relevant for the present universe, and in particular it can modify the equation of state of the vacuum  it to depart from $-1$ at present (cf. \hyperref[sec:EoS-QVacuum]{Sec.\,\ref{sec:EoS-QVacuum}}). The higher order terms in the last two lines, in contrast,  might be relevant only for the very early universe, in principle. However, these terms involve  time derivatives and hence vanish for $H=$const. This fact will have implications for our discussion of RVM-inflation in \hyperref[sec:RVM-inflation]{Sec.\,\ref{sec:RVM-inflation}}, since inflation can be shown to exist in this framework for $H=$const., cf. \hyperref[sec:RVM-inflation]{Sec.\,\ref{sec:RVM-inflation}}. So at the end of the day, the higher order terms in the last two lines of Eq.\,\eqref{eq:PressureEnergyVacuum} become irrelevant both at low and high energies in this framework. The consequence is that the EoS of the quantum vacuum stays very close to $-1$  during inflation, in contrast to the vacuum EoS in subsequent eras of the cosmic evolution  (cf. \hyperref[sec:EoS-QVacuum]{Sec.\,\ref{sec:EoS-QVacuum}}).

\subsection{Trace Anomaly}
It is a very well known result that if a field theory has a classical action which is conformally invariant, then the trace of the classical EMT vanishes exactly. For this  it is necessary to work with a massless field, otherwise the presence of a mass breaks the symmetry since it introduces a fixed length scale. For instance, for a massless scalar field,
\begin{align}
    \lim\limits_{\xi \to 1/6} \lim\limits_{m_\phi \to 0 } T_{\rm Cl.} \left(\phi\right)=0.
\end{align}
 This follows immediately from \eqref{KG} and \eqref{EMTScalarField}. However, it is also true that this simple result does not hold when one takes into account the quantum effects from  the scalar field and constitutes the scalar part of the conformal anomaly,\cite{Birrell:1982ix}.  This follows after a careful study of the diverging part of the vacuum effective action, $W_{\rm eff}^{\rm Div}$, in which $W_{\rm eff}$ was defined in  Eq.\,\eqref{eq:DefWeff}. The part   $W_{\rm eff}^{\rm Div}$ is not conformally invariant for an arbitrary number of spacetime dimensions $n$ (although   $W_{\rm eff}$  is so in the massless limit), except for the case $n=4$. As a consequence, $W_{\rm eff}^{\rm Div}$ receives a finite payoff for $n\to 4$ owing to the existing pole $1/(n-4)$ in it. Correspondingly, the  VEV of the on-shell EMT receives a nontrivial contribution in the massless limit coming from the divergent part of the effective action, even in the case $\xi=1/6$:
 \begin{align}\label{TraceLimitScalar}
   \lim\limits_{m_\phi\to 0}\lim\limits_{\xi \to 1/6}  \left\langle T^{\delta \phi} \right\rangle =-\lim\limits_{m_\phi\rightarrow 0} m_\phi^2 \left\langle \delta\phi^2\right\rangle\,.
 \end{align}
 The term $\left\langle \delta\phi^2\right\rangle$ contains some elements of $4th$ adiabatic order proportional to $1/m_\phi^2$, so that the corresponding limit results in a finite contribution.
 The same idea applies in the fermionic case,
 \begin{align}\label{TraceLimitFermion}
    \lim_{m_\psi \to 0}\left\langle T^{\delta\psi}\right\rangle=-\lim_{m_\psi \to 0} m_\psi \left\langle \bar{\psi} {\psi}\right\rangle\,.
 \end{align}
 Here the term $\left\langle \bar{\psi} {\psi}\right\rangle$ contains $4th$ adiabatic order terms that are proportional to $1/m_\psi$ which make the limit non-trivial.
 Technically speaking \eqref{TraceLimitScalar} and \eqref{TraceLimitFermion} are not yet what we call the {\it trace anomaly} or {\it conformal anomaly} . This is due to the fact that the total effective action is conformally invariant and the corresponding EMT is traceless, so the part of the trace associated with the finite and divergent parts should be equal but with opposite sign in the conformal limit\,\cite{Birrell:1982ix}. The anomaly is generated from the finite part, so its actual value for the scalar field case is
 \begin{align}\label{AnomalyScalarField}
   \left\langle T^{\delta \phi} \right\rangle^{\rm  Anom.} =-\lim\limits_{m_\phi\rightarrow 0}  \left\langle T^{\delta \phi} \right\rangle = \frac{1}{480\pi^2 a^4}\left(4\mathcal{H}^2\mathcal{H}^\prime -\mathcal{H}^{\prime\prime\prime}\right)=\frac{1}{2880\pi^2}\left(R^{\mu\nu}R_{\mu\nu}-\frac{R^2}{3}+\Box R\right)\,,
 \end{align}
 where the conversion of the anomaly result into an invariant expression in the last step can be performed using the formulae of \hyperref[sec:appendixA]{Appendix\,\ref{sec:appendixA}}.  This result was explicitly verified in the calculation of \cite{Moreno-Pulido:2022phq}.  We remark that  for an arbitrary curved background the expression for the conformal anomaly is more involved\,\cite{Birrell:1982ix}. However,  since the spatially flat  FLRW spacetime  is conformally flat  (i.e. conformal to Minkowski space)  the contribution from the Weyl tensor  vanishes identically and hence also  its square (entering the anomaly).  Additional terms beyond $4th$ adiabatic order decouple when $m_\phi \rightarrow \infty$, satisfying the Appelquist-Carazzone decoupling theorem \cite{Appelquist:1974tg}. These terms are not finite in the massless limit, and hence do not take part of the anomaly.

 In practice we have derived the  anomaly \,\eqref{AnomalyScalarField}  from  the unrenormalized trace of the vacuum  EMT  for scalar fields,  $\left\langle T^{ \delta \phi} \right\rangle$, which is given in full detail  in \cite{Moreno-Pulido:2022phq}.
  The corresponding conformal anomaly for fermions can be similarly extracted from the unrenormalized  $ \left\langle T^{ \delta\psi} \right\rangle$  and it is a bit cumbersome as well, so we shall spare details here.  We limit ourselves to provide the final result. Once more we can recognize the expression of the anomaly as a linear combination of finite terms of adiabatic order 4  which are independent of the mass scale. We find
  \begin{align}\label{AnomalyFermionField}
    \left\langle T^{\delta \psi} \right\rangle^{\rm  Anom.} =-\lim\limits_{m_\psi\rightarrow 0}  \left\langle T^{
    \delta \psi} \right\rangle = \frac{1}{240\pi^2 a^4}\left(7\mathcal{H}^\prime\mathcal{H}^2 -3\mathcal{H}^{\prime\prime\prime}\right)=\frac{11}{2880\pi^2}\left( R^{\mu\nu}R_{\mu\nu}-\frac{R^2}{3}+\frac{6}{11}\Box R\right)\,.
 \end{align}
One natural question is related with the physical consequences of the conformal anomaly. It is well-known that it is a valuable theoretical concept  encoding essential information on the VEV of the renormalized EMT\cite{Birrell:1982ix}, although it need not be itself part of the observable quantities of the renormalized theory. There are some attempts in the literature to remove the anomaly by particular prescriptions or definitions of the renormalized EMT \cite{brown1978energy}. This is also the case of the renormalization procedure employed in this work, as defined in \eqref{RenormalizedEMTScalar} and \eqref{RenormalizedEMTFermion}, where the anomaly has no physical effects. The reason is that the on-mass-shell VEV of the EMT is subtracted at an arbitrary scale, $M$,  up to $4th$ adiabatic order. Since the anomaly is of $4th$ adiabatic order and it is independent of the mass of the fields and, of course, also from the arbitrary renormalization point, it gets cancelled exactly in our ARP renormalization procedure. Alternatively, one can think in terms of the effective action. Indeed, in \cite{Moreno-Pulido:2022phq}, we defined the renormalized effective lagrangian density off-shell at an arbitrary scale $M$,
\begin{align}\label{RenormalizedEffLagrangian}
L_{W}^{\rm Ren}(M)\equiv L_W (m)-L_W^{\rm Div}(M)
\end{align}
and it was shown by expanding it through  the DeWitt-Schwinger series  that  it eventually leads exactly to the same renormalized EMT defined by \eqref{RenormalizedEMTScalar}.  This result was obtained explicitly for a scalar field $\phi$  and can be repeated for fermions, although we shall not provide details here. Now the anomaly is related with the divergent part of the effective Lagrangian, corresponding to the lowest adiabatic orders (up to  $4h$ order).  As a consequence  it gets once more exactly cancelled in \eqref{RenormalizedEffLagrangian}  analogously  to the subtraction of the EMT.

As previously indicated, the anomaly part of the trace is contained in the unrenormalized trace of the EMT (even though the anomaly itself is a finite part of it).  In our framework, however,  the anomaly cancels since the anomaly is independent of the mass scale and our  renormalized VEV of the EMT is defined through a subtraction of its value at two different scales, see equations \eqref{RenormalizedEMTScalar} and \eqref{RenormalizedEMTFermion}.  Thus the conformal anomaly is not involved in the renormalized expressions for the vacuum energy density and pressure in our framework.  Despite it  not having  physical consequences in our approach, the explicit calculation of the anomaly is certainly useful as a nontrivial cross-check of our intermediate results.

%\newpage
\section{Combined fermionic and bosonic contributions}\label{sec:CombinedBandF}
Let us now determine the combined vacuum contributions from a multiplicity of non-interacting fermionic and bosonic degrees of freedom. As defined before (cf. \hyperref[sec:RGEforVED]{Sec.\,\ref{sec:RGEforVED}} ), we consider $N_{\rm f}$ species of fermion fields with  masses $m_{\psi,\ell}$  ($\ell\in\{1,2,\dots , N_{\rm f}\}$), and $N_{\rm s}$ scalar field species with  masses $m_{\phi, j}$ ($j\in \{1,2,\dots, N_{\rm s}\}$).

\subsection{Running vacuum from an arbitrary number of quantized matter fields}
The renormalized expression of the vacuum energy density is, in that case,
\begin{align}\label{VEDMulticomponent}
    \rho_{\rm vac}(M,H)=\rho_\Lambda (M)+\frac{\sum_{j=1}^{N_{\rm s}} \left\langle T_{00}^{\delta \phi_j}\right\rangle_{\rm ren} (M,H)+\sum_{\ell=1}^{N_{\rm f}}\left\langle T_{00}^{\delta \psi_\ell}\right\rangle_{\rm ren} (M,H)}{a^2}\,.
\end{align}
Exactly as we did in \hyperref[sec:RenZPEfermions]{Sec.\,\ref{sec:RenZPEfermions}},  we need to  subtract Einstein's equations at two different
energy scales $M$ and $M_0$ in order to obtain the running of the couplings with the change of the scale.  We find:
\begin{equation}
\begin{split}
&\sum_{j=1}^{N_{\rm s}} \left(\left\langle T_{00}^{\delta \phi_j}\right\rangle_{\rm ren} (M,H)-\left\langle T_{00}^{\delta \phi_j}\right\rangle_{\rm ren} (M_0,H)\right)+\sum_{\ell=1}^{N_{\rm f}}\left(\left\langle T_{00}^{\delta \psi_\ell}\right\rangle_{\rm ren} (M,H)-\left\langle T_{00}^{\delta \psi_\ell}\right\rangle_{\rm ren} (M_0,H)\right)\\
&=\sum_{j=1}^{N_{\rm s}}\Bigg[\frac{a^2}{128\pi^2}\left(-M^4+M_0^4+4m_{\phi_j}^2 \left(M^2-M_0^2\right)-2m_{\phi_j}^4\ln\frac{M^2}{M_0^2}\right)\\
&+\frac{3\left(\xi_j-\frac{1}{6}\right)a^2H^2}{16\pi^2}\left(M^2-M_0^2-m_{\phi_j}^2 \ln \frac{M^2}{M_0^2}\right)+\frac{9\left(\xi_j-\frac{1}{6}\right)^2 a^2}{16\pi^2}\left(\dot{H}^2-2\ddot{H}H-6H^2\dot{H}\right)\ln \frac{M^2}{M_0^2}\Bigg]\\
&+\sum_{\ell=1}^{N_{\rm f}}\Bigg[\frac{a^2}{32\pi^2}\left(M^4-M_0^4-4m_{\psi_\ell}^2 \left(M^2-M_0^2\right)+2m_{\psi_\ell}^4\ln\frac{M^2}{M_0^2}\right)+\frac{a^2H^2}{16\pi^2}\left(M_0^2-M^2+m_{\psi_\ell}^2 \ln \frac{M^2}{M_0^2}\right)\Bigg]\\
&=\left(\rho_\Lambda (M)-\rho_\Lambda (M_0)\right)g_{00}+\left(\frac{1}{8\pi G{(M)}}-\frac{1}{8\pi G(M_0)}\right)G_{00}+\left(a_1 (M)-a_1(M_0)\right)\leftidx{^{(1)}}{\!H}_{00}\,.
\end{split}
\end{equation}
Notice the appearance of the $00th$ component of $\leftidx{^{(1)}}{\!H}_{\mu\nu}$, which is a HD  tensor of  ${\cal O}(H^4)$,  hence of adiabatic order $4$, see \hyperref[sec:appendixA]{Appendix\,\ref{sec:appendixA}}.  Its presence in the generalized Einstein's GR equations is indispensable for renormalization purposes and  constitutes  a UV completion of the field equations.  No additional HD tensors are needed for conformally flat spacetimes\,\cite{Birrell:1982ix}.  In our case, $\leftidx{^{(1)}}{\!H}_{\mu\nu}$  is necessary for the renormalization of the  short-distance effects produced by the quantum fluctuations of the scalar fields,  as these  involve  ${\cal O}(H^4)$  corrections.  However, as previously indicated  in \hyperref[sec:RenZPEfermions]{Sec.\,\ref{sec:RenZPEfermions}}, the renormalized EMT for fermions does not contain ${\cal O}(H^4)$ terms.  By using the expression of $\leftidx{^{(1)}}{\!H}_{00}$ in \hyperref[sec:appendixA]{Appendix\,\ref{sec:appendixA}} we can recognize the tensorial structure of the various terms, and from it we can pin down immediately the running of the couplings:
\begin{equation}\label{eq:deltaLambdaquartic}
\begin{split}
  \rho_\Lambda (M)-\rho_\Lambda (M_0)&=\frac{1}{128\pi^2}\left(-4N_{\rm f}+N_{\rm s}\right)\left(M^4-M_0^4\right)+\frac{1}{32\pi^2}\left(4\sum_{\ell=1}^{N_{\rm f}}m_{\psi_\ell}^2 -\sum_{j=1}^{N_{\rm s}}m_{\phi_j}^2\right)\left(M^2-M_0^2\right)\\
    &+\frac{1}{64\pi^2}\left(-4\sum_{\ell=1}^{N_{\rm f}} m_{\psi_\ell}^4+\sum_{j=1}^{N_{\rm s}}m_{\phi_j}^4\right)\ln \frac{M^2}{M_0^2}\,,
\end{split}
\end{equation}
\begin{equation}\label{eq:RunningKappa}
\begin{split}
  \frac{1}{8\pi G (M)}-\frac{1}{8\pi G(M_0)}&=\frac{1}{48\pi^2}\left(-N_{\rm f}+3\sum_{j=1}^{N_{\rm s}}\left(\xi_j-\frac{1}{6}\right)\right)\left(M^2-M_0^2\right)\\
  &+\frac{1}{48\pi^2}\left(\sum_{\ell=1}^{N_{\rm f}}m_{\psi_\ell}^2 -3\sum_{j=1}^{N_{\rm s}}\left(\xi_j-\frac{1}{6}\right)m_{\phi_j}^2 \right)\ln \frac{M^2}{M_0^2}\,,
\end{split}
\end{equation}
\begin{equation}
\begin{split}
  a_1 (M)- a_1 (M_0)&=-\frac{1}{32\pi^2}\sum_{j=1}^{N_{\rm s}}\left(\xi_j-\frac{1}{6}\right)^2\ln\frac{M^2}{M_0^2}\,.
\end{split}
\end{equation}
From the above formulas we can now use  Eq.\,\eqref{VEDMulticomponent} to find out the difference between the values of the VED at two different scales:
\begin{align}\label{rho vac 2}
\begin{split}
    \rho_{\rm vac} (M,H)-\rho_{\rm vac} (M_0,H_0)&=\frac{3}{16\pi^2}H^2\sum_{j=1}^{N_{\rm s}}\left(\xi_j-\frac{1}{6}\right)\left(M^2-m_{\phi_j}^2+m_{\phi_j}^2 \ln \frac{m_{\phi_j}^2}{M^2}\right)\\
    &-\frac{3}{16\pi^2}H_0^2\sum_{j=1}^{N_{\rm s}}\left(\xi_j-\frac{1}{6}\right)\left(M_0^2-m_{\phi_j}^2+m_{\phi_j}^2 \ln \frac{m_{\phi_j}^2}{M_0^2}\right)\\
    &+\frac{1}{16\pi^2}H^2\sum_{\ell=1}^{ N_{\rm f}}\left(-M^2+m_{\psi_\ell}^2-m_{\psi_\ell}^2 \ln \frac{m_{\psi_\ell}^2}{M^2}\right)\\
    &-\frac{1}{16\pi^2}H_0^2\sum_{\ell=1}^{ N_{\rm f}}\left(-M_0^2+m_{\psi_\ell}^2-m_{\psi_\ell}^2 \ln \frac{m_{\psi_\ell}^2}{M_0^2}\right)\\
    &+\frac{9}{16\pi^2}\left(2H\ddot{H}+6H^2\dot{H}-\dot{H}^2\right)\sum_{j=1}^{ N_{\rm s}}\left(\xi_j-\frac{1}{6}\right)^2\ln \frac{m_{\phi_j}^2}{M^2}\\
    &-\frac{9}{16\pi^2}\left(2H_0\ddot{H}_0+6H_0^2\dot{H}_0-\dot{H}_0^2\right)\sum_{j=1}^{ N_{\rm s}}\left(\xi_j-\frac{1}{6}\right)^2\ln \frac{m_{\phi_j}^2}{M_0^2}\\
    &+\frac{\sum_{j=1}^{ N_{\rm s}} \left\langle T_{00}^{\delta \phi_j}\right\rangle_{\rm ren}^{(6)} (M,H)+\sum_{\ell=1}^{ N_{\rm f}}\left\langle T_{00}^{ \psi_\ell}\right\rangle_{\rm ren}^{(6)} (M,H)}{a^2}\\
    &-\frac{\sum_{j=1}^{ N_{\rm s}} \left\langle T_{00}^{\delta \phi_j}\right\rangle_{\rm ren}^{(6)} (M_0,H_0)+\sum_{\ell=1}^{ N_{\rm f}}\left\langle T_{00}^{ \psi_\ell}\right\rangle_{\rm ren}^{(6)} (M_0,H_0)}{a^2}+\dots\\
\end{split}
\end{align}
In the last line, the dots collectively represent all the terms of adiabatic 8 or beyond, which are not considered in our analysis. Notice that in the previous expression we have used the important relation \eqref{eq:deltaLambdaquartic}, which is essential to cancel the quartic mass contributions from the matter fields.

Following the same prescription that we used before to derive equations  \eqref{VacuumEnergyDensityScalarField} and \eqref{eq:nueffBososns} for a single scalar field,  we may implement now the scale settings $M=H$ and $M_0=H_0$ in order to compare the evolution of the VED between these two points, in the present case involving the full contributions from all the matter fields. For simplicity, let us call $\rv(H)\equiv\rho_{\rm vac} (H,H)$ and $\rv(H_0)\equiv\rho_{\rm vac} (H_0,H_0)$ when using the above expression \eqref{rho vac 2}. The expansion history times $H$ and $H_0$ can be arbitrary, of course, but for obvious reasons we choose  $H_0=H(t_0)$ to be the value of the Hubble function at the present time, $t_0$, and $H=H(t)$ a value at a point in our past ($t<t_0$).  Therefore, the running of the VED between these two points can be expressed as follows:
\begin{align*}
\begin{split}
    \rho_{\rm vac} (H)-\rho_{\rm vac} (H_0)&=\frac{3}{16\pi^2}H^2\sum_{j=1}^{N_{\rm s}}\left(\xi_j-\frac{1}{6}\right)\left(H^2-m_{\phi_j}^2+m_{\phi_j}^2 \ln \frac{m_{\phi_j}^2}{H^2}\right)\\
    &-\frac{3}{16\pi^2}H_0^2\sum_{j=1}^{N_{\rm s}}\left(\xi_j-\frac{1}{6}\right)\left(H_0^2-m_{\phi_j}^2+m_{\phi_j}^2 \ln \frac{m_{\phi_j}^2}{H_0^2}\right)\\
    &+\frac{1}{16\pi^2}H^2\sum_{\ell=1}^{ N_{\rm f}}\left(-H^2+m_{\psi_\ell}^2-m_{\psi_\ell}^2 \ln \frac{m_{\psi_\ell}^2}{H^2}\right)\\
    \end{split}
\end{align*}
\begin{align}\label{eq:DiffHHH0H0}
\begin{split}
    &\phantom{aaaaaaaaaaaaaaaaa}-\frac{1}{16\pi^2}H_0^2\sum_{\ell=1}^{ N_{\rm f}}\left(-H_0^2+m_{\psi_\ell}^2-m_{\psi_\ell}^2 \ln \frac{m_{\psi_\ell}^2}{H_0^2}\right)\\
    &\phantom{aaaaaaaaaaaaaaaaaa}+\frac{9}{16\pi^2}\left(2H\ddot{H}+6H^2\dot{H}-\dot{H}^2\right)\sum_{j=1}^{ N_{\rm s}}\left(\xi_j-\frac{1}{6}\right)^2\ln \frac{m_{\phi_j}^2}{H^2}\\
    &\phantom{aaaaaaaaaaaaaaaaa}-\frac{9}{16\pi^2}\left(2H_0\ddot{H}_0+6H_0^2\dot{H}_0-\dot{H}_0^2\right)\sum_{j=1}^{ N_{\rm s}}\left(\xi_j-\frac{1}{6}\right)^2\ln \frac{m_{\phi_j}^2}{H_0^2}\\
    &\phantom{aaaaaaaaaaaaaaaaa}+\frac{\sum_{j=1}^{ N_{\rm s}} \left\langle T_{00}^{\delta \phi_j}\right\rangle_{\rm ren}^{(6)} (H,H)+\sum_{\ell=1}^{ N_{\rm f}}\left\langle T_{00}^{ \psi_\ell}\right\rangle_{\rm ren}^{(6)} (H,H)}{a^2}\\
    &\phantom{aaaaaaaaaaaaaaaaa}-\frac{\sum_{j=1}^{ N_{\rm s}} \left\langle T_{00}^{\delta \phi_j}\right\rangle_{\rm ren}^{(6)} (H_0,H_0)+\sum_{\ell=1}^{ N_{\rm f}}\left\langle T_{00}^{ \psi_\ell}\right\rangle_{\rm ren}^{(6)} (H_0,H_0)}{a^2}\,.\\
\end{split}
\end{align}
Obviously,  if the point $H$ is in the nearby past we can neglect all the ${\cal O}(H^4)$ terms generated in the above expression since they are much smaller than the ${\cal O}(H^2)$ contributions.  We will do this in the next section, where we  study in more detail the low energy regime, in particular the late time universe where we live.  Let us however clarify that the ${\cal O}(H^2)$  terms are dominant not only for the late time universe around our time, but in actual fact for the entire post-inflationary regime.

Finally,  we can extract the running of the gravitational constant from eq.\,\eqref{eq:RunningKappa}, with the following result:
\begin{equation}\label{G(M)}
G(M)=\frac{G(M_0)}{1+\frac{G(M_0)}{2\pi}\left(\sum\limits_{j=1}^{N_{\rm s}}\left(\xi_j-\frac{1}{6}\right)-\frac{N_{\rm f}}{3}\right)(M^2-M_0^2)+\frac{G(M_0)}{2\pi}\left(\sum\limits_{\ell=1}^{N_{\rm f}}\frac{m_{\psi_\ell}^2}{3} -\sum\limits_{j=1}^{N_{\rm s}}\left(\xi_j-\frac{1}{6}\right)m_{\phi_j}^2 \right)\ln \frac{M^2}{M_0^2}}\,.
\end{equation}
We  discuss the running of $\rv$ and  $G$ in terms of $H$ in the next section.
%%%%%%%%%%%%%%%%%%%%%%%%

\subsection{The low energy regime: evolution of $\rv$ and $G$  in the present universe}\label{sec:RVMpresentUniverse}

Of paramount importance is the evolution of the VED and of the gravitational coupling $G$  in the low energy regime, especially around our time.  Therefore, following our prescription, we evaluate \eqref{eq:DiffHHH0H0} for the late universe, when the dominant powers of $H$ are the $H^2$ ones. Such an expression  then boils down to
\begin{align}\label{LowEnergyRegime}
    \rho_{\rm vac}(H)=\rho_{\rm vac}(H_0)+\frac{3\nu_{\rm eff}(H)}{8\pi}\mpl^2\left(H^2-H_0^2\right)+\mathcal{O}(H^4)\,,
\end{align}
where the function $\nu_{\rm eff}(H)$ is defined as follows:
\begin{align}\label{eq:nueffALLFIELDS}
    \begin{split}
\nu_{\rm eff} (H)&=\frac{1}{2\pi}\Bigg[\sum_{j=1}^{N_{\rm s}}\left(\xi_j-\frac{1}{6}\right)\frac{m_{\phi_j}^2}{\mpl^2}\left(\ln\frac{m_{\phi_j}^2}{H_0^2}-1\right)-\frac{1}{3}\sum_{\ell=1}^{N_{\rm f}}\frac{m_{\psi_\ell}^2}{\mpl^2}\left(\ln\frac{m_{\psi_\ell}^2}{H_0^2}-1\right)\\
    &+\frac{H^2}{H^2-H_0^2}\ln\frac{H^2}{H_0^2}\left(\frac{1}{3}\sum_{\ell=1}^{N_{\rm f}}\frac{m_{\psi_\ell}^2}{\mpl^2 }-\sum_{j=1}^{N_{\rm s}}\left(\xi_j-\frac{1}{6}\right)\frac{m_{\phi_j}^2}{\mpl^2}\right)\Bigg]\,.
    \end{split}
\end{align}
It is indeed a function evolving with the Hubble rate,  but is almost constant since the dependence on $H$ is very mild, as we shall make manifest  in a moment.
%Alternatively, it can be written as
%\begin{eqnarray}\label{eq:nueffALLFIELDS}
%  \begin{split}
 %   \nu_{\rm eff}\label{nueff(H)} (H)&=\frac{1}{2\pi}\Bigg[\sum_{j=1}^{N_{\rm s}}\left(\xi_j-\frac{1}{6}\right)\frac{m_{\phi_j}^2}{\mpl^2}\left(\ln\frac{m_{\phi_j}^2}{H^2}-1\right)- % \frac{1}{3}\sum_{\ell=1}^{N_{\rm f}}\frac{m_{\psi_\ell}^2}{\mpl^2}\left(\ln\frac{m_{\psi_\ell}^2}{H^2}-1\right)\\
 %   &+\frac{H_0^2}{H^2-H_0^2}\ln\frac{H^2}{H_0^2} \left(\frac{1}{3}\sum_{\ell=1}^{N_{\rm f}}\frac{m_{\psi_\ell}^2}{\mpl^2 }-\sum_{j=1}^{N_{\rm s}}\left(\xi_j-\frac{1}{6}\right)\frac{m_{\phi_j}^2}{\mpl^2}\right)\Bigg].
%    \end{split}
%\end{eqnarray}
Let us emphasize that the $\mathcal{O}(H^4)$ terms correcting the \textit{r.h.s.} of Eq.\,\eqref{LowEnergyRegime} are completely irrelevant for the current universe, and hence  they can be safely  ignored for the  FLRW regime, that is to say,  during the entire period  following the inflationary stage (cf.  next section). Therefore, equation \eqref{LowEnergyRegime} should actually be relevant for the full cosmological evolution that is accessible (directly or indirectly) to our physical measurements and observations.

It is convenient to define the parameter
\begin{align}\label{epsilon}
    \epsilon \equiv \frac{1}{2\pi}\left(\sum_{j=1}^{N_{\rm s}}\left(\xi_j-\frac{1}{6}\right)\frac{m_{\phi_j}^2}{\mpl^2}-\frac{1}{3}\sum_{\ell=1}^{N_{\rm f}}\frac{m_{\psi_\ell}^2}{\mpl^2}\right)\,.
\end{align}
This parameter is connected to the $\beta$-function \eqref{eq:Totalbetafunctions}  at low energies. Indeed, when we consider $M=H$ in the low energy regime, it is obvious that $H^2\ll m^2$ for any particle mass, and hence Eq.\,\eqref{eq:Totalbetafunctions} reduces to
 \begin{equation}\label{eq:epsiloncoeffbeta}
\beta_{\rv}=-\frac{3}{4\pi}\,\epsilon\,\mpl^2\,H^2\,,
  \end{equation}
within  a very good approximation. Quite obviously  we can see that  $\epsilon$ plays the role of coefficient of the low-energy $\beta$-function of the VED.  However, the eventual  coefficient that effectively  controls the final evolution of the VED is actually enhanced with respect to $\epsilon$ by a big logarithmic factor. To see this,  let us take the current limit ($H\to H_0$) of the function \eqref{eq:nueffALLFIELDS}:
\begin{align}\label{nueff0}
\nu_{\rm eff}^0 \equiv \lim\limits_{H\rightarrow H_0}\nu_{\rm eff}(H)=\frac{1}{2\pi}\left[\sum_{j=1}^{N_{\rm s}}\left(\xi_j-\frac{1}{6}\right)\frac{m_{\phi_j}^2}{\mpl^2}\left(\ln \frac{m_{\phi_j}^2}{H_0^2}-2\right)-\frac{1}{3}\sum_{\ell=1}^{N_{\rm f}}\frac{m_{\psi_\ell}^2}{\mpl^2}\left(\ln \frac{m_{\psi_\ell}^2}{H_0^2}-2\right)\right]\,.
\end{align}
A simple rearrangement now shows that  we can rephrase \eqref{eq:nueffALLFIELDS}  in terms of $\epsilon$ and $\nueff^0$:
\begin{align}\label{Approxnueff(H)}
    \nu_{\rm eff}(H)=\nu_{\rm eff}^0+ \left(1-\frac{H^2}{H^2-H_0^2}\ln \frac{H^2}{H_0^2}\right)\epsilon\,.
\end{align}
This  formula is exact, but in practice some simplifications are perfectly possible.  For example, consider the big logarithms $\ln m_i^2/H_0^2$ (with $m_i$ any particle mass, boson or fermion)  involved in $\nueff^0$ but not in $\epsilon$.   For any known massive particle, we have  $\ln m_i^2/H_0^2\gg 1$, this being true even for the lightest neutrinos (recall that $H_0\sim 10^{-42}$ GeV$=10^{-30}$ meV). Typically $ \ln m_i^2/H_0^2={\cal O}(100)$ in all cases.  But as a matter of fact the only relevant contributions to $\nueff(H)$  come from the heavy massive particles  that belong to a GUT at a characteristic scale  $M_X\sim 10^{16}$GeV.  For these particles  (whether bosons or fermions, with masses $m_i\sim M_X$) we have   $m_i^2/\mpl^2\sim M_X^2/\mpl^2$ and this number is not so small since it may thrust the value of $\nueff$  up to  $\nueff\sim 10^{-3}$,  if one takes into account the large multiplicities of heavy fields existing in a typical GUT. This was first estimated  long ago in \cite{Sola:2007sv}.
Thus,  it is natural to expect $\left|\nu_{\rm eff}^0\right| \gg \left|\epsilon\right|$.
 It follows that we can safely neglect the term proportional to  $\epsilon$ in \eqref{Approxnueff(H)}.
It also means that we can neglect the very mild time-dependence of $\nu_{\rm eff}(H)$   and replace  it  with the constant coefficient $\nu_{\rm eff}^0$ in which $H$ is evaluated at the current time, $H=H_0$.   By the same token we can also ignore the numerical additive terms accompanying the big logarithms in \eqref{nueff0}.  All in all, in very good approximation the evolution of the vacuum energy density can be described through  the formula
\begin{align}\label{eq:VEDtotapprox}
\rho_{\rm vac}(H)= \rho_{\rm vac}(H_0)+\frac{3\nu_{\rm eff}}{8\pi}\mpl^2 (H^2-H_0^2)\,,
\end{align}
with an effective $\nueff\simeq\nueff^0$ given by
\begin{align}\label{nueffapprox}
\nu_{\rm eff} =\frac{1}{2\pi}\left[\sum_{j=1}^{N_{\rm s}}\left(\xi_j-\frac{1}{6}\right)\frac{m_{\phi_j}^2}{\mpl^2}\,\ln \frac{m_{\phi_j}^2}{H_0^2}-\frac{1}{3}\sum_{\ell=1}^{N_{\rm f}}\frac{m_{\psi_\ell}^2}{\mpl^2}\,\ln \frac{m_{\psi_\ell}^2}{H_0^2}\right]\,.
\end{align}
As it turns out, in practice  it all amounts to replace $\nueff(H)\to\nueff$ in \eqref{LowEnergyRegime} since the result still retains a great degree of accuracy.
From the previous two equations, coefficient $\nueff$ is seen to play the role of $\beta$-function  for the running vacuum directly as a function of $H$.  If we compare \eqref{nueffapprox} with \eqref{epsilon} we can see that $\nueff$ and $\epsilon$ are roughly `proportional'  through a big log:
\begin{equation}\label{eq:relationepsilonnueff}
  \nueff\sim \epsilon \ln \frac{m_i^2}{H_0^2}\sim {\cal O}(100)\,\epsilon\,.
\end{equation}
Despite there being a summation over different masses, and hence such a proportionality not being strict,  the above relation is nevertheless true in order of magnitude.  The presence of the  big log factor in $\nueff$ makes the running of the VED faster than the tiny value of $\epsilon$ might convey at first sight.  On the other hand, as we shall see below, it is $\epsilon$ alone that controls the (much softer) running of the gravitational coupling $G$, which does not receive any enhancement from big log factors.

Equations \eqref{eq:VEDtotapprox} and \eqref{nueffapprox} suffice to study the behavior of the VED near our time\footnote{It is apparent that for one single neutral scalar field and no fermion field the above expressions   reduce to the formulas \eqref{VacuumEnergyDensityScalarField} and \eqref{eq:nueffBososns}, as should be expected.}.
 These simplified  formulas have been previously used  de facto to fit the value of $\nueff$ from  the latest cosmological data, see e.g.\,\cite{SolaPeracaula:2021gxi,SolaPeracaula:2023swx} and references therein.  Here, however, we provide for the first time the full theoretical structure behind this parameter in the QFT context from the quantum effects induced by an arbitrary number of quantized matter fields.  The typical fitting value obtained in the mentioned reference is $\nueff\sim 10^{-3}$ and positive, which is well within the said expectations.  This phenomenological determination  picks up of course the net outcome from the various quantum matter fields involved in \eqref{nueffapprox}, which at this point  cannot be discriminated in an individual way.

Finally, insofar as  the running gravitational constant is concerned, it can be written using the same renormalization scale as follows:
\begin{equation}\label{eq:GHExactllfields}
G(H)=\frac{G_N}{1+\frac{1}{2\pi}\left(\sum\limits_{j=1}^{N_{\rm s}}\left(\xi_j-\frac{1}{6}\right)-\frac{N_{\rm f}}{3}\right)\frac{H^2-H_0^2}{\mpl^2}-\epsilon\ln \frac{H^2}{H_0^2}}\,.
\end{equation}
{The former expression can be derived straightforwardly from eq.\,\eqref{G(M)} by setting $M=H$ and $M_0=H_0$, with $H_0$ being the current value of the Hubble function, and we have defined  $G_N\equiv G(M_0)$  (the current value of the gravitational coupling). We follow  exactly the same recipe as for the VED.
In the low energy regime, where $H^2\ll \mpl^2$, we can approximate with high accuracy the former expression by just
\begin{equation}\label{eq:GHAllfields}
G(H)=\frac{G_N}{1-\epsilon\ln \frac{H^2}{H_0^2}}\,,
\end{equation}
where $\epsilon$ receives contributions from all the matter fields, see Eq.\,\eqref{epsilon}. Recall from \eqref{eq:relationepsilonnueff} that $|\epsilon|\ll|\nueff|$, and also that  the running of  $G(H)$ is logarithmic, in contrast to the running of $\rv(H)$ which is quadratic in $H^2$ at low energies.  Therefore, the running of $G$ is much lesser than that of the VED. It may, however, be interesting to note that when $H$ approaches $\mpl$ the term $\sim H^2/\mpl^2$ in the denominator of the more accurate formula Eq.\,\eqref{eq:GHExactllfields} could be dominant over the logarithmic one.  If the multiplicity of matter  fields is large enough, such a term could make the gravitational coupling to evolve asymptotically free at very large energies when we approach the Planck scale.  Until that point,  $G$ increases at high energies for $\epsilon>0$. Beyond that point, decreases.

As an additional cross-check, we can see that the running of the vacuum energy density and of the  gravitational coupling  are compatible through the Bianchi identity. This can be translated into a local energy exchange between the vacuum fluid and the background gravitational field due to the quantum fluctuations. A deeper insight on the local (covariant) energy conservation and the Bianchi identity can be found in the previous works \cite{Moreno-Pulido:2022phq,Moreno-Pulido:2022upl}, where the reader may find a detailed derivation of the logarithmic evolution law, that is to say, equation\,\eqref{eq:GHAllfields}, in the simpler scenario of one real scalar field.
In fact, one finds that the $\beta$-function \eqref{eq:RGEVED1} for the VED running is crucially involved also in the local conservation law of the VED, which can be written in two alternative ways\cite{Moreno-Pulido:2022phq}:
 \begin{equation}\label{eq:NonConserVED2}
\dot{\rho}_{\rm vac}+3H\left(\rv+P_{\rm vac}\right)=\frac{\dot{M}}{M}\,\beta_{\rv}=- \frac{\dot{G}}{G}\,\frac{3H^2}{8\pi G}\,.
\end{equation}
The first equality expresses the fact that the non-conservation of the VED is due to both the running of $\rv$ with $M$  (i.e. the fact that $\beta_{\rv}\neq 0$)  and also to the cosmic time dependence of $M$ (viz. $\dot{M}\neq 0$), whereas the second equality is a direct reflex of the Bianchi identity in Einstein's equations with variable $\rv$ and $G$, and hence provides a link between the time variation of the VED and that of the gravitational coupling $G$.
 The former equation does not depend on the number or nature of the fields involved, and holds as long as the  matter components are covariantly conserved on their own. The non-conservation of $\rv$, however, preserves the Bianchi identity thanks to the corresponding running of the gravitational coupling.  This does not preclude, however, that one can still formulate scenarios where matter can exchange energy with $\rv$, but we do not address this situation here.
 Taking the leading terms of  $\beta_{\rv}$ from the \textit{r.h.s.} of \eqref{eq:Totalbetafunctions} for the present universe, and setting $M=H$,  we immediately obtain the following differential equation
\begin{equation}\label{MixedConservationApprox3}
\frac{1}{G^2}\frac{dG}{dt}=\frac{2\epsilon}{G_N} \frac{\dot{H}}{H}\,,
\end{equation}
where $\epsilon$ is the full expression \eqref{epsilon} involving the contributions from all the matter fields.
Its solution is precisely Eq.\,\eqref{eq:GHAllfields}, as can be readily checked.

It is also interesting to note from the above formulas that this framework predicts a (mild) cosmic time variation of the ``fundamental constants'', such as the gravitational coupling $G$ and $\rv$, as a function of  $H(t)$. Whence, it  implies a small evolution of these `constants' with the cosmological expansion. The possibility for such a variation is not new. It has long been  discussed in the literature\,\cite{Uzan:2010pm} and is still a hot matter of debate and of intensive test by different groups, see also \cite{Barontini:2021mvu} and the ample bibliography provided in them. Specific theoretical models accounting for such a possible variation are manifold, and in some cases they imply a time-dependence of the running couplings and masses in the particle and nuclear physics world, see e.g. \cite{Calmet:2001nu,Calmet:2017czo}. While most of the proposals are based on strict particle physics scenarios, in particular on GUT's, testing the evolution of the VED in curved spacetime  is a novel feature suggested in our framework, which was actually put forward on more phenomenological grounds sometime ago  in \cite{Fritzsch:2012qc,Fritzsch:2015lua,Fritzsch:2016ewd}. The QFT calculations presented in the current work provide indeed a solid theoretical support to these same ideas but from first principles. Recall the golden rule in this arena: when one ``fundamental constant'' varies, then all of them vary!

The formulas discussed above concern important epochs of the cosmological expansion such as the radiation-dominated epoch, matter-dominated epoch and the current epoch in which the vacuum energy resurfaces and became finally dominant over matter. During the entire FLRW regime the dominant power of the Hubble rate in the VED is $H^2$ or $\dot{H}$ (which are of the same adiabatic order). The terms with powers of $H$ (or of equal adiabatic order)  higher than $H^2$ (indicated by $\mathcal{O}(H^4)$ in \eqref{LowEnergyRegime}) acquire real relevance much early on in the expansion history  since only during the most primitive stages of the universe we encounter  a truly  high energy scenario. As previously anticipated, an interesting feature regarding these higher powers is that they can provide us with  a possible mechanism for inflation. Namely, if the early cosmic era possesses a short period where $H$ remains approximately constant and very large (typically near a GUT scale), the universe may go through a phase of exponential expansion in which the VED starts from a huge  value which then quickly decays into radiation and triggers the ordinary FLRW regime. This situation is possible also in the RVM framework, and is called `RVM-inflation'\cite{Moreno-Pulido:2022phq,Sola:2015rra}. See also \cite{Mavromatos:2020kzj} for a `stringy' version.  We  reassess RVM-inflation in the next section in the extended QFT context of the present considerations, where we now have both scalar and fermion fields.

\subsection{Inflation from running vacuum}\label{sec:RVM-inflation}

 It was noted in \cite{Moreno-Pulido:2022phq} that the quantum effects computed from the adiabatic expansion lead to higher powers of the Hubble rate and its derivatives, which are irrelevant for the current universe but capable to bring about inflation in the very early universe. They are characterized by a short period where $H$=const., provided this constant value is, of course, very large, namely around a characteristic GUT scale. The regime $H$=const. in our case  is totally  unrelated to the ground state  of a scalar field potential and therefore this new mechanism does not require  any \textit{ad hoc} inflaton field.  As said, it  is called  `RVM-inflation'.  Here we consider the contribution from the fermions fields and provide a formula for the dominant term of the energy density receiving contributions from an arbitrary number of non-minimally coupled scalar fields and also an arbitrary number of fermions fields.  The payoff from the latter stems from setting $H=$const in Eq.\,\eqref{EMT00},  where we can see that all the time derivatives of the Hubble rate vanish except for a single term which is proportional to $H^6/m_\psi^2$.   The contribution from a non-minimally coupled scalar field was computed in \cite{Moreno-Pulido:2022phq} and here we just combine it with that of fermions assuming any number of both species.  Overall, we find that the total VED involving the contributions from bosons and fermions  at very high energies (hence relevant for triggering RVM-inflation in the very early universe) can be put in the following fashion:
 \begin{equation}\label{eq:RVMinflationCombined}
 \begin{split}
\rv^{\rm inf}=&C_{\rm inf}H^6+F(\dot{H}, \ddot{H},\vardot{3}{H}...),
\end{split}
\end{equation}
where
\begin{equation}\label{eq:Cinf}
C_{\rm inf}\equiv\frac{1}{80\pi^2}\left\{\sum_ j^{N_{\rm s}}\frac{1}{m_{\phi_j}^2}\left[\left(\xi_j-\frac16\right)-\frac{2}{63}-360\left(\xi_j-\frac16\right)^3\right] - \frac{31}{252}\sum_\ell^{N_{\rm f}} \frac{1}{m_{\psi_\ell}^2}\right\}\,.
\end{equation}
The terms collected in the function $ F(\dot{H}, \ddot{H},\vardot{3}{H}...)$ depend on different combinations of powers of $H$ involving derivatives of $H$ in all cases, and hence they all vanish for $H=$const. Thus   $F=0$ for $H=$const.  Overall we see that the dominant contribution is of the form $\rv^{\rm inf}\propto H^6$ with a complicated coefficient $C_{\rm inf}$ which depends on the number of scalar and fermions fields, their masses, multiplicities and also on the non-minimal couplings of the different scalars. In the case of fermions this contribution is seen to be negative-definite, whereas in the case of the scalars it can be positive.
 Let us note that during the inflationary period the EoS of the quantum vacuum is essentially $-1$, with very tiny deviations caused by terms which depend on the various time derivatives of the Hubble rate. To the extent that the condition $H$=const. is fulfilled these deviations are extremely small, see Eq.\,\eqref{eq:PressureEnergyVacuum}.   In the next section we shall see that, in contrast,  the EoS of the quantum vacuum in the present time can deviate from $-1$ by a small amount which is not as negligible as in the very early universe and therefore could be detected and even mimic quintessence behavior.

The solution of the cosmological equations proceeds along the same lines as in \cite{Moreno-Pulido:2022phq}, except that now the fermionic contribution is also included but it only modifies the specific coefficient of $H^6$. Therefore, one finds again that a short period of inflation can be generated with $H\approx$ const. and subsequently the vacuum decays quickly into radiation\,\cite{Moreno-Pulido:2022phq}:
\begin{equation}\label{eq:SolHaInflation}
H(a)=\frac{H_I}{\left(1+\hat{a}^8\right)},
\end{equation}
\begin{eqnarray}\label{rhodensities}
\rho_r(\hat{a})=\rI\,{\hat{a}^8}\left(1+\hat{a}^8\right)^{-\frac{3}{2}}\,\,, \ \ \ \ \ \ \ \ \
\rv(\hat{a})={\rI} \left(1+\hat{a}^{8}\right)^{-\frac{3}{2}}\,,
\end{eqnarray}
in which $\hat{a}\equiv a/\astar$,  with  $\astar$ is the transition point from the regime of vacuum dominance into one of radiation dominance, which can be estimated to be around $\astar\sim 10^{-30}$  (see Eq.\,\eqref{eq:astar} below). Moreover,  $H_I$ and $\rho_I$ are the value of $H$ and $\rho_{\rm vac}$, respectively, at the beginning of inflation, with $\rho_I= C_{\rm inf}H_I^6$. Applying the Friedmann equation, we find
\begin{equation}\label{eq:HIdef}
    H_I=\left(\frac{3}{8\pi G_I C_{\rm inf}}\right)^{1/4}\,,
\end{equation}
\begin{equation}\label{eq:rhoIdef}
    \rho_I=\frac{3}{8\pi G_I}H_I^2=\frac{3}{8\pi G_I}\left(\frac{3}{8\pi G_I C_{\rm inf}}\right)^{1/2}=C_{\rm inf}^{-1/2}\left(\frac{3}{8\pi G_I }\right)^{3/2}\,,
\end{equation}
where $G_I\equiv G(H_I)$ is the gravitational coupling at $H=H_I$, the latter being value of the Hubble parameter at the inflationary era.  Needless to say, the difference between $G_I$ and the usual $G_N$ is not very important here since the running of $G$ is logarithmic, and hence the effect is very small as compared to the fast evolution of the $H^6$ term, so in practice we can neglect the running of $G$ for these considerations.   To trigger inflation in an effective way, we must have a positive coefficient $C_{\rm inf}>0$. In the light of Eq.\,\eqref{eq:Cinf} we can see that this is perfectly possible since the couplings $\xi_j$ and masses of the fields can take a variety of values that make this possible, as can be shown in a devoted study that will be presented elsewhere.

The masses of the relevant fields involved must be very large,  say around a typical GUT scale, $M_X\sim 10^{16}$ GeV. This may not be obvious at first sight. A naive interpretation of the higher order terms of the VED, which are related with the $6th$ adiabatic order of the ZPE (see equation \eqref{EMT00}), may give the erroneous impression that the relevant masses are to be the lightest possible ones, but this is by no means true since in such a situation the adiabatic expansion would break down. On the other hand, the analysis through the Friedmann equations reveals the correct dependency of the VED and of the Hubble function on the masses during the inflationary regime. From equation \eqref{eq:Cinf} it is obvious that $C_{\rm inf}^{-1/2}\propto m_{\phi,\psi}$, where the notation stands for a linear combination of the typical masses of the matter fields. The inflationary parameters \eqref{eq:HIdef}-\eqref{eq:rhoIdef}, therefore, depend on a positive power of $m_{\phi,\psi}$, and as a result the process of RVM-inflation is actually dominated by the heaviest masses, in contrast to naive expectations; namely, masses $m_{\phi,\psi}\sim M_X\sim 10^{16}$GeV of order of a typical GUT, as mentioned above. It follows that the same heavy masses which may generate a mild (but non-negligible) quadratic running $\sim H^2$ of the VED (with a coefficient $\nu_{\rm eff}\sim 10^{-3}$) in the late universe  can also be responsible for driving fast inflation in the primeval stages of the cosmological evolution.  To see this feature more explicitly, let us recall that the differential equation driving the Hubble function in the presence of a high power $H^6$ in the VED\,\eqref{eq:RVMinflationCombined} reads \cite{Lima:2013dmf,Perico:2013mna,SolaPeracaula:2019kfm}
\be\label{Hdot2}
\dot H + \frac{3}{2} \, (1 + \omega_m) \, H^2 \, \Big( 1  - \frac{H^4}{H_I^4} \Big) =0\,,
\ee
where $\omega_m=1/3$ is the EoS of matter in the relativistic epoch, and $H_I$ is given in \eqref{eq:HIdef}. We have neglected the influence of the term $H^2$ and also of the constant term in the very early universe. It is obvious from \eqref{Hdot2}  that there is a constant solution $H=H_I$ to that equation, which is precisely the one which triggers the inflationary period.  From this observation one can then solve equation \eqref{Hdot2} exactly to find Eq.\,\eqref{eq:SolHaInflation}. The latter shows clearly the departure of $H$ from $H_I$ when $\hat{a}>1$ (i.e. $a>a_*$). The inflationary phase actually occurs during the short period when the departure remains small, namely when $H$ remains approximately constant, $H\simeq H_I$. During such period the $F$-term on the \textit{r.h.s.} of Eq.\,\eqref{eq:RVMinflationCombined} just vanishes,
 $F(\dot{H}, \ddot{H},\vardot{3}{H}...)=0$, since all dependence on $H$ is through time derivatives.
From equations \eqref{eq:HIdef} and \eqref{eq:rhoIdef}, we find that the order of magnitude of the physical scales involved in RVM-inflation is the following:
\begin{equation}\label{eq:HIrhoI}
H_I\sim\left(M_X\,\mpl\right)^{1/2}\sim 10^{17}\,{\rm GeV}\,,\ \ \ \ \ \ \ \  \rho_I\sim M_X\,\mpl^3\sim\left(10^{18} {\rm GeV}\right)^4\,,
\end{equation}
up to numerical coefficients and  multiplicity factors, of course. Thus,  if the masses of the relevant matter fields lie in the expected range for a GUT, the right order of magnitude for the relevant physical parameters at the inflationary epoch can be obtained.  Keeping in mind  that the mechanism of RVM-inflation can also be motivated in `stringy' scenarios\,\cite{Mavromatos:2020kzj,Mavromatos:2021urx,Mavromatos:2021sew,Basilakos:2020qmu,Basilakos:2019acj,Mavromatos:2022gtr,Gomez-Valent:2023hov,Mavromatos:2023opj},  it should be natural to expect RVM-infation in the range between the GUT scale and the Planck scale.  This is exactly what the above estimates suggest in order of magnitude.

One more observation is in order. One can easily check from \eqref{rhodensities} that for $\hat{a}\gg 1$ (i.e.  $a\gg\astar$) we retrieve the standard decaying behavior of  radiation, $\rho_r(a)\sim a^{-4}$. This condition enforces the following relation between $\rho_r^0$ (the current value of radiation energy density) and $\rho_I$ (the energy density at the inflationary time):
\begin{equation}
     \rho_{\rm I}\approx\rho_{\rm r}^0 \astar^{-4}\,.
\end{equation}
Following the line of the previous estimations, it yields an equality point between vacuum energy and radiation around the value
\begin{equation}\label{eq:astar}
\astar=
\left(\Omega_r ^0\,\frac{\rho_c^0}{\rho_I}\right)^{\frac{1}{4}}\simeq \left(10^{-4}\,\frac{10^{-47}}{10^{72}}\right)^{1/4}\sim 10^{-30}\,,
\end{equation}
where $\rho_c^0\sim 10^{-47}$ GeV$^4$ is the current critical density.
In the meantime the vacuum energy becomes negligible and does not disturb primordial BBN, see  \cite{Moreno-Pulido:2022phq,Moreno-Pulido:2022upl}  for more details.  See also \cite{Lima:2013dmf,Perico:2013mna,Sola:2015rra,SolaPeracaula:2019kfm}   for interesting phenomenological applications prior to the QFT treatment of RVM-inflation,  first presented in\cite{Moreno-Pulido:2022phq}.

Finally, we should stress that RVM-inflation is genuinely different from e.g. Starobinsky's inflation\,\cite{Starobinsky:1980te}, as explained in detail in \cite{Sola:2015rra}. While it may be natural to conceive that a consistent inflationary model of the very early Universe
should be a good candidate for an effective theory of quantum gravity, at least at energies much less than the Planck scale, RVM-inflation reveals itself as one such possible candidate, all the more if we take into account that  a low-energy `stringy' version of RVM-inflation has been also identified and sharing most of the virtues of the current QFT formulation\,\cite{Mavromatos:2020kzj,Mavromatos:2021urx}.

%\newpage
\subsection{Equation of state of the quantum vacuum}\label{sec:EoS-QVacuum}
The quantum effects of the fields have an imprint on the vacuum equation of state, which is not exactly the traditional one $P_{\rm vac}=-\rho_{\rm vac}$.  From the expressions of the renormalized energy density and pressure of the vacuum that have been obtained in the previous sections and considering their generalization to an arbitrary number of fermions and scalars, we arrive at the following expression for the EoS of the quantum vacuum:
%\begin{equation*}
 %   \begin{split}
 %       \omega_{\rm vac}(H)&=\frac{P_{\rm vac}(H)}{\rho_{\rm vac}(H)}\\
 %       &=-1+\frac{1}{8\pi^2 \rho_{\rm vac}(H)}\Bigg\{\sum_{\ell=1}^{N_{\rm f}}  \Bigg[\frac{\dot{H}}{3}\left(H^2-m_{\psi_\ell}^2+m_{\psi_\ell}^2\ln \frac{m_{\psi_\ell}^2}{H^2}\right)\Bigg]\\
 %       &\phantom{xxxxxxxxxxxxxxxxx}+\sum_{j=1}^{N_{\rm s}}\Bigg[\left(\xi_j-\frac{1}{6}\right)\dot{H}\left(m_{\phi_j}^2-H^2-m_{\phi_j}^2 \ln \frac{m_{\phi_j}^2}{H^2}\right)\\
  %      &\phantom{xxxxxxxxxxxxxxxxxx}-3\left(\xi_j-\frac{1}{6}\right)^2\left(6\dot{H}^2+3H\ddot{H}+\vardot{3}{H}\right)\ln \frac{m_{\phi_j}^2}{H^2}\Bigg]\Bigg\}+\mathcal{O}\left(\frac{H^6}{m^2}\right) \\
  %       \end{split}
%\end{equation*}
\begin{equation}\label{eq:EoScombined}
    \begin{split}
    \omega_{\rm vac}(H)   &=-1+\frac{1}{8\pi^2\rho_{\rm vac}(H)}\Bigg\{\Bigg[\frac{1}{3}\sum_{\ell=1}^{N_{\rm f}} \left(H^2-m_{\psi_\ell}^2+m_{\psi_\ell}^2\ln \frac{m_{\psi_\ell}^2}{H^2}\right)\\
        &\phantom{xxxxxxxxxxxxxx}+\sum_{j=1}^{N_{\rm s}}\left(\xi_j-\frac{1}{6}\right)\left(m_{\phi_j}^2-H^2-m_{\phi_j}^2 \ln\frac{m_{\phi_j}^2}{H^2}\right)\Bigg]\dot{H}\\
        &\phantom{xxxxxxxxxxxxxx}-3\left[\sum_{j=1}^{N_{\rm s}}\left(\xi_j-\frac{1}{6}\right)^2\ln \frac{m_{\phi_j}^2}{H^2}\right]\left(6\dot{H}^2+3H\ddot{H}+\vardot{3}{H}\right)\Bigg\}+\mathcal{O}\left(H^6\right)\,.
    \end{split}
\end{equation}
Here  $\mathcal{O}\left(H^6\right)$ stands for the terms of adiabatic order $6$ or higher,  such as  $H^6/m^2$, $\ddot{H}^2/m^2,\dots$ ($m=m_{\psi_j}, m_{\phi_\ell}$ ).
%coming from $\langle T_{00}^{\delta \psi_\ell} \rangle^{(6)}_{\rm ren}(H,H),$ $\langle T_{00}^{\delta \phi_\ell}\rangle^{(6)}_{\rm ren}(H,H),$ $\langle T_{11}^{\delta \psi_\ell}\rangle^{(6)}_{\rm ren}(H,H)$,  $\langle T_{11}^{\delta \phi_\ell}\rangle^{(6)}_{\rm ren}(H,H)$ and higher order.
All these higher order terms can be neglected during the postinflationary era. %so that we can perfectly avoid them in what follows.
The above  relation shows that the quantum vacuum EoS is of dynamical nature. As we can see there is a deviation from the rigid value $-1$, which is the traditional EoS ascribed to the cosmological constant in the $\Lambda$CDM framework. The correction terms due to both bosonic and fermionic fields are small in the present era, in comparison with the constant term $-1$, but need not be negligible since the particle masses involved can be from a typical GUT, and hence one can estimate that the effective parameter  $\nu_{\rm eff}$  -- the parameter defined in Eq.\,\eqref{nueffapprox} --could reach up to $ 10^{-3}$\cite{Sola:2007sv}.  Furthermore, if we focus only on the ${\cal O}(\dot{H})\sim {\cal O}(H^2) $ terms relevant for the current universe and the radiation epoch we may also neglect the higher order adiabatic terms of ${\cal O}(H^4)$  in the last lines of Eq.\,\eqref{eq:EoScombined}.  Following the steps of \cite{Moreno-Pulido:2022upl}, the EoS can finally be written in a rather compact form as a function of the cosmological redshift:
\begin{equation} \label{ApproximateEos}
\wv= -1+\frac{\left[\nu_{\rm eff}+\epsilon\left(1-\ln E^2 (z) \right)\right] \left[\Omega_{\rm m}^0 (1+z)^3+\frac{4\Omega_{\rm r}^0}{3}(1+z)^4\right]}{\Omega_{\rm vac}^0+\nu_{\rm eff}\left(-1+E^2 (z)\right)-\epsilon\left(1-E^2(z)+E^2(z) \ln E^2(z) \right)}\,,%+{\cal O}\left(\nueff^2\right)\,,
\end{equation}
where $\nu_{\rm eff}$  contains the combined effects from fermions and bosons, see Eq.\,\eqref{nueffapprox}, and we have defined the normalized Hubble rate with respect to the present time ($H_0$):
\begin{align}\label{eq:Equadrat}
  E^2 (z)\equiv\frac{H(z)}{H_0}= \Omega_{\rm vac}^0+\Omega_{\rm m}^0 \left(1+z\right)^3+\Omega_{\rm r}^0 \left(1+z\right)^4\,.
\end{align}
Here  $\Omega_{\rm vac}^0=\rho_{\rm vac}^0/\rho_{\rm c}^0 \approx 0.7$, $\Omega_{\rm m}^0=\rho_{\rm m}^0/\rho_{\rm c}^0 \approx 0.3$ and $\Omega_{\rm r}^0=\rho_{\rm r}^0/\rho_{\rm c}^0 \approx 10^{-4}$ are the current fractions of vacuum energy, dust-like matter and radiation, respectively.   The EoS  formula may be further simplified if we neglect the effect of  the small coefficient $\epsilon$  in Eq.\,\eqref{Approxnueff(H)}.  This is justified since $\big|\nu_{\rm eff}\big| \gg \big| \epsilon \big|$ owing to the logarithmic extra terms $\ln m^2/H_0^2$ contained in $\nu_{\rm eff}^0$, which can typically be of $\mathcal{O}(100)$, see \eqref{eq:relationepsilonnueff}. Thus, to within a very good approximation, we can write
\begin{equation}\label{ApproximateEos2}
\wv \simeq -1+\nu_{\rm eff} \,\frac{\Omega_{\rm m}^0 (1+z)^3+\frac{4\Omega_{\rm r}^0}{3}(1+z)^4}{\Omega_{\rm vac}^0+\nu_{\rm eff} \left(-1+E^2 (z)\right)}\,.
\end{equation}
Notice that the term proportional to $\nueff$ in the denominator cannot be neglected at large $z$ since it becomes dominant. In this case, the EoS takes on the form
\begin{equation}\label{EosMDE}
\wv \approx -1+ \,\frac{\Omega_{\rm m}^0 (1+z)^3+\frac{4\Omega_{\rm r}^0}{3}(1+z)^4}{E^2 (z)}\ \ \  \ (z\gg 1)\,,
\end{equation}
where $\nueff$ has cancelled.
For example, for $z$ large enough but within the matter-dominated epoch the dominant term in equation \eqref{eq:Equadrat} is the $\sim (1+z)^3$ one, and we can see from \eqref{EosMDE} that the vacuum EoS then mimics matter since $\wv\simeq 0$. Similarly, at much larger values of $z$ already in the radiation-dominated epoch, where $\sim (1+z)^4$ is the dominant term, then $\wv\simeq 1/3$ and the vacuum imitates radiation. Such a `chameleonic' behavior of the quantum vacuum was first noticed in \cite{Moreno-Pulido:2022upl}, and in fact
the formula for the vacuum EoS that we have found here is a generalization for an arbitrary number of fermion and boson fields of the expression previously found in \cite{Moreno-Pulido:2022upl}.
Last but not least, the evolution of the vacuum EoS in the late universe is no less remarkable and striking. From \eqref{ApproximateEos2} we find
\begin{equation}\label{eq:EoSDeviation}
\wv(z) \simeq   -1+\nueff \frac{\Omega_{\rm m}^0}{\Omega_{\rm vac}^0}(1+z)^3\ \ \ \ \ \ \ \ \ \  \ \  \ (z\lesssim 5)\,.
\end{equation}
In this approximation we recover once more the form \eqref{EqStateScalar}, but in this case with $\nueff$ involving the contributions from all the quantized matter fields.

\section{\jtext{Signatures of the running vacuum and application to the  $\sigma_8$ and $H_0$ tensions}}\label{sec:signatures}

In this final section before presenting our conclusions, we would like to provide a taste of the possible phenomenological consequences of the running vacuum as predicted in the QFT framework described in this work and in \cite{Moreno-Pulido:2020anb, Moreno-Pulido:2022phq}.  We refer the reader to the already existing  literature,  see particularly the references\,\cite{SolaPeracaula:2021gxi,SolaPeracaula:2023swx}, which constitute the latest in a series of studies which have repeatedly  put the RVM to the test over the last years. Some older  phenomenological works include e.g. \cite{Gomez-Valent:2014rxa,Sola:2015wwa,Sola:2016jky,SolaPeracaula:2016qlq,Sola:2017znb,SolaPeracaula:2017esw,Gomez-Valent:2018nib,Gomez-Valent:2017idt} and references therein.

We have seen in Sec. \ref{sec:RVMpresentUniverse} that the dynamical evolution of the VED in the late universe adopts the form \eqref{LowEnergyRegime}.  However,   when considering the tensorial structure of the vacuum in curved spacetime,  one may expect generalized forms of  Eq.\, \eqref{VacuumEMT}, namely involving  nontrivial geometric terms which are not possible in Minkowski  spacetime but are certainly available in a curved background.  Typically we can expect  a generalization of the form
 \begin{equation}\label{eq:GeneralVDE1}
 \left\langle T_{\mu\nu}^{\rm vac} \right\rangle= -\rL g_{\mu\nu} +\left\langle T_{\mu\nu}^{\delta\phi} \right\rangle + \alpha_1 R g_{\mu\nu} +\alpha_2 R_{\mu\nu}+\mathcal{O}(R^2)\,.
\end{equation}
Here  $\alpha_i$ are coefficients of dimension $+2$ in natural units.  For the physics of the late universe, we may obviously discard  the influence from the higher order terms $\mathcal{O}(R^2)$  such as $R^2$, $R_{\mu\nu}R^{{\mu\nu}},\dots$   General covariance indeed leads to a generic form as the above one for $\left\langle T_{\mu\nu}^{\rm vac} \right\rangle$  and specific examples have been discussed in  the literature, see e.g.\,\cite{Maggiore:2010wr,Bilic:2011zm}. Interestingly, these generalized structures are also expected in the aforementioned stringy version of the RVM, see\,\cite{Mavromatos:2020kzj,Mavromatos:2021urx} and also in the recent work \cite{Gomez-Valent:2023hov}.  Notice that in the FLRW metric,  $R$ and $R_{\mu\nu}$  lead  both to $H^2 $ and $\dot{H}$ terms, which are of the same adiabatic order.   It is  therefore natural to expect that a more general structure for the VED takes on the following form\,\cite{Moreno-Pulido:2022phq}:
\begin{equation}\label{eq:RVMvacuumdadensity}
\rv(H) = \frac{3}{8\pi G_N}\left(c_{0} + \nueff{H^2+\tilde{\nu}_{\rm eff}\dot{H}}\right)+{\cal O}(H^4)\,,
\end{equation}
where again the ${\cal O}(H^4)$ terms will be neglected for the present universe, and where coefficients $\nueff$ and $\tilde{\nu}_{\rm eff}$ must be fitted to the observational data\footnote{As explained in\,\cite{Moreno-Pulido:2022phq}, the more general form \eqref{eq:RVMvacuumdadensity} is to be expected not only from the extended geometric structure of the the vacuum \eqref {eq:GeneralVDE1}, but also because the vacuum pressure (being dynamical in the QFT framework)  introduces $\dot{H}$ terms explicitly in the low energy regime. This was already so for scalar fields\,\cite{Moreno-Pulido:2020anb,Moreno-Pulido:2022phq} and it is also the case with fermion fields, as can be seen in  Eq.\,\eqref{eq:PressureEnergyVacuum}.  Coefficients $\nueff$ and  $\tilde{\nu}_{\rm eff}$, therefore, are generally present and both receive contributions from the quantized bosons and fermion matter fields.  }.   To adopt this wider perspective  is particularly important at this point where we want to assess the potential phenomenological implications of the QFT framework elaborated here.  The calculations in this paper attest for the fact that the coefficients $\nu$ and $\tilde{\nu}$ are nonvanishing, and this result alone is very important to substantiate the prediction of vacuum dynamics  from the theoretical point of view.  However, as our calculation has also shown, these coefficients depend on the masses of the quantized  bosons and fermions and also of the non-minimal couplings $\xi_i$.  It is therefore obvious that the overall values of these coefficients cannot be predicted at this point, except roughly  in order of magnitude, and hence in practice they can only be determined accurately from fitting the model to observations.

%%%%%%%%%%%%%%%%%%%%%%%%%%%%%%%%%%%%%%%%%%%%%%%%%%%%%%%%%%%%%%%%%%%%%%%%%%%%%%%%%%%%%%%%
\begin{figure}[t]
\begin{center}
\includegraphics[scale=1.0]{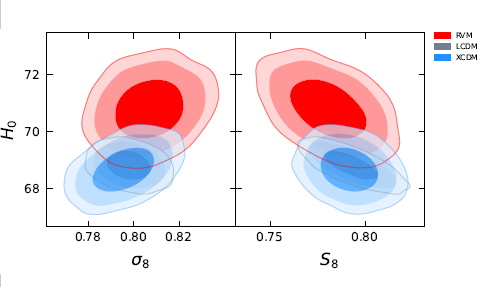}
\end{center}
\caption{Contours  at $1\sigma$,  $2\sigma$ and $3\sigma$ c.l.  in the ($\sigma_8$-$H_0$) and ($S_8$-$H_0$) planes corresponding to the RVM,  $\CC$CDM and XCDM models for the data sets  SnIa+BAO+$H(z)$+LSS+CMB mentioned in the text.  It can be seen that the RVM is quite effective in alleviating the $H_0$ tension and at the same time it reduces  the $S_8$ one. The generic DE parameterization XCDM is unable to do a comparable job.}
\label{Fig1}
\end{figure}
%%%%%%%%%%%%%%%%%%%%%%%%%%%%%%%%%%%%%%%%%%%%%%%%%%%%%%%%%%%%%%%%%%%%%%%%%%%%%%%%%%%%%%%%%%%%%%%%%%%%%%

To simplify the phenomenological analysis  we shall adhere to the approach of \cite{SolaPeracaula:2021gxi,SolaPeracaula:2023swx}, in which   $\tilde{\nu}_{\rm eff}=\nueff/2$.  This reduces the number of parameters to one and the VED  adopts the simpler and suggestive  form
\begin{equation}\label{eq:RRVM}
\rv(H) =\frac{3}{8\pi{G_N}}\left(c_0 + \frac{\nu}{12} {R}\right)\equiv \rv({ R})\,,
\end{equation}
in which  ${R} = 12H^2 + 6\dot{H}$ is the curvature scalar.   We  assume that the matter is locally conserved (as in the standard $\CC$CDM) and  hence the change in the  VED with the expansion is compensated for by the mild evolution of  the gravitational coupling $G$, see Eq.\,\eqref{eq:GHAllfields}\footnote{Notice that while no interaction between matter and vacuum is considered here, in the phenomenological analyses of \cite{SolaPeracaula:2021gxi,SolaPeracaula:2023swx} such an option has also been contemplated and gives rise to a different type of scenarios which we shall not address in the current study.}.
In other words, the VED evolves together with $G$ such that the Bianchi identity can be satisfied, just as explained in Sec.\,\ref{sec:RVMpresentUniverse}, see Eq.\,\eqref{eq:NonConserVED2}.  Following  \cite{SolaPeracaula:2023swx}  one can solve the model, but only numerically since an exact analytic expression is not possible owing to the logarithmic evolution $G=G(H)$ given in Eq.\,\eqref{eq:GHAllfields}.  However, it will suffice to display here the VED dynamics to within linear order in the small parameter $\nueff$:
\begin{equation}\label{eq:VDEm}
\rv(a)=\left(\frac{\Omega_{\rm  vac}^0}{\Omega_m^0}-\frac14\,\nueff\right) \rho_m^0+\frac14\nueff\rho_m^{0}a^{-3}+\mathcal{O}(\nueff^2)\,,
\end{equation}
where as usual  $\Omega_i^0=(8\pi G_N) \rho_i^0/(3 H_0^2)$  are the current cosmological parameters(in FLRW spacetime with flat three-dimensional hypersurfaces).
The rigid cosmological constant case is recovered only for  $\nueff=0$, and then  $\rv=\frac{3 H_0^2}{8\pi G_N} \Omega_{\rm vac}^0=\rvo=$const.  This is, of course,  the situation in the $\CC$CDM model.  However, for nonvanishing $\nueff$ the vacuum energy  exhibits  a mild dynamics since this parameter is small.  This dynamics, together with that of $G(H)$, can help to alleviate the $H_0$ and $\sigma_8$ tensions mentioned in the introduction\,\cite{Perivolaropoulos:2021jda,Abdalla:2022yfr,Dainotti:2023yrk}. This can be seen in the contour plots shown in Fig.\,\ref{Fig1}.  In it, we display  the  $1\sigma$,  $2\sigma$ and $3\sigma$ c.l. contours  in the $(\sigma_8, H_0)$  and $(S_8, H_0)$  planes for the $\CC$CDM, the  RVM with vacuum energy density \eqref{eq:RRVM} and the general XCDM (also called $w$CDM) parameterization of the dark energy\,\cite{Turner:1997npq},  in which the DE density evolves as $a^{-3(1+w)}$, where $w$ is the EoS parameter (assumed to be constant).  In Fig.\,\ref{Fig1}, $S_8 \equiv \sigma_8\sqrt{\Omega_{\rm m}^0/0.3}$ is the usual parameter employed to express the $\sigma_8$ tension since it is the one directly obtained from the  measurements of weak gravitational lensing at low redshift ($z<1$)\cite{DiValentino:2020vvd}.   The observational data sets used  in this analysis involve type Ia supernovae, baryonic acoustic oscillations, cosmic chronometers, large scale structure (LSS)  formation data (on $f\sigma_8$) and the cosmic microwave background observations from Planck 2018 (i.e. the data string SnIa+BAO+$H(z)$+LSS+CMB).  For the plots in Fig.\,\ref{Fig1}, we have taken  the Baseline+$H_0$ dataset of  Ref.\,\cite{SolaPeracaula:2023swx}, which uses the full Planck 2018  TT,TE,EE+lowE likelihood \cite{Planck:2018vyg}. We refer the reader to the aforesaid  reference \cite{SolaPeracaula:2023swx} for a detailed definition of these data  as well as for the methodology used in the fitting analysis.
In the light of the results presented in  Fig.\,\ref{Fig1} ,  it is fair to say  that  the RVM may alleviate significantly the  $H_0$  tension and also to some extend the $S_8$ one (which in any case is far less acute), whereas the generic XCDM parameterization cannot improve the situation of the $\CC$CDM.  From that figure, we may clearly appreciate that  $H_0$ tends to higher values for the RVM as compared to both the $\CC$CDM and the XCDM, and $S_8$ tends to be smaller. Specifically,  we find that the existing  $5\sigma$ tension\cite{Perivolaropoulos:2021jda,Abdalla:2022yfr}  between the measurements of the local value of $H_0$ and from the early universe (CMB)  subsides at the residual level of $<2\sigma$.  This is achieved at no cost for the  $S_8$ tension which does not increase and in fact is slightly reduced.  The results presented in this section are  suggestive of the potential of the RVM  for  improving the description of the cosmological data and in particular to relieve  the $H_0$ and $S_8$ tensions, see \cite{SolaPeracaula:2021gxi,SolaPeracaula:2023swx}, for more details.   The QFT formulation of the dynamical vacuum energy presented in this paper may therefore help in finding  a possible fundamental explanation for the aforementioned tensions.

\jtext{On the other hand, as mentioned previously in the introduction, perhaps the most straightforward quantitative  explanation of the $H_0$ tension (to the best of our knowledge), is the one based on taking into account the cosmological look-back time considered at the redshift where measurements are performed, see the proposal \cite{Capozziello:2023ewq}. While it is conceivable that new physics may not be necessary for the particular issue of the $H_0$ tension, if the interpretation put forward by these authors  is correct, one cannot exclude (as the authors themselves admit)  that new physics beyond the $\CC$CDM model may still be necessary  in order to understand dark energy on fundamental grounds,  as well as dark matter and of course also the ultimate mechanism of inflation in the early universe, all of which suggest to go beyond GR standard cosmology.  In point of fact,  in the framework of the current paper and the preceding ones\,\cite{Moreno-Pulido:2020anb, Moreno-Pulido:2022phq,Moreno-Pulido:2022upl}, we have  provided some solid hints for  a possible fundamental  explanation of the DE and inflation within the context of QFT in curved spacetime.  Amazingly, the same fundamental framework seems to provide a helpful hand to ameliorate the situation with the $H_0$ and $\sigma_8$ tensions.}

Finally, we consider another potential phenomenological implication of the  QFT framework presented here concerning the vacuum energy.  In Sec. \ref{sec:EoS-QVacuum} we have studied the effective EoS of the running vacuum and we have seen that depending on the sign of $\nueff$ it can behave as effective quintessence or effective phantom DE.  In other words, we have shown that the EoS of the quantum vacuum departs from the classical rigid value $\wv=-1$ when the quantum effects are taken into account.  In Fig.\,\ref{Fig2}, we plot the  EoS formula \eqref{ApproximateEos} and we use typical values of $\nueff$ obtained in numerical fits of the RVM (all of them satisfying the condition $|\nueff|\ll1$, as suggested by the QFT calculation).  The remaining basic cosmological parameters are fixed from the best-fit values of \cite{Planck:2018vyg}. We show the results for the two possible signs of $\nueff$ since a final phenomenological analysis of the QFT version of the RVM is not available yet, and hence there is no a priori preference for a sign at this point.  As we can see,  an effective quintessence or phantom behaviour near our time may emerge from a fundamental QFT origin without necessity of invoking real quintessence or phantom fields.  In Fig.\ref{Fig2}, we display the EoS of the quantum vacuum at low energy and also its evolution across the matter-dominated epoch up to the incipient radiation-dominated era. As first advanced in \cite{Moreno-Pulido:2022phq,Moreno-Pulido:2022upl}, the effective EoS of the running vacuum mimics the dominant matter component at high redshift, this being independent of the sign of $\nueff$ since this parameter cancels at large redshift, see Eq.\,\ref{EosMDE}. In the case $\nueff<0$ there appears a fake singularity in the past caused by the fact that the VED becomes zero at this point, but of course it is not a real singularity since neither the density nor the pressure of the vacuum become singular at any point.  This phenomenon is akin to different forms of the DE studied previously in the literature, see e.g. \cite{Sola:2005et,Shafieloo:2005nd,Basilakos:2013vya}.

%%%%%%%%%%%%%%%%%%%%%%%%%%%%%%%%%%%%%%%%%%%%%%%%%%%%%%%%%%%%%%%%%%%%%%%%%%%%%%%%%%%%%%%%
\begin{figure}[t]
\begin{center}
\includegraphics[scale=0.55]{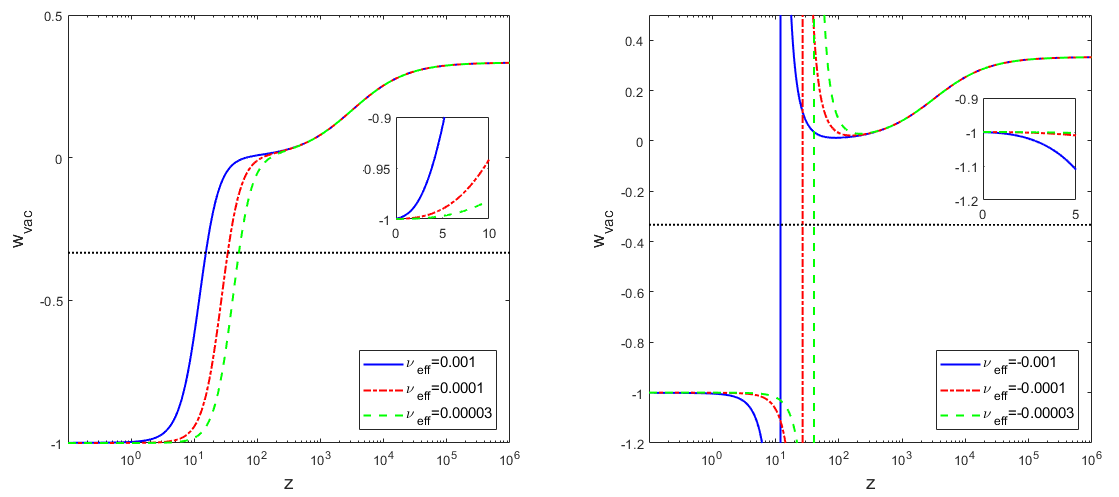}
\end{center}
\caption{Evolution of the vacuum EoS \eqref{ApproximateEos} with the  redshift,  including the quantum effects computed in this work.  We test the behavior for positive (left plot) and negative (right plot) values of the parameter $\nueff$ \jtext{which controls the vacuum dynamics. It  satisfies $|\nueff|\ll 1$ in all cases.}  The running vacuum mimics quintessence for $\nueff>0$, whereas it mimics phantom dark energy for $\nueff<0$.  In the latter case the vacuum density vanishes at some point in the past, and this appears as a vertical asymptote in the EoS plot, but no real singularity occurs in any physical observable.}
\label{Fig2}
\end{figure}
%%%%%%%%%%%%%%%%%%%%%%%%%%%%%%%%%%%%%%%%%%%%%%%%%%%%%%%%%%%%%%%%%%%%%%%%%%%%%%%%%%%%%%%%%%%%%%%%%%%%%%

To summarize, the running vacuum mimics the EoS of the dominant component at a given time of the cosmic evolution, and at present is dynamical. This was first confirmed for scalar fields in \cite{Moreno-Pulido:2022phq,Moreno-Pulido:2022upl}.  In the present work,  we have found  that the formal structure of the EoS does not change when we evaluate the fermionic contribution and therefore similar qualitative behaviors occur.  Overall, we find that  $\wv$ is  $-1$ during inflation; stays  close to 1/3 in the radiation dominated epoch, and then close to 0 in the matter dominated epoch.  Furthermore, and most importantly, at present it mimics quintessence  for $\nueff>0$ and phantom DE for $\nueff<0$.  These situations are realized in different phenomenological versions of the RVM, but the QFT version put forward here will be analyzed in more detail elsewhere in the future.  Be as it may,  a solid and remarkable result is that the quantum EoS deviates from the traditional  value $\wv=-1$ of the classic vacuum.  This is perhaps the most outstanding consequence of the quantum vacuum behavior.  Finally, we note that the vacuum  EoS  in the remote future will be the same as in the very early (inflationary)  universe, namely  $\wv\to -1$, since the cosmic expansion asymptotes towards a new inflationary period.  Only in the de Sitter epochs the quantum vacuum retrieves its conventional EoS $w=-1$.

%\newpage
\section{Conclusions}\label{sec:Conclusions}

In this paper, we have evaluated the contributions to the vacuum energy density (VED) from the quantized matter fields in a semiclassical gravity approach. By using a regularization technique of Quantum Field Theory (QFT) in curved spacetime called adiabatic regularization and making use of a specific (off-shell) subtraction prescription amply tested in previous works\cite{Moreno-Pulido:2020anb, Moreno-Pulido:2022phq,Moreno-Pulido:2022upl}, we have been able to calculate the mode functions and the renormalized zero-point energy (ZPE)  from spin-1/2 quantum fields in a FLRW background up to sixth adiabatic order. Together with the contribution from the $\rL$ term in the Einstein-Hilbert action, we have obtained the properly renormalized VED. Since the corresponding computation for scalar fields had already been accounted for in the aforementioned works, we have put forward here the combined contribution to the VED  from an arbitrary number of quantized matter fields. We did not consider interactions among them, however, as the free field calculation in curved spacetime is already rather cumbersome in itself. One interesting difference between the expression of the ZPE of these two types of fields is that in the fermionic case the terms of fourth adiabatic order (viz. those involving  four time derivatives) are not present.  The final result  is that the overall VED of the quantized matter fields upon adiabatic renormalization appears to be a soft dynamical quantity with the cosmological evolution. This is a most remarkable outcome of the present study.  More specifically, the VED shows up in the form of an expansion in powers of the Hubble rate $H$ and its time derivatives, all these powers being of even adiabatic order, a property which is fully consistent (and expected) from the general covariance of the theory.  Such a series expansion appears to take the canonical form of the running vacuum model (RVM),  see \cite{Sola:2013gha,Peracaula:2022vpx} and references therein.  This means, in particular, that the leading quantum effects obtained for the late universe are of second adiabatic order, thus  $\sim H^2$ and $\sim\dot{H}$. Obviously, this may have consequences for the present universe, and these consequences have been tested in previous phenomenological works. What is more, these quantum effects turn out to have a positive impact on a possible solution to the $\CC$CDM tensions, \jtext{see  Sec.\,\hyperref[sec:signatures]{\ref{sec:signatures}}) and the existing studies \cite{SolaPeracaula:2021gxi,SolaPeracaula:2023swx,Gomez-Valent:2014rxa,Sola:2015wwa,Sola:2016jky,SolaPeracaula:2016qlq,Sola:2017znb,SolaPeracaula:2017esw,Gomez-Valent:2018nib,Gomez-Valent:2017idt}}\footnote{\jtext{In the most recent study \,\cite{SolaPeracaula:2023swx},  fully updated information  is provided on the phenomenological performance of the RVM and its implications on the current $\CC$CDM tensions.}}

\jtext{In this comprehensive work},  we have also discussed some of the theoretical difficulties in trying to renormalize the cosmological term, $\CC$, and its relation with the VED. To start with, it should be emphasized that these are two different concepts that can only be properly related in non-flat spacetime. If $\CC$ is taken to be the physically measured value of the cosmological term at present, $\CC_{\rm phys}$, then its relation with the current VED is $\rvo=\CC_{\rm phys}/(8\pi G_N)$. However, at a formal QFT level these quantities have to be derived from a gravitational action in curved spacetime and a lot more of care needs to be exercised. Leaving for the moment quantum gravity considerations for a better  future (viz. for when, hopefully, the quantum treatment of the gravitational field becomes possible),  the more pedestrian renormalization of $\rv$ in QFT in curved spacetime proves to be already quite helpful at present\cite{Moreno-Pulido:2020anb, Moreno-Pulido:2022phq}. It shows, for example -- and the present works attest once more for this fact --  that the renormalized VED in the FLRW background is a mild dynamical quantity evolving with the cosmic expansion, and hence that $\rvo$ is just its value at present. There is in fact no such thing as a rigid, everlasting,  cosmological constant in the context of  QFT in the FLRW background. In general, $\rv=\rv(H)$ is a function of the Hubble rate and its time derivatives.  What we call the `cosmological constant'  $\Lambda$  appears in our framework as the nearly sustained value of the renormalized quantity  $8\pi G(H)\rho_{\rm vac}(H)$  around (any)  given epoch $H$, where $G(H)$ is the renormalized gravitational coupling, which is also running, although very mildly (logarithmically) with $H$, i.e. $G=G(\ln H)$.  At present, $G(H_0)=G_N$ and $\rv(H_0)=\rvo$, and this defines $\CC_{\rm phys}=8\pi G_N\rvo$ in a precise way  in QFT in  FLRW spacetime (within our renormalization framework).

The longstanding and widespread confusion in the literature about the notion of cosmological constant, $\CC$, and that of  vacuum energy (density), $\rv$, has prevented from achieving a proper treatment of the renormalization of these quantities in cosmological spacetime.  In particular, the attempts to relate these concepts in the context of flat spacetime calculations  are meaningless and their repeated iteration has been highly  counterproducing\,\cite{Peracaula:2022vpx}.

In the simplified scenario considered here, where only interactions with the gravitational background are allowed, the VED is the sum of two contributions, a parameter in the effective action, $\rho_\Lambda$, and the ZPE of the quantized fields. After renormalization, the VED depends on a  scale $M$, and the setting $M=H$ at the end of the calculation allows us to compare the VED at different epochs of the cosmic history, in a manner similar to the standard association made of the renormalization point with a characteristic energy scale of a given process in ordinary gauge theories. Thus the difference between the VED values  at any two points of the cosmological expansion, say $H(t_1)$ and $H(t_2)$,  provides a smooth running of the VED. Remarkably, such an evolution turns out to be free from the undesirable $\sim m^4$ contributions that emerge from the quantized matter fields in other frameworks. As a result there is no fine tuning involved in the evolution of the VED in the present calculation.
The VED, in fact, adopts the standard form of the RVM, which  in the late universe reads $\rho_{\rm vac}(H_2)= \rho_{\rm vac} (H_1)+3\nu_{\rm eff}/(8\pi) \mpl^2 (H_2^2-H_1^2)$, where $H_1$ and $H_2$ can be, for example, the current value, $H_0$, and another value $H$ near our past.  \jtext{The small parameter  $\nu_{\rm eff}$ is  related to the $\beta$-function of the renormalization group running of the VED, whose value has been explicitly computed in this work from the fluctuations of the quantized matter fields. As illustrated in Sec.\,\hyperref[sec:signatures]{\ref{sec:signatures}}, depending on the sign of $\nu_{\rm eff}$  the VED can mimic  quintessence or phantom-like behavior.}

Much earlier in the cosmic history, the higher powers of $H$ (larger than $H^2$ and of even adiabatic  order to preserve covariance) took their turn  and could be relevant for the inflationary regime, in the sense that they had the capacity to trigger inflation through a mechanism that has been called `RVM-inflation'\cite{Moreno-Pulido:2020anb, Moreno-Pulido:2022phq,Moreno-Pulido:2022upl}. While the scalar field contribution to this inflationary mechanism had been computed in the previous references,  in this work we have accounted for the spin-1/2 fermionic contribution and combined the two types of effects for an arbitrary matter content. In both cases (scalar and fermion fields)  the sixth order adiabatic  terms ${\cal O}(H^6)$  had to be computed.
Finally, the renormalized vacuum fluid's pressure, $\Pv$, has been determined  using the same QFT techniques as for $\rv$. Equipped with these nontrivial results the equation of state (EoS) of the quantum vacuum can be computed from first principles.  The entire contribution from quantized matter fields (bosons and fermions) can be encoded in the effective $\nueff$ parameter. We find that the EoS function $\wv=\Pv/\rv$  deviates  from the traditional result  -1, a fact which is  worth emphasizing. This is true in most of the cosmological history, especially after inflation (which is the only period in our past where the vacuum EoS stayed very close to $-1$). It is no less noteworthy, as previously mentioned, that in the late universe, and most particularly near our time, the vacuum EoS behaves as quintessence for $\nu_{\rm eff}>0$,  the latter being the sign most often preferred by the existing phenomenological fits to the overall cosmological data -- see \cite{SolaPeracaula:2021gxi,SolaPeracaula:2023swx}, for example. For higher and higher redshifts during the FLRW regime, the vacuum EoS mimics the equation of state of the dominant matter component (relativistic or non-relativistic) at the corresponding epoch. Such a peculiar behavior of the running vacuum energy density  was referred to as ``chameleonic'' in\,\cite{Moreno-Pulido:2022upl}. The tracking of the EoS of matter by the vacuum ceases to exist in the late universe, where the DE epoch breaks through and  $\wv$ behaves \jtext{as effective quintessence or phantom}, the reason being that the EoS is then in the process  to asymptote towards $-1$ in the remote future, exactly as it was in the primeval inflationary time. In fact, the inflationary process in the late universe is eventually resumed, but very slowly.

Overall, by combining the results from an arbitrary number of  quantized matter fields we find that  the main cosmic running of $\rv$ depends on the  quadratic terms in the boson and fermion masses times the square of the Hubble function, i.e. $\sim m_\psi^2 H^2$ and $\sim m_\phi^2 H^2$. These effects are obviously much softer than the naively expected (hard) contributions $\sim  m_\psi^4$ and  $\sim m_\phi^4$.  As remarked, the soft terms have been profusely tested in phenomenological works on the RVM existing in the literature, \jtext{see e.g.\,\cite{SolaPeracaula:2021gxi,SolaPeracaula:2023swx,Gomez-Valent:2014rxa,Sola:2015wwa,Sola:2016jky,SolaPeracaula:2016qlq,Sola:2017znb,SolaPeracaula:2017esw} -- the latest being \cite{SolaPeracaula:2023swx}}. The QFT effects that we have computed here and in the preceding studies \cite{Moreno-Pulido:2020anb, Moreno-Pulido:2022phq,Moreno-Pulido:2022upl} provide a solid theoretical underpinning of those phenomenological analyses. They even bring to light new relevant features, such as the dynamical character  of the EoS of the quantum vacuum, which is unprecedented in the literature to the best of our knowledge. In particular, they suggest that if in future cosmological observations we can collect clear signs that the EoS of the dark energy deviates from $-1$, such a feature could be explained by the running vacuum. It therefore opens the possibility that such observations (if confirmed)  may be accounted for from  fundamental properties of QFT attributable to the fluctuations of the quantized matter fields in curved spacetime rather than to the existence of \textit{ad hoc} quintessence fields and the like.  This could be an extremely  interesting smoking gun of this approach. The EoS dynamics is prompted here by the virtual quantum effects produced by quantized fermion and boson fields, the same kind of effects which trigger a smooth evolution of the vacuum energy density in cosmological spacetime. As it turns out from the above considerations, in the RVM framework  the need for fundamental quintessence fields and also for inflaton fields subsides dramatically, for they can both be replaced by the running effect of the quantum vacuum.

%\newpage
On pure cosmological/observational grounds, the physical outcome of the theoretical framework presented here can be summarized in the following way.  The renormalization of the vacuum energy density in QFT in the FLRW background  leads to the  non-constancy of the `cosmological constant', $\CC$, in Einstein's equations  and predicts a slow time variation of the  vacuum energy density and of its  equation of state at the present time, which departs slightly from the traditional EoS value $-1$ for the vacuum.   This conclusion emerges from explicit QFT calculations in our approach and may point to a possible explanation for a wide range of cosmological problems that have been dealt with phenomenologically in the past in terms of \textit{ad hoc} quintessence or phantom fields. In our context, the currently measured cosmological `constant' is neither mimicked nor supplanted by any ersatz entity from the already too crammed black box of the dark energy.   We could simply phrase it in a nutshell as follows: the quantum vacuum shows up here as if it were a form of dynamical dark energy,  but it is (quantum) vacuum after all.  In fact, it is the same vacuum producing inflation in the early universe by means of higher (even) powers of the Hubble rate $H^n$ beyond $n=2$,  and still leaving  a smoothly evolving cosmological term in the late universe through the lowest possible power  compatible with general covariance, which is   $H^2$.  Formally, all these powers of the Hubble rate emerge as quantum effects  that are part of the effective action of vacuum. Today's physical cosmological term  appears  as  a quantity directly connected with  the  (properly renormalized)  vacuum energy density in QFT in curved spacetime:  $\Lambda_{\rm phys}=8\pi G\rv$.   As a running parameter,  it  is sensitive to the fluctuations  of the quantized matter fields.  Taking into account that in our framework  the scale  of renormalization is linked  with the cosmological expansion, represented by the Hubble rate, it turns out that $\Lambda_{\rm phys}$,  despite it appearing  as an approximately rigid term during  a typical  cosmic span around any given epoch, it is actually a physical observable in evolution during the entire cosmic history.   Its time  variation,  which is ultimately of quantum origin,  is very small at present but it helps  in relieving the current tensions of the $\CC$CDM and  it might eventually provide an explanation for the cosmic acceleration observed in our Universe from first principles.

\vspace{0.3cm}

\newpage

\subsection*{Acknowledgements}

Two of us (CMP and JSP) are funded by projects  PID2019-105614GB-C21 and FPA2016-76005-C2-1-P (MINECO, Spain), 2017-SGR-929 (Generalitat de Catalunya) and CEX2019-000918-M (ICCUB).  CMP is also partially supported  by  the fellowship 2019 FI$_{-}$B 00351. The work of JSP is  also partially supported by the COST Association Action CA18108  ``Quantum Gravity Phenomenology in the Multimessenger Approach  (QG-MM)''. CMP and JSP also acknowledge participation in the COST Action CA21136 - Addressing observational tensions in cosmology with systematics and fundamental physics (CosmoVerse), supported by COST (European Cooperation in Science and Technology). SC is supported by the Transilvania Fellowship Program, 2022.
CMP and JSP are  very  grateful  to A.  G\' omez-Valent and  J. de Cruz P\'erez  for  the fruitful collaboration over the years in the task of understanding the RVM and its manifold phenomenological implications. \\

\subsection*{Special acknowledgment dedicated to Harald Fritzsch }
We would like to dedicate this comprehensive work on Dark Energy in Quantum Field Theory to the memory of Professor Harald Fritzsch. Those of us (JSP) who  have worked with him over the years  in subjects relating Particle Physics and Cosmology, and participated in  very successful conferences organized by him in so many places in the world (and in particular  in  NTU-Singapore for  a long time),  will never forget such an immense privilege.  Reference \cite{Fritzsch:2012qc}, for instance,  is very much connected with the current work. We have lost a great man and  a giant scientist  who played a key leading role in  understanding our universe. He was indeed among the proposers of Quantum Chromodynamics (QCD),  the gauge theory of strong interactions; namely,  the fundamental physical  theory  which ultimately explains  the structure of matter in the universe from first principles.  Thanks for everything Harald, for your generosity and for your beautiful and powerful science!

\newpage

\appendix
\section{Appendix: Conventions and Useful Formulas}\label{sec:appendixA}
In this work, natural units are used. That is $\hbar=c=1$ and $G_N=1/\mpl^2$, with the usual Planck mass given numerically  as  $\mpl\approx 1.22 \times 10^{19}$ GeV.
Our framework is a Friedmann-Lemaître-Robertson-Walker (FLRW) background with null spatial curvature. The conventions are as in \cite{Moreno-Pulido:2022phq}, i.e.  $(+,+,+)$ in the Misner-Thorne-Wheeler notation\,\cite{Misner:1973prb}:
\begin{enumerate}
    \item[$\bullet$] $g_{\mu\nu}=a^2(\tau) {\rm diag}(-,+,+,+)$, with $\tau$ denoting conformal time.
    \item[ $\bullet$] $R^\lambda_{\mu \nu \sigma}=\partial_\nu \Gamma^\lambda_{\mu\sigma}+\Gamma^\rho_{\mu\sigma}\Gamma^\lambda_{\rho \nu}+\dots$, with $R_{\mu\nu}=R^\lambda_{\mu\lambda\nu}$ and $R=R_{\mu\nu}g^{\mu\nu}$.
    \item[$\bullet$] The Einstein field equations are  $G_{\mu\nu}+\Lambda g_{\mu\nu}=8\pi G T_{\mu\nu}$, with $G_{\mu\nu}\equiv R_{\mu\nu}-\frac{1}{2}R g_{\mu\nu}$.
\end{enumerate}
For derivatives, the notations $()^\prime=d/d\tau$ and $\dot{()}=d/dt$ are used. In particular,  $H=\dot{a}/a$,  $\mathcal{H}\equiv a^\prime / a=a H$,   hence  $a'=a \mathcal{H}=a^2 H$,   $a''=a^3(2 H^2+\dot{H})$ etc.  (see Appendix A.1 of  \cite{Moreno-Pulido:2022phq}).

In this case, the non-vanishing Christoffel symbols in the conformal frame are
\begin{align}
    \Gamma_{00}^0=\mathcal{H},\qquad \Gamma_{ij}^0 =\mathcal{H}\delta_{ij},\qquad \Gamma_{i0}^j=\mathcal{H}\delta_i^{j}.
\end{align}
On the other hand, the Ricci scalar is
\begin{align}
    R=6\frac{a^{\prime\prime}}{a^3}=\frac{6}{a^2}\left(\mathcal{H}^\prime+\mathcal{H}^2\right)=6(2H^2+\dot{H}),
\end{align}
and the $00th$ components of the Ricci and Einstein tensors are
\begin{align}
    R_{00}=-3a^2\left(H^2+\dot{H}\right), \qquad G_{00}=3a^2H^2.
\end{align}
The renormalization program requires taking into account higher derivative (HD) terms in Einstein's equations\,\cite{Birrell:1982ix}. In the particular case of FLRW spacetime, it is enough to consider just one of the characteristic higher order tensors, $\leftidx{^{(1)}}{\!H}_{\mu\nu}$, given by the metric functional derivative of $R^2$ in the effective action (we refer once more to Appendix A.1 of  \cite{Moreno-Pulido:2022phq} for more details):
\begin{align}
\leftidx{^{(1)}}{\!H}_{\mu\nu}=\frac{1}{\sqrt{-g}}\frac{\delta}{\delta g^{\mu\nu}}\int d^4 x \sqrt{-g}\,R^2 =-2\nabla_\mu\nabla_\nu R+2g_{\mu\nu}\Box R-\frac{1}{2}g_{\mu\nu}R^2+2RR_{\mu\nu}.
\end{align}
Its $00th$ component is
\begin{align}
    \leftidx{^{(1)}}{\!H}_{00}=-18a^2 \left(\dot{H}^2-2H\ddot{H}-6H^2\dot{H}\right)
\end{align}
and its $11th$ component is
\begin{align}
    \leftidx{^{(1)}}{\!H}_{11}=-a^2 \left(108H^2 \dot{H}+54\dot{H}^2+72H\ddot{H}+12\vardot{3}{H}\right).
\end{align}
We will also need the invariants
\begin{equation}\label{eq:RmunusquareBoxR}
 R^{\mu\nu} R_{\mu\nu}=\frac{12}{a^4}\left({\cH^\prime}^2+\cH^\prime\cH^2+\cH^4\right)\,,\qquad\Box R=-\frac{6}{a^4}\left(\cH^{\prime\prime\prime}-6\cH'\cH^2\right)\,,
\end{equation}
which hold good for flat three-dimensional FLRW spacetime.

For gamma matrices (in flat spacetime), the standard Dirac basis is chosen for our calculations with spin-1/2 fermions:
\begin{align}
\gamma^0=\begin{pmatrix}
I & 0\\
0 & -I
\end{pmatrix}\qquad \gamma^k =\begin{pmatrix}
0 & \sigma_k\\
-\sigma_k & 0
\end{pmatrix}\,,
\end{align}
 where $\sigma_k$ ($k=1,2,3$)  are the usual Pauli matrices. In terms of the above $\gamma^\alpha$, the curved spacetime $\gamma$-matrices read $\underline{\gamma}^\mu(x)=e^{\mu}_{\,\alpha}(x)\gamma^\alpha $, where $e^{\mu}_{\,\alpha}(x) $ is the vierbein (cf. \hyperref[sec:QuantizFermions]{Sec.\ref{sec:QuantizFermions}}).

%\newpage

\section{Appendix: Adiabatic expansion of the spin-1/2 field modes}\label{sec:appendixB}

In the main text (cf. \hyperref[sec:QuantizFermions]{Sec.\,\ref{sec:QuantizFermions}}) we have presented an iterative procedure which allows us to determine the two types of field modes $h_k^{\rm I}$ and $h_k^{\rm II}$ which are necessary to construct the $2$-component spinor fields. They are functions of both momentum ($k$) and conformal time ($\tau$) and have the following structure:
\be\label{ansatz2}
\begin{split}
&	h^{\rm I}_k(\tau)=\sqrt{\frac{\omega_k-aM}{2\omega_k}}F(\tau)\, e^{-i\int^\tau  W_k (\tilde{\tau}) d\tilde{\tau}},\\
&	h^{\rm II}_k(\tau)=\sqrt{\frac{\omega_k+aM}{2\omega_k}}G(\tau)\,e^{-i\int^\tau  W_k (\tilde{\tau}) d\tilde{\tau}}\,,
\end{split}
\ee
where
\begin{equation}\label{ExpansionF}
F \equiv1+F^{(1)}+F^{(2)}+F^{(3)}+\cdots,
\end{equation}
\begin{equation}\label{ExpomegakansionG}
G \equiv 1+G^{(1)}+G^{(2)}+G^{(3)}+\cdots,
\end{equation}
\begin{equation}\label{ExpansionW}
W_k \equiv \omega_k+\omega_k^{(2)}+\omega_k^{(4)}+\omega_k^{(6)}+\cdots
\end{equation}
Here $W_k$ is a real function playing an analogous role to the effective frequency introduced (with the same notation) in the scalar field case, Eq.\,\eqref{eq:phaseintegral}. The superscript $(n=1,2,3,...)$ indicates that the corresponding quantity is of adiabatic order $n$. The modes \eqref{ansatz2} are constrained to satisfy the normalization condition
\begin{equation}\label{NormalizationConditionMode}
    |h_{k}^{\rm I}|^2+|h_{k}^{\rm II}|^2=1.
\end{equation}
Some of the notation is similar to that of\,\cite{Landete:2013axa, Landete:2013lpa, delRio:2014cha}, although we use conformal metric and different conventions, and moreover we deal with  FLRW spacetime rather than with de Sitter (where the EMT takes a simpler form).  In addition, as explained in the main text, we perform the renormalization at the arbitrary scale $M$  (not at the on-shell point). This is important in order to test the scaling dependence of the renormalized VED, which is the main aim of our calculation.

In what follows we use the notation  $\omega_k  \equiv \sqrt{k^2+M^2 a^2}$.  Recall that unless it is explicitly noted the mass scale involved is the off-shell point $M$.  When the subtraction \eqref{RenormalizedEMTFermion} is implemented within our  renormalization procedure, one just sets $M=m_\psi$ in the subtracted part. The mass $m_\psi$ can be conveniently expanded in even adiabatic orders as
\begin{equation}\label{expansionm}
m_\psi=\sqrt{M^2+ \Delta^2}=M+\frac{\Delta^2}{2M}-\frac{\Delta^4}{8M^3}+\frac{\Delta^6}{16 M^5}+\dots
\end{equation}
After completing  $\ell \geq 1$ steps in the process described in \hyperref[sec:QuantizFermions]{Sec.\,\ref{sec:QuantizFermions}}, we end up with Eq.\,\eqref{GeneralSolution} in the main text, which depends on the following expression:
\begin{equation}\label{GeneralSolutionProduct}
    \Omega_k \Omega_{k,1}\cdots \Omega_{k,\ell-1} = \omega_k+\omega_k^{(1)}+\omega_k^{(2)}+\dots+\omega_k^{(2\ell-1)}+\dots
\end{equation}
where $\omega_k^{(j)}$ represents a function of adiabatic order $j$.
Functions $W_k (\tau,M)$, $F(\tau,M)$ and $G(\tau,M)$ in the ansatz \eqref{ansatz2} can be estimated with the help of this product. However the following clarifications may be necessary to better understand this process, together with the explanations already given in the main text, see \hyperref[sec:QuantizFermions]{Sec.\,\ref{sec:QuantizFermions}}:
%\newpage
%
\begin{itemize}
	\item For $\ell=1$ we only have performed one iterative step, and at this point we need to deal with the square root of
\begin{equation}\label{ExampleOmegak}
\begin{split}
\Omega_k^2=\omega_k^2-i\sigma+a^2\Delta^2=\omega_k^2-iMa^\prime+a^2\Delta^2-\frac{ia\Delta^2}{2M}\frac{a^\prime}{a}+\dots
\end{split}	
\end{equation}
as defined in \eqref{adiabatic order}. The dots ``..." in \eqref{ExampleOmegak} account for terms of adiabatic order 4 and beyond. The square root of the previous result yields
\begin{equation}\label{ExampleOmegakSquareRoot}
\begin{split}
	\Omega_k = &\omega_k+\omega_k^{(1)}+\frac{a^2\omega_k}{8} \left[ \frac{M^2}{\omega_k^4}\left(\frac{a^\prime}{a}\right)^2+\frac{4\Delta^2}{\omega_k^2} \right]\\
	&-\frac{iaM}{16\omega_k}\frac{a^\prime}{a}\left[\frac{4\Delta^2}{M^2}-\frac{4a^2\Delta^2}{\omega_k^2}-\frac{a^2 M^2 }{\omega_k^4}\left(\frac{a^\prime}{a}\right)^2\right]+\dots
\end{split}
\end{equation}
with
\be
\omega_k^{(1)}\equiv-\frac{iaM}{2\omega_k}\frac{a^\prime}{a}\,.
\ee
From the \textit{r.h.s} of \eqref{ExampleOmegakSquareRoot}, the first two terms, $\omega_k$ and $\omega_k^{(1)}$, are used in the first order approximation of the modes (see \,\eqref{h_I(1)} and previous equations in the main text).

Now suppose that we proceed with a further step in the iterative process, $\ell=2$. We have to deal with the square root of
\begin{equation}\label{ExampleOmegak1}
			\Omega_k^2 \Omega_{k,1}^2 =\left(\omega_k^2-iMa^\prime+a^2\Delta^2-\frac{ia\Delta^2 }{2M}\frac{a^\prime}{a}+\dots\right)\left( 1+\epsilon_2\right)\,.\\	
\end{equation}
The introduction of
\begin{equation}\label{Epsilon2expansion}
		\epsilon_2=\epsilon_2^{(2)}+\epsilon_2^{(3)}+\dots\,,
\end{equation}
whose expression can be seen in \eqref{Epsilon2}, does not alter the $0th$  nor the $1st$ orders of \eqref{ExampleOmegak} and \eqref{ExampleOmegakSquareRoot} since $\epsilon_2$ is a sum of terms of adiabatic order $2$ and higher:
\begin{equation}\label{ExampleOmegak1SquareRoot}
\begin{split}
\Omega_k\Omega_{k,1}=&\omega_k+\omega_k^{(1)}+\frac{\omega_k}{2} \epsilon_2^{(2)}+\frac{a^2 \Delta^2}{2\omega_k}+\frac{a^2 M^2}{8\omega_k^3}\left(\frac{a^\prime}{a}\right)^2+\frac{\omega_k}{2}\epsilon_2^{(3)}-\frac{iaM}{4\omega_k}\epsilon_2^{(2)}\frac{a^\prime}{a}\\
&-\frac{ia\Delta^2}{4M\omega_k}\frac{a^\prime}{a}+\frac{ia^3 M\Delta^2 }{4\omega_k^3}\frac{a^\prime}{a}+\frac{ia^3M^3}{16\omega^5_k}\left(\frac{a^\prime}{a}\right)^3+\dots
\end{split}
\end{equation}
Nonetheless, the $2nd, 3rd,\dots$ adiabatic orders of \eqref{ExampleOmegak1SquareRoot} do not coincide with the ones of  \eqref{ExampleOmegakSquareRoot}. Similarly, by going to the next iterative step, $\ell=3$, implies working with the square root of the product
\begin{equation}\label{ExampleOmegak2}
		\Omega_k^2\Omega_{k,1}^2\Omega_{k,2}^2=\left(\omega_k^2-iMa^\prime+a^2\Delta^2+\dots\right)\left( 1+					\epsilon_2\right)\left(1+\epsilon_4\right)\\	
\end{equation}
with	
\begin{equation}\label{Epsilon4expansion}
		\epsilon_4=\epsilon_4^{(4)}+\epsilon_4^{(5)}+\dots
\end{equation}
The introduction of $\epsilon_4$ does not alter neither the 0{\it th}, 1{\it st}, 2{\it nd} nor the 3{\it rd} adiabatic orders of \eqref{ExampleOmegak1} or \eqref{ExampleOmegak1SquareRoot}, since $\epsilon_4$ is a sum of terms of adiabatic order $4$ and higher. By the same token, the 4{\it th} and 5{\it th} adiabatic orders and beyond in \eqref{ExampleOmegak1} do not coincide with the ones in \eqref{ExampleOmegak2}. Similar considerations apply to the square root of these quantities.

We can sum up this by saying that for each iterative step we can compute two consecutive adiabatic orders of \eqref{GeneralSolutionProduct} that will not be altered by the subsequent steps. Then, after $\ell$ steps, the $0,1,\dots, 2\ell-1$ adiabatic orders of the product \eqref{GeneralSolutionProduct} are trustworthy for the estimation of the modes.

%\newpage

 \item The \textit{r.h.s.}  of \eqref{GeneralSolutionProduct} has a pure imaginary part conformed by the odd orders, precisely those which do no take part in the computation of $W_k$:
    \begin{equation}\label{GeneralSolutionExponentialAdiabOrder}
        \begin{split}
            &-i\int^\tau \left(\omega_k+\omega_k^{(1)}+\omega_k^{(2)}+\dots+\omega_k^{(2\ell-1)}\right)d\tilde{\tau}\\
            &=-i\int^\tau \left(W_k^{\left(0-2(\ell-1)\right)}\right)d\tilde{\tau}-i\int^\tau \left(\omega_k^{(1)}+\omega_k^{(3)}+\dots+\omega_k^{(2\ell-1)}\right)d\tilde{\tau}\,.
        \end{split}
    \end{equation}
However, because of the factor $-i$ in front of the integral, the imaginary terms in the integrand become real and are then necessary for the computation of $F$ and $G$ in \eqref{ansatz2}.

\item We did not specify the limits of integration in \eqref{GeneralSolution} for the following reasons. On the one hand, even terms take part in the pure imaginary exponential of \eqref{ansatz2}. Now because the imaginary exponential does not appear in the final result of the relevant quantities that we compute in the main text (since they cancel in the products of a function times its complex conjugate) one might   wrongly be led to conclude that $W_k$ does not influence the calculation of the EMT. This would, however, be incorrect since the derivatives of the modes $h_k^{\rm I,II}$ are present in these calculations. On the other hand, after integrating the odd terms without specifying the limits in the integral, there exists some residual freedom in the form of a set of functions of the momentum only (i.e. not depending on time). These are called $f_k^{(0)},g_k^{(0)},f_k^{(1)}, g_k^{(1)},\dots$ where the superscript indicates the adiabatic order. If our goal is to compute an adiabatic expansion of the modes up to 6{\it th} order  we need 7 arbitrary constants for $h^{\rm I}$, namely $f_k^{(0)},\dots,f_k^{(6)}$. Similarly for $h^{\rm II}$,  which we denote $g_k^{(0)},\dots,g_k^{(6)}$.
	
\item From Eqs. \eqref{h1 and h2 eq} and \eqref{dc eq of spinors}, it is clear that $h_k^{\rm I}(\tau,M)=h_{k}^{\rm II}(\tau,-M)$ so $F(\tau,M)=G(\tau,-M)$ and $f_k^{(n)}(M)=g_k^{(n)}(-M)$.

\end{itemize}

With this considerations in mind let's us put forward the adiabatic expansion of $W_k$ explicitly.
As said, $W_k$ can be specified though the even terms of \eqref{GeneralSolutionProduct}. We are interested to compute at least up to 6{\it th} adiabatic order (that means, at least, $\ell = 4$ steps). Therefore we find:
\begin{equation}\label{WkAdiabaticOrdersExpansion}
W_k^{(0-6)}(\tau)=\omega_k+\omega_k^{(2)}+\omega_k^{(4)}+\omega_k^{(6)}\,,
\end{equation}
with
\begin{equation}\label{Wk0th}
\omega_k^{(0)}=\omega_k\,,
\end{equation}
\vspace{0.5cm}
\begin{equation}\label{Wk2nd}
\omega_k^{(2)}=\frac{a^2\Delta^2}{2\omega_k}-\frac{ a^2 M^2}{8 \omega_k^3}\left(\frac{a^\prime}{a}\right)^2+\frac{5 a^4 M^4}{8\omega_k^5}\left(\frac{a^\prime}{a}\right)^2-\frac{a^2 M^2}{4\omega_k^3}\frac{a^{\prime\prime} }{a}\,,
\end{equation}
\vspace{0.5cm}
\begin{equation}\label{Wk4th}
\begin{split}
\omega_k^{(4)}&=-\frac{a^4\Delta^4}{8\omega_k^3}+\left(-\frac{25 a^6 M^4}{16\omega_k^7}+\frac{23 a^4 M^2}{16\omega_k^5}-\frac{a^2}{8\omega_k^3}\right)\Delta^2 \left(\frac{a^\prime}{a}\right)^2
+\left(\frac{3a^4M^2}{8\omega_k^5}-\frac{ a^2}{4\omega_k^3}\right)\Delta^2  \frac{a^{\prime \prime}}{a}\\
&-\left(\frac{1105a^8 M^8}{128\omega_k^{11}}-\frac{267a^6 M^6}{64\omega_k^9}+\frac{21 a^4 M^4}{128\omega_k^7}\right)\left(\frac{a^\prime}{a}\right)^4+\left(\frac{221a^6 M^6}{32\omega_k^9}-\frac{57 a^4 M^4}{32\omega_k^7}\right)\left(\frac{a^\prime}{a}\right)^2 \frac{a^{\prime\prime}}{a}\\
&-\left(\frac{19a^4 M^4 }{32\omega_k^7}-\frac{a^2 M^2}{32\omega_k^5}\right)\left(\frac{a^{\prime\prime}}{a}\right)^2-\left(\frac{7a^4M^4}{8\omega_k^7}-\frac{a^2 M^2}{16\omega_k^5}\right)\frac{a^{\prime}}{a}\frac{a^{\prime\prime\prime}}{a}+\frac{a^2 M^2}{16\omega_k^5}\frac{a^{\prime\prime\prime\prime}}{a}\,,
\end{split}
\end{equation}
\vspace{0.5cm}
\begin{equation*}
\begin{split}
\omega_k^{(6)}&=\frac{a^6 \Delta^6}{16\omega_k^5}+\left(\frac{175a^8 M^4}{64\omega_k^9}-\frac{215 a^6 M^2 }{64\omega_k^7}+\frac{13a^4}{16\omega_k^5}\right) \Delta^4  \left(\frac{a^\prime}{a}\right)^2-\left(\frac{15a^6 M^2 }{32\omega_k^7}-\frac{3a^4}{8\omega_k^5}\right)\Delta^4 \frac{a^{\prime\prime}}{a}\\
&+\left(\frac{12155 a^{10} M^8}{256 \omega_k^{13}}-\frac{6823a^8 M^6}{128\omega_k^{11}}+\frac{3351 a^6 M^4 }{256\omega_k^9}-\frac{21 a^4 M^2}{64\omega_k^7}\right)\Delta^2 \left(\frac{a^{\prime}}{a}\right)^4\\
&+\left(\frac{133 a^6  M^4 }{64\omega_k^9}-\frac{81 a^4  M^2}{64\omega_k^7}+\frac{a^2}{32\omega_k^5}\right)\Delta^2 \left(\frac{a^{\prime\prime}}{a}\right)^2\\
\end{split}
\end{equation*}
 \begin{equation}\label{Wk6th}
\begin{split}
\phantom{xxxxxx}&-\left(\frac{1989a^8 M^6}{64\omega_k^{11}}-\frac{1725 a^6 M^4}{64\omega_k^9}+\frac{57a^4 M^2}{16\omega_k^7}\right) \Delta^2 \left(\frac{a^{\prime}}{a}\right)^2 \frac{a^{\prime\prime}}{a}\\
&+\left(\frac{49a^6 M^4}{16\omega_k^9}-\frac{61 a^4  M^2 }{32\omega_k^7}+\frac{a^2}{16\omega_k^5}\right)\Delta^2 \frac{a^\prime}{a} \frac{a^{\prime\prime\prime}}{a}-\left(\frac{5a^4   M^2 }{32\omega_k^7}-\frac{a^2}{16\omega_k^5}\right)\Delta^2 \frac{a^{\prime\prime\prime\prime}}{a}\\
&+\left(\frac{414125 a^{12} M^{12}}{1024 \omega_k^{17}} - \frac{
 338935 a^{10} M^{10}}{1024 \omega_k^{15}} + \frac{56271 a^8 M^8 }{1024 \omega_k^{13}} - \frac{869 a^6 M^6}{1024 \omega_k^{11}}\right)\left(\frac{a^\prime}{a}\right)^6\\
& -\left(\frac{248475 a^{10} M^{10}}{512 \omega_k^{15}}-\frac{73457 a^8 M^8 }{256 \omega_k^{13}}
+ \frac{12087 a^6 M^6 }{512 \omega_k^{11}}\right)\left(\frac{a^{\prime}}{a}\right)^4 \frac{a^{\prime\prime}}{a} \\
&+ \left(\frac{34503 a^8 M^8}{256 \omega_k^{13}} - \frac{3225 a^6 M^6 }{64 \omega_k^{11}}+ \frac{249 a^4 M^4 }{256 \omega_k^9}\right)\left(\frac{a^{\prime}}{a}\right)^2\left(\frac{a^{\prime\prime}}{a}\right)^2\\
 &- \left(\frac{631 a^6 M^6 }{128 \omega_k^{11}} -\frac{109 a^4 M^4 }{128 \omega_k^9}\right)\left(\frac{a^{\prime\prime}}{a}\right)^3 + \left(\frac{
 1055 a^8 M^8 }{16 \omega_k^{13}} -\frac{3171 a^6 M^6}{128 \omega_k^{11}}+\frac{69 a^4 M^4}{128 \omega_k^9}\right) \left(\frac{a^\prime}{a}\right)^3 \frac{a^{\prime\prime\prime}}{a} \\
 &-\left(\frac{ 1391 a^6 M^6}{64 \omega_k^{11}}-\frac{245 a^4 M^4 }{64 \omega_k^9}\right)\frac{a^{\prime}}{a} \frac{a^{\prime\prime}}{a}\frac{ a^{\prime\prime\prime}}{a} +\left(\frac{69 a^4 M^4}{128 \omega_k^9} -\frac{a^2 M^2}{128 \omega_k^7}\right)\left(\frac{a^{\prime\prime\prime}}{a}\right)^2 \\
 &-\left(\frac{815 a^6 M^6}{128 \omega_k^{11}}-\frac{149  a^4 M^4 }{128 \omega_k^9}\right)\left(\frac{a^\prime}{a}\right)^2 \frac{a^{\prime\prime\prime\prime}}{a}+\left(\frac{ 55 a^4 M^4}{64 \omega_k^9} -\frac{a^2 M^2}{64 \omega_k^7}\right)\frac{a^{\prime\prime}}{a} \frac{a^{\prime\prime\prime\prime}}{a}\\
 & +\left(\frac{ 27 a^4 M^4 }{64 \omega_k^9}-\frac{ a^2 M^2}{64 \omega_k^7}\right)\frac{a^\prime}{a} \frac{a^{ \prime\prime\prime\prime\prime}}{a} -\frac{ a^2 M^2 }{64 \omega_k^7}\frac{a^{\prime\prime\prime\prime\prime\prime}}{a}\,.
\end{split}
\end{equation}
The odd terms in the expansion \eqref{GeneralSolutionProduct} yield a real exponential contribution in the integrals involved in \eqref{ansatz2} and hence do not contribute to the expansion of $W_k$ in \eqref{ExpansionW}, but are nevertheless necessary to compute the amplitude of the modes. Notice that after computing the integral, the adiabatic order decreases by one unit, so in order to estimate the amplitude up to 6{\it th} order is mandatory to compute up to $\omega_k^{(7)}$. The corresponding integrals for these terms are listed below:
\begin{equation}\label{Integral1st}
	-i\int{\omega_k^{(1)}d\tau}=\int^\tau \left[-\frac{ M a^\prime}{2\omega_k}\right]d\tilde{\tau}=\log\left(\frac{\omega_k-aM}{\omega_k+aM}\right)^{1/4}\,,
\end{equation}
\begin{equation}\label{Integral3rd}
	\begin{split}
		-i\int{\omega_k^{(3)}d\tau}=&\Delta^2\int^\tau \left[\frac{ a^2 a^\prime  M}{4\omega_k^3}-\frac{i a^\prime }{4M\omega_k}\right]d\tilde{\tau}\\
		&+\int^\tau \left[\frac{25  a^2 M^5 {a^\prime}^3}{16\omega_k^7}-\frac{5   M^3 {a^\prime}^3}{16\omega_k^5}-\frac{ a M^3 a^\prime 				a^{\prime\prime}}{\omega_k^5}+\frac{Ma^{\prime\prime\prime}}{8\omega_k^3}\right]d\tilde{\tau}\\
		&=-\frac{a\Delta^2}{4M\omega_k}+\frac{aM}{8\omega_k^3}\frac{a^{\prime\prime}}{a}-\frac{5a^3 M^3 }{16\omega_k^5}\left(\frac{a^\prime}{a}\right)^2\,,
\end{split}
\end{equation}
\vspace{0.5cm}
\begin{equation}\label{Integral5th}
	\begin{split}
	-i\int^\tau \omega_k^{(5)}d\tilde{\tau}	&=\Delta^4\int^\tau \left[\frac{-3  a^4  M a^\prime}{16\omega_k^5}+\frac{a^2  a^\prime}{8 M\omega_k^3}+\frac{ a^\prime}{16 M^3 \omega_k}\right]d\tilde{\tau}\phantom{aaaaaaaaaaaaaaaaaaaaaaaaaaaaaaaaaaaaaa}\\
&+\Delta^2\int^\tau \bigg[-\frac{175 a^4  M^5 {a^\prime}^3}{32\omega_k^9}+\frac{75  a^2  M^3 {a^\prime}^3}{16 \omega_k^7}-\frac{15  M {a^\prime}^3}{32  \omega_k^5}+\frac{5a^3  M^3 a^\prime a^{\prime \prime}}{2 \omega_k^7}-\frac{3 a  M a^\prime a^{\prime\prime}}{2\omega_k^5}\\
&\phantom{aaaaaaaaa}-\frac{3 a^2 M a^{\prime\prime\prime}}{16\omega_k^5}+\frac{a^{\prime\prime\prime}}{16M\omega_k^3}\Bigg]d\tilde{\tau}\\
\phantom{aaaaaaaaa}&+\int^\tau\Bigg[-\frac{12155  a^4 M^9 {a^\prime}^5}{256 \omega_k^{13}}+\frac{3453 a^2 M^7 {a^\prime}^5}{128 \omega_k^{11}}-\frac{399  M^5{a^\prime}^5}{256 \omega_k^9}+\frac{1547  a^3 M^7 {a^\prime}^3 a^{\prime\prime}}{32\omega_k^{11}}\\
&\phantom{aaaaaaa}-\frac{543  a M^5 {a^\prime}^3 a^{\prime\prime}}{32\omega_k^9}-\frac{575  a^2 M^5 a^\prime {a^{\prime\prime}}^2}{64\omega_k^9}+\frac{89 M^3 a^\prime {a^{\prime\prime}}^2}{64\omega_k^7}-\frac{417 a^2 M^5 {a^\prime}^2 a^{\prime\prime\prime}}{64\omega_k^9}\\
&\phantom{aaaaaaa}+\frac{63  M^3 {a^\prime}^2 a^{\prime\prime\prime}}{64\omega_k^7}+\frac{33  a M^3 a^{\prime\prime} a^{\prime\prime\prime}}{32\omega_k^7}+\frac{19 a M^3 a^\prime a^{\prime\prime\prime\prime}}{32\omega_k^7}-\frac{M {a^{\prime\prime\prime\prime\prime}}}{32 \omega_k^5}\Bigg]d\tilde{\tau}\\
&=\left(\frac{a^3}{16 M \omega_k^3}+\frac{a}{16M^3 \omega_k}\right)\Delta^4+\left(-\frac{3a^3 M}{16\omega_k^5}+\frac{a}{16M \omega_k^3}\right)\Delta^2 \frac{a^{\prime\prime}}{a}\\
&+\left(\frac{25 a^5 M^3}{32 \omega_k^7}-\frac{15 a^3 M }{32 \omega_k^5}\right)\Delta^2 \left(\frac{a^\prime}{a}\right)^2+\left(-\frac{221a^5M^5}{64\omega_k^9}+\frac{35a^3 M^3}{64\omega_k^7}\right) \frac{a^{\prime \prime}}{a} \left(\frac{a^\prime}{a}\right)^2\\
&+\left(\frac{1105 a^7 M^7 }{256\omega_k^{11}}-\frac{399 a^5 M^5}{256 \omega_k^9}\right)  \left(\frac{a^\prime}{a}\right)^4+\frac{19a^3 M^3 }{64\omega_k^7}\left(\frac{a^{\prime\prime}}{a}\right)^2+\frac{7 a^3  M^3}{16 \omega_k^7}\frac{a^\prime}{a} \frac{a^{\prime\prime  \prime}}{a}-\frac{a M}{32\omega_k^5}\frac{a^{\prime\prime\prime\prime}}{a}\,,
	\end{split}
\end{equation}
\vspace{0.5cm}
\begin{equation*}
\begin{split}
-i \int^\tau \omega_k^{(7)}d \tilde{\tau}=&\Delta^6 \int^\tau \left[\frac{5  a^6 M a^\prime}{32\omega_k^7}-\frac{3 a^4  a^\prime}{32 M \omega_k^5}-\frac{ a^2 a^\prime}{32 M^3 \omega_k^3}-\frac{  a^\prime}{32 M^5 \omega_k}\right]d\tilde{\tau}\\
		&+\Delta^4 \int^\tau \Bigg[\frac{1575  a^6  M^5 {a^\prime}^3}{128 \omega_k^{11}}-\frac{1925  a^4  M^3 {a^\prime}^3}{128 \omega_k^9}		+	\frac{525  a^2  M {a^\prime}^3}{128 \omega_k^7}-\frac{15   {a^\prime}^3 }{128 M \omega_k^5}-\frac{35  a^5  M^3 a^\prime a^{\prime \prime}}{8 \omega_k^9} \\
		&  \phantom{aaaaaaaaa} + \frac{15  a^3  M a^\prime a^{\prime\prime}}{4\omega_k^7}-\frac{3 a  a^\prime a^{\prime				\prime}}{8 M \omega_k^5}+\frac{15  a^4  M a^{\prime\prime\prime}}{64 \omega_k^7}-\frac{3 a^2  a^{\prime\prime \prime}}{32 M \omega_k^5}-\frac{  a^{\prime \prime  \prime}}{64 M^3 \omega_k^3}\Bigg]d\tilde{\tau}\\
		&+\Delta^2\int^\tau \Bigg[\frac{158015  a^6  M^9 {a^\prime}^5}{512 \omega_k^{15}}-\frac{185361  a^4  M^7 {a^\prime}^5}{512\omega_k^{13}}+\frac{51933  a^2  M^5 {a^\prime}^5}{512 \omega_k^{11}}\\
		&\phantom{aaaaaaaaa}-\frac{1995   M^3 {a^\prime}^5}{512 \omega_k^9}-\frac{17017 a^5  M^7 {a^\prime}^3 a^{\prime \prime}}{64\omega_k^{13}}+\frac{3929 a^3  M^5 {a^\prime}^3 a^{\prime\prime}}{16 \omega_k^{11}}\\
  &\phantom{aaaaaaaaa}-\frac{2715 a  M^3 {a^\prime}^3 a^{\prime \prime}}{64 \omega_k^9}+\frac{5175  a^4  M^5 a^\prime {a^{\prime\prime}}^2}{128 \omega_k^{11}}-\frac{1749  a^2  M^3 a^\prime {a^{\prime\prime}}^2}{64\omega_k^9}\\
		&\phantom{aaaaaaaaa}+\frac{267   M a^\prime {a^{\prime\prime}}^2}{128 \omega_k^{7}}+\frac{3753  a^4  M^5 {a^\prime}^2 a^{\prime\prime\prime}}{128\omega_k^{11}}-\frac{1263  a^2  M^3 {a^\prime}^2 a^{\prime\prime \prime}}{64\omega_k^9}\\
  &\phantom{aaaaaaaaa}+\frac{189   M {a^\prime}^2 a^{\prime\prime\prime}}{128\omega_k^7}-\frac{231  a^3  M^3 a^{\prime\prime}a^{\prime\prime\prime}}{64\omega_k^9}+\frac{99  a  M a^{\prime\prime}a^{\prime \prime \prime}}{64 \omega_k^7}-\frac{133 a^3  M^3 a^\prime a^{\prime\prime\prime\prime}}{64\omega_k^9}\\
		&\phantom{aaaaaaaaa}+\frac{57  a  M a^\prime a^{\prime\prime\prime\prime}}{64 \omega_k^7}+\frac{5  a^2  		M {a^{\prime\prime\prime\prime\prime}}}{64\omega_k^7}-\frac{  {a^{\prime\prime\prime\prime\prime}}}{64 M \omega_k^5}\Bigg]d\tilde{\tau}\\
		\end{split}
\end{equation*}
\begin{equation}\label{Integral7th}
	\begin{split}
	&\phantom{aaaaaaaaaaaaa}+\int^\tau \Bigg[\frac{7040125  a^6 M^{13}{a^\prime}^7}{2048 \omega_k^{19}}-\frac{6664175  a^4 M^{11}{a^\prime}^7}{2048\omega_k^{17}}+\frac{1429935  a^2 M^9 {a^\prime}^7}{2048 \omega_k^{15}}-\frac{39325  M^7 {a^\prime}^7}{2048 \omega_k^{13}}\\
		&\phantom{aaaaaaaaaaaaaaaaaaa}-\frac{1242375  a^5 M^{11}{a^\prime}^5 a^{\prime\prime}}{256 \omega_k^{17}}+\frac{449881  a^3 M^9 {a^\prime}^5 a^{\prime\prime}}{128 \omega_k^{15}}-\frac{112779  a M^7 {a^\prime}^5 a^{\prime\prime}}{256 \omega_k^{13}}\\
  &\phantom{aaaaaaaaaaaaaaaaaaa}+\frac{945489  a^4 M^9 {a^\prime}^3 {a^{\prime\prime}}^2}{512\omega_k^{15}}-\frac{30273  a^2 M^7 {a^\prime}^3 {a^{\prime\prime}}^2}{32\omega_k^{13}}+\frac{24435  M^5 {a^\prime}^3 {{a^{\prime\prime}}^2}}{512 \omega_k^{11}}\\
  &\phantom{aaaaaaaaaaaaaaaaaaa}-\frac{10361  a^3 M^7 a^\prime {a^{\prime\prime}}^3}{64\omega_k^{13}}+\frac{1639  a M^5 a^\prime {a^{\prime\prime}}^3}{32\omega_k^{11}}-\frac{90425  a^3 M^7 {a^\prime}^2 a^{\prime\prime}a^{\prime\prime\prime}}{256 \omega_k^{13}}\\
		&\phantom{aaaaaaaaaaaaaaaaaaa}+\frac{687335  a^4 M^9 {a^\prime}^4 a^{\prime\prime\prime}}{1024 \omega_k^{15}}-\frac{173583  a^2 M^7 {a^\prime}^4 a^{\prime\prime\prime}}{512 \omega_k^{13}}+\frac{17259  M^5 {a^\prime}^4 a^{\prime\prime\prime}}{1024 \omega_k^{11}}\\
		&\phantom{aaaaaaaaaaaaaaaaaaa}+\frac{27923  a M^5 {a^\prime}^2a^{\prime\prime}a^{\prime\prime\prime}}{256\omega_k^{11}}+\frac{4675  a^2 M^5 {a^{\prime \prime}}^2a^{\prime\prime\prime}}{256 \omega_k^{11}}-\frac{649  M^3 {a^{\prime \prime}}^2 a^{\prime\prime\prime}}{256\omega_k^9}\\
		&\phantom{aaaaaaaaaaaaaaaaaaa}+\frac{3403  a^2 M^5 a^\prime {a^{\prime \prime \prime}}^2}{256\omega_k^{11}}-\frac{447  M^3 a^\prime {a^{\prime\prime\prime}}^2}{256\omega_k^9}-\frac{17405  a^3 M^7 {a^\prime}^3 a^{\prime\prime\prime\prime}}{256\omega_k^{13}}+\frac{5215  a M^5 {a^\prime}^3 a^{\prime\prime			\prime\prime}}{256 \omega_k^{11}}\\
		&\phantom{aaaaaaaaaaaaaaaaaaa}+\frac{2701  a^2 M^5 a^\prime a^{\prime\prime}a^{\prime\prime\prime\prime}}{128 \omega_k^{11}}-\frac{349  M^3 a^\prime a^{\prime\prime}a^{\prime\prime\prime\prime}}{128 \omega_k^9}-\frac{31 a M^3 a^{\prime\prime\prime}a^{\prime\prime\prime\prime}}{32\omega_k^9}+\frac{1301 a^2 M^5 {a^\prime}^2 {a^{\prime\prime\prime\prime\prime}}}{256 \omega_k^{11}}\\
        &\phantom{aaaaaaaaaaaaaaaaaaa}-\frac{159 M^3 {a^\prime}^2 {a^{\prime\prime\prime\prime\prime}}}{256 \omega_k^9}-\frac{41 a M^3 a^{\prime\prime}a^{\prime\prime\prime\prime\prime}}{64 \omega_k^9}-\frac{17 a M^3 a^\prime a^{\prime\prime\prime\prime\prime\prime}}{64 \omega_k^9}+\frac{M a^{\prime\prime\prime\prime\prime\prime\prime}}{128\omega_k^{7}}\Bigg]d\tilde{\tau}\\
		&\phantom{aaaaaaaaaaaaa}=-\left(\frac{a^5}{32M \omega_k^5}+\frac{a^3}{48M^3\omega_k^3}+\frac{a}{32M^5 \omega_k}\right)\Delta^6\\
        &\phantom{aaaaaaaaaaaaa}-\left(\frac{175a^7 M^3}{128 \omega_k^9}-\frac{75 a^5 M}{64\omega_k^7}+\frac{15 a^3 }{128 M \omega_k^5}\right)\Delta^4 \left(\frac{a^\prime}{a}\right)^2\\
		&\phantom{aaaaaaaaaaaaa}+\left(\frac{15 a^5 M}{64\omega_k^7}-\frac{3a^3}{32 M \omega_k^5}-\frac{a}{64 M^3 \omega_k^3}\right)\Delta^4 \frac{a^{\prime \prime}}{a}-\left(\frac{133 a^5 M^3}{128\omega_k^9}-\frac{57a^3 M}{128\omega_k^7}\right)\Delta^2  \left(\frac{ a^{\prime\prime}}{a}\right)^2\\
        &\phantom{aaaaaaaaaaaaa}-\left(\frac{12155 a^9 M^7}{512\omega_k^{13}}-\frac{11326 a^7 M^5 }{512 \omega_k^{11}}+\frac{1995 a^5 M^3}{512 \omega_k^9}\right)\Delta^2  \left(\frac{a^\prime}{a}\right)^4\\
		&\phantom{aaaaaaaaaaaaa}+\left(\frac{1989 a^7 M^5}{128\omega_k^{11}}-\frac{1350 a^5 M^3 }{128 \omega_k^9}+\frac{105 a^3 M}{128\omega_k^7}\right)\Delta^2 \frac{a^{\prime\prime}}{a}\left(\frac{a^\prime}{a}\right)^2\\
        &\phantom{aaaaaaaaaaaaa}-\left(\frac{49 a^5 M^3}{32 \omega_k^9}-\frac{21 a^3 M}{32\omega_k^7}\right) \Delta^2 \frac{a^{\prime\prime\prime}}{a}\frac{a^{\prime}}{a}+\left(\frac{5a^3 M}{64 \omega_k^7}-\frac{a}{64 M \omega_k^5}\right)\Delta^2 \frac{a^{\prime\prime\prime\prime}}{a}\\
		&\phantom{aaaaaaaaaaaaa}-\left(\frac{414125a^{11}M^{11}}{2048 \omega_k^{17}}-\frac{459355 a^9 M^9 }{3072\omega_k^{15}}+\frac{39325 a^7 M^7}{2048 \omega_k^{13}}\right) \left(\frac{a^\prime}{a}\right)^6-\frac{55 a^3 M^3  }{128\omega_k^9}\frac{a^{\prime\prime\prime\prime}}{a}\frac{a^{\prime\prime}}{a}\\
        &\phantom{aaaaaaaaaaaaa}-\left(\frac{34503a^7 M^7}{512\omega_k^{13}}-\frac{11037 a^5 M^5  }{512 \omega_k^{11}}\right)\left(\frac{a^{ \prime \prime}}{a}\frac{a^\prime}{a}\right)^2-\frac{27a^3 M^3  }{128\omega_k^9}\frac{a^{\prime\prime\prime\prime\prime}}{a}\frac{a^\prime}{a}+\frac{a M  }{128\omega_k^7}\frac{a^{\prime\prime\prime\prime\prime\prime}}{a}\\
        &\phantom{aaaaaaaaaaaaa}-\left(\frac{1055 a^7 M^7}{32\omega_k^{13}}-\frac{330 a^5 M^5}{32\omega_k^{11}}\right) \frac{a^{\prime\prime\prime}}{a}\left(\frac{a^\prime}{a}\right)^3+\left(\frac{815 a^5 M^5}{256\omega_k^{11}}-\frac{105 a^3 M^3}{256\omega_k^9}\right)\frac{a^{\prime\prime\prime\prime}}{a}\left(\frac{a^\prime}{a}\right)^2\\
		&\phantom{aaaaaaaaaaaaa}+\left(\frac{248475 a^9 M^{9}}{1024\omega_k^{15}}-\frac{64863 a^7 M^7 }{512 \omega_k^{13}}+\frac{6699 a^5 M^5}{1024 \omega_k^{11}}\right)\frac{a^{\prime\prime}}{a}\left(\frac{a^{\prime}}{a}\right)^4-\frac{69a^3 M^3}{256\omega_k^9}\left(\frac{a^{\prime\prime\prime}}{a}\right)^2\\	
		&\phantom{aaaaaaaaaaaaa}+\left(\frac{631 a^5 M^5}{256\omega_k^{11}}-\frac{271 a^3 M^3}{768 \omega_k^9}\right)\left(\frac{a^{\prime\prime}}{a}\right)^3+\left(\frac{1391 a^5 M^5}{128\omega_k^{11}}-\frac{189 a^3 M^3}{128\omega_k^9}\right)\frac{a^{\prime\prime\prime}}{a}\frac{a^{\prime\prime}}{a}\frac{a^{\prime}}{a}\,.
	\end{split}
\end{equation}
\newpage
 Use of Mathematica\,\cite{Mathematica} has been helpful to work out the above integrals. The computational strategy consists in using a sufficiently general ansatz for the structure of the result that is compatible with the adiabaticity order of the calculation, and then solve for the coefficients (form factors)  of the ansatz from pure algebraic manipulations. The procedure has been illustrated with a specific example in \hyperref[sec:QuantizFermions]{Sec.\,\ref{sec:QuantizFermions}}.

Finally, by applying the normalization condition \eqref{NormalizationConditionMode} for the modes at each order, there exists a constraint to fix  the residual freedom mentioned earlier, which is parametrized by the time independent factors $f_k$ and $g_k$ (only depending on the momentum $k$):

\begin{align}\label{NormalizationConditions}
\begin{split}
&\mathbb{R}e f^{(1)}_k=0\,,\\
&\left( \mathbb{I}m f^{(1)}_k \right)^2+\sqrt{2k}\mathbb{R}e f^{(2)}_k=0\,,\\
&2\mathbb{I}m f^{(2)}_k\mathbb{I}m f^{(1)}_k+\sqrt{2k}\mathbb{R}e f^{(3)}_k=0\,,\\
&\left|f^{(2)}_k\right|^2+2\mathbb{I}m f^{(1)}_k\mathbb{I}m f^{(3)}_k+\sqrt{2k}\mathbb{R}e f^{(4)}_k=0\,,\\
&2\mathbb{I}m f^{(1)}_k\mathbb{I}m f^{(4)}_k+2\mathbb{I}m f^{(2)}_k\mathbb{I}m f^{(3)}_k+2\mathbb{R}e f^{(2)}_k\mathbb{R}e f^{(3)}_k+\sqrt{2k}\mathbb{R}e f^{(5)}_k=0\,,\\
&2\mathbb{I}m f^{(1)}_k\mathbb{I}m f^{(5)}_k+2\mathbb{I}m f^{(2)}_k\mathbb{I}m f^{(4)}_k+\left|f^{(3)}_k\right|^2+2\mathbb{R}e f^{(2)}_k\mathbb{R}e f^{(4)}_k+\sqrt{2k}\mathbb{R}e f^{(6)}_k=0\,.
\end{split}
 \end{align}
Notice that, imposing the former conditions is equivalent to claim that
\begin{equation}\label{NormalizationConditionII}
	\Bigg| 1+\sqrt{\frac{2}{k}}\left(f_k^{(1)}+f_k^{(2)}+f_k^{(3)}+f_k^{(4)}+f_k^{(5)}+f_k^{(6)}\right) \Bigg|^2\approx 1\,,
\end{equation}
where the departure from 1 are just terms of adiabatic order 7 or bigger. Similarly for the functions $g_k$. It can be shown that the observables made out of quadratic terms in the modes $h_k^{\rm I},h_{k}^{\rm II}$ (such as e.g. the EMT), depend on the particular values of $f_k$ in the form \eqref{NormalizationConditionII}. It follows that they are not actual degrees of freedom if they satisfy the conditions \eqref{NormalizationConditions}. It is then safe to set particular values for the functions as long as quantities are computed up to $6th$ adiabatic order. The simplest solution satisfying the normalization conditions \eqref{NormalizationConditions} is $f_k^{(1)}=f_k^{(2)}=f_k^{(3)}=f_k^{(4)}=f_k^{(5)}=f_k^{(6)}=0$ and it is the chosen option for the formulas shown in the rest of this Appendix.

Equipped with these results we are able to calculate the different orders of $F\left(\tau,M\right)$ up to $6th$ order. A comparison between the general equation\,\eqref{GeneralSolution} and the ansatz\,\eqref{ansatz2}, the different orders of $F$ are conformed by the rightful combinations of terms of the denominator $\sqrt{\Omega_k\Omega_{k,1}\Omega_{k,2}\Omega_{k,3}}$ and the real factors of the  exponential $\exp\left(-i\int^\tau \Omega_k\Omega_{k,1}\Omega_{k,2}\Omega_{k,3} d\tilde{\tau}\right)$.
Now the different orders of $F$ are
\begin{equation}\label{F1st}
    \begin{split}
		&F^{(1)}(M)=\frac{i a M}{4\omega_k^2}\frac{a^\prime}{a}\,,
    \end{split}
\end{equation}
\vspace{0.5cm}
\begin{equation}\label{F2nd}
    \begin{split}
		&F^{(2)}(M)=\left(-\frac{a^2}{4\omega_k^2}-\frac{a }{4 M \omega_k}\right)\Delta^2-\left(\frac{5 a^4 M^4 }{16 \omega_k^6}+ \frac{5 a^3 M^3 }{16 \omega_k^5 }
		+\frac{M^2 a^2}{32 \omega_k^4}\right) \left(\frac{a^ \prime}{a}\right)^2+ \left(\frac{a^2 M^2 }{8\omega_k^4} + \frac{a M}{8\omega_k^3}\right) \frac{a^{\prime\prime}}{a}\,,
    \end{split}
\end{equation}
\vspace{0.5cm}
\begin{equation}\label{F3rd}
    \begin{split}
		F^{(3)}(M)&=i\left(-\frac{5 M a^3 }{16 \omega_k ^4}-\frac{a^2}{16 \omega_k ^3}+\frac{a}{8 M \omega_k ^2}\right)\frac{a^\prime}{a}\Delta^2+i\left(-\frac{65 M^5 			a^5}{64 \omega_k^8}-\frac{5 M^4 a^4}{64 \omega_k^7}+\frac{21 M^3 a^3}{128 \omega_k^6}\right)\left(\frac{a^\prime}{a}\right)^3\\
		&+i\left(\frac{19 M^3 a^3}{32 \omega_k^6}+\frac{M^2 a^2}{32 \omega_k^5}\right)\frac{a'a''}{a^2}-i\frac{a M}{16 \omega_k ^4}\frac{a^{\prime\prime\prime}}{a}\,,
    \end{split}
\end{equation}
\vspace{0.5cm}
\begin{equation}\label{F4th}
    \begin{split}
		F^{(4)}(M)&=\left(\frac{5a^4}{32 \omega_k^4}+\frac{a^3}{8M\omega_k^3}+\frac{a^2}{32 M^2 \omega_k^2}+\frac{a}{16 M^3 \omega_k}\right)\Delta^4\\
		&+\left(\frac{65a^6M^4}{64\omega_k^8}+\frac{15a^5 M^3}{16\omega_k^7}-\frac{61 a^4 M^2}{128\omega_k^6}-\frac{59 a^3 M}{128\omega_k^5}-\frac{a^2}{32\omega_k^4}			\right)\left(\frac{a^\prime}{a}\right)^2\Delta^2\\
		&+\left(-\frac{9a^4 M^2}{32\omega_k^6}-\frac{a^3 M}{4\omega_k^5}+\frac{3a^2}{32\omega_k^4}+\frac{a}{16M\omega_k^3}\right)\frac{a^{\prime \prime}}{a}\Delta^2\\
		&+\left(\frac{2285 a^8 M^8}{512\omega_k^{12}}+\frac{565a^7 M^7}{128 \omega_k^{11}}-\frac{349 a^6 M^6}{256\omega_k^{10}}-\frac{793 a^5 M^5}{512\omega_k^9}-				\frac{85 a^4 M^4}{2048 \omega_k^8}\right)\left(\frac{a^\prime}{a}\right)^4\\
		&+\left( -\frac{457 a^6 M^6}{128\omega_k^{10}}-\frac{113 a^5 M^5}{32\omega_k^9}+\frac{113 a^4 M^4 }{256\omega_k^8}+\frac{139 a^3M^3}{256 \omega_k^7}\right)				\left(\frac{a^\prime}{a}\right)^2\frac{a^{\prime\prime}}{a}\\
		&+\left(\frac{41a^4 M^4}{128\omega_k^8}+\frac{5a^3 M^3}{16\omega_k^7}-\frac{a^2 M^2}{128\omega_k^6}\right)\left(\frac{a^{\prime\prime}}{a}\right)^2+					\left(\frac{7 a^4 M^4 }{16 \omega_k ^8}+\frac{7 a^3 M^3 }{16 \omega_k^7}+\frac{a^2 M^2}{64 \omega_k^6}\right)\frac{a^\prime}{a}\frac{a^{\prime\prime\prime}}{a}\\
		&-\left(\frac{a^2 M^2}{32\omega_k^6}+\frac{a M}{32\omega_k^5}\right)\frac{a^{\prime\prime\prime\prime}}{a}\,,
    \end{split}
\end{equation}
\vspace{0.5cm}
\begin{equation}\label{F5th}
    \begin{split}
		F^{(5)}(M)&=i\left(\frac{45a^5 M}{128\omega_k^6}+\frac{3a^4}{32\omega_k^5}-\frac{19a^3}{128 M \omega_k^4}-\frac{a^2}{64M^2 \omega_k^3}-\frac{a}						{32M^3\omega_k^2}\right)\frac{a^\prime}{a}\Delta^4\\
		&+i\left(\frac{1105 a^7 M^5}{256 \omega_k^{10}}+\frac{35a^6 M^4}{64\omega_k^9}-\frac{1563 a^5M^3}{512\omega_k^8}-\frac{101 a^4 M^2}{512\omega_k^7}+\frac{63 			a^3 M}{256\omega_k^6}\right)\left(\frac{a^\prime}{a}\right)^3\Delta^2\\
		&+i\left(-\frac{247a^5 M^3}{128\omega_k^8}-\frac{15 a^4 M^2}{64\omega_k^7}+\frac{113 a^3 M}{128\omega_k^6}+\frac{a^2}{32\omega_k^5}\right)\frac{a^\prime}{a}				\frac{a^{\prime\prime}}{a}\Delta^2\\
		&+i\left(\frac{9 a^3 M}{64\omega_k^6}+\frac{a^2}{64\omega_k^5}-\frac{a}{32M \omega_k^4}\right)\frac{a^{\prime\prime\prime}}{a}\Delta^2\\
		&+\left(\frac{57125 a^9 M^9}{2048\omega_k^{14}}+\frac{715 a^8 M^8}{512\omega_k^{13}}-\frac{7657 a^7 M^7}{512 \omega_k^{12}}-\frac{903 a^6 M^6}							{2048\omega_k^{11}}+\frac{6511a^5 M^5}{8192\omega_k^{10}}\right)\left(\frac{a^\prime}{a}\right)^5\\
		&+i\left(-\frac{14167 a^7 M^7}{512\omega_k^{12}}-\frac{301 a^6 M^6}{256\omega_k^{11}}+\frac{9273 a^5 M^5}{1024\omega_k^{10}}+\frac{161 a^4 M^4}{1024\omega_k^9}			\right)\frac{a^{\prime\prime}}{a}\left(\frac{a^\prime}{a}\right)^3\\
		&+i\left(\frac{2525 a^5 M^5}{512\omega_k^{10}}+\frac{19 a^4 M^4}{128\omega_k^9}-\frac{361 a^3 M^3}{512\omega_k^8}\right)\frac{a^\prime}{a}								\left(\frac{a^{\prime\prime}}{a}\right)^2\\
		&+i\left(\frac{933 a^5 M^5}{256\omega_k^{10}}+\frac{33 a^4 M^4}{256\omega_k^9}-\frac{255 a^3M^3}{512\omega_k^8}\right)\left(\frac{a^\prime}{a}							\right)^2\frac{a^{\prime\prime\prime}}{a}\\
		&+i\left(-\frac{69 a^3 M^3}{128\omega_k^8}-\frac{M^2 a^2}{128\omega_k^7}\right)\frac{a^{\prime\prime}}{a}\frac{a^{\prime\prime\prime}}{a}-i\left(\frac{41a^3 M^3}{128\omega_k^8}+\frac{M^2 a^2}{128\omega_k^7}\right)\frac{a^{\prime}}{a}\frac{a^{\prime\prime\prime\prime}}{a}+i\frac{a M}{64\omega_k^6}\frac{a^{\prime\prime\prime\prime\prime}}{a}\,,
		\end{split}
\end{equation}
\vspace{0.5cm}
\begin{equation}\label{F6th}
    \begin{split}
		F^{(6)}(M)&=\left(-\frac{15 a^6 }{128\omega_k^6}-\frac{11 a^5}{128 M \omega_k^5}-\frac{3 a^4}{128M^2 \omega_k^4}-\frac{5a^3}{128 M^3 \omega_k^3}-\frac{a^2}{64 			M^4 \omega_k^2}-\frac{a}{32M^5 \omega_k}\right)\Delta^6\\
		&+\left(-\frac{1105 a^8 M^4}{512\omega_k^{10}}-\frac{965 a^7 M^3}{512\omega_k^9}+\frac{1713 a^6 M^2}{1024\omega_k^8}+\frac{715 a^5 M}{512\omega_k^7}-\frac{149 			a^4}{1024\omega_k^6}-\frac{57 a^3}{512 M \omega_k^5}\right)\left(\frac{a^\prime}{a}\right)^2\Delta^4\\
		&+\left(\frac{117 a^6 M^2}{256\omega_k^8}+\frac{97 a^5 M}{256\omega_k^7}-\frac{55 a^4}{256\omega_k^6}-\frac{33 a^3 }{256 M\omega_k^5}-\frac{a^2}{128 M^2 				\omega_k^4}-\frac{a}{64 M^3 \omega_k^3}\right)\frac{a^{\prime\prime}}{a}\Delta^4\\
		&+\left(-\frac{57125 a^{10} M^8}{2048 \omega_k^{14}}-\frac{54265 a^9 M^7}{2048\omega_k^{13}}+\frac{24479 a^8 M^6}{1024\omega_k^{12}}+\frac{47405 a^7 M^5}				{2048\omega_k^{11}}-\frac{28887 a^6 M^4}{8192 \omega_k^{10}}\right.\\
		&\left. \phantom{aaa}-\frac{31635 a^5  M^3}{8192 \omega_k^9}-\frac{85 a^4 M^2 }{1024 \omega_k^8}\right)\left(\frac{a^\prime}{a}\right)^4\Delta^2\\
        &+\left(\frac{9597 a^8 M^6}{512 \omega_k^{12}}+\frac{9045 a^7 M^5}{512 \omega_k^{11}}-\frac{11985 a^6 M^4}{1024 \omega_k^{10}}-\frac{5619 a^5 M^3}{512\omega_k^9}+\frac{765 a^4 M^2}{1024\omega_k^8}\right.\\
        &\phantom{aaa}\left.+\frac{417 a^3 M}{512\omega_k^7}\right)\left(\frac{a^\prime}{a}\right)^2\frac{a^{\prime\prime}}{a}\Delta^2+\left(\frac{13 a^4 M^2}{128 \omega_k^8}+\frac{3a^3 M}{32\omega_k^7}-\frac{3a^2}{128\omega_k^6}-\frac{a}{64M \omega_k^5}\right)\frac{a^{\prime\prime\prime\prime}}{a}\Delta^2\\
        &+\left(-\frac{697 a^6 M^4}{512\omega_k^{10}}-\frac{641 a^5 M^3}{512\omega_k^9}+\frac{301 a^4 M^2}{512 \omega_k^8}+\frac{241 a^3 M}{512 \omega_k^7}-\frac{a^2}{128 \omega_k^6}\right)\left(\frac{a^{\prime\prime}}{a}\right)^2\Delta^2\\
        &+\left(-\frac{119 a^6 M^4}{64\omega_k^{10}}-\frac{7 a^5 M^3}{4\omega_k^9}+\frac{183 a^4 M^2}{256\omega_k^8}+\frac{167 a^3 M}{256 \omega_k^7}+\frac{a^2}{64\omega_k^6}\right)\frac{a^\prime}{a}\frac{a^{\prime\prime\prime}}{a}\Delta^2\\
		&+\left(-\frac{1690275 a^{12}M^{12}}{8192\omega_k^{18}}-\frac{1678975 a^{11}M^{11}}{8192 \omega_k^{17}}+\frac{2377685 a^{10} M^{10}}{16384 \omega_k^{16}}+				\frac{1231405 a^9 M^9}{8192 \omega_k^{15}}\right.\\
		&\phantom{aaa}\left.-\frac{519009a^8 M^8}{32768 \omega_k^{14}}-\frac{627179 a^7 M^7}{32768 \omega_k^{13}}-\frac{13989 a^6 M^6}{65536 \omega_k^{12}}\right)\left(\frac{a^\prime}{a}\right)^6+					\left(\frac{a^2 M^2}{128\omega_k^8}+\frac{aM}{128\omega_k^7}\right)\frac{a^{\prime\prime\prime\prime\prime\prime}}{a}\\
        &+\left(\frac{1014165a^{10}M^{10}}{4096 \omega_k^{16}}+\frac{1007385 a^9 M^9}{4096 \omega_k^{15}}-\frac{250133 a^8 M^8}{2048\omega_k^{14}}-\frac{521273 a^7M^7}{4096 \omega_k^{13}}\right.\\
        &\phantom{aaa}\left.+\frac{74799 a^6 M^6}{16384\omega_k^{12}}+\frac{106819 a^5 M^5}{16384 \omega_k^{11}}\right)\frac{a^{\prime\prime}}{a}\left(\frac{a^\prime}{a}\right)^{4}-\left(\frac{27 a^4 M^4}{128\omega_k^{10}}+\frac{27 a^3 M^3}{128\omega_k^9}+\frac{a^2 M^2}{256\omega_k^8}\right)\frac{a^\prime}{a}\frac{a^{\prime\prime\prime\prime\prime}}{a}\\
		&-\left(\frac{141309 a^8 M^8}{2048\omega_k^{14}}+\frac{140205 a^7 M^7}{2048\omega_k^{13}}-\frac{85737 a^6 M^6}{4096 \omega_k^{12}}-\frac{44397 a^5 M^5}{2048\omega_k^{11}}-\frac{441 a^4 M^4}{4096 \omega_k^{10}}\right)\left(\frac{a^\prime}{a} \frac{a^{\prime\prime}}{a}\right)^2\\
		&+\left(\frac{2643 a^6 M^6}{1024\omega_k^{12}}+\frac{2603 a^5 M^5}{1024\omega_k^{11}}-\frac{403 a^4 M^4}{1024\omega_k^{10}}-\frac{363 a^3 M^3}{1024\omega_k^9}\right)\left(\frac{a^{\prime\prime}}{a}\right)^3\\
		&+\left(-\frac{8545 a^8 M^8}{256 \omega_k^{14}}-\frac{4255 a^7 M^7}{128\omega_k^{13}}+\frac{9721 a^6 M^6}{1024\omega_k^{12}}+\frac{10541 a^5 M^5}						{1024\omega_k^{11}}+\frac{277 a^4 M^4}{2048 \omega_k^{10}}\right)\frac{a^{\prime\prime\prime}}{a}\left(\frac{a^\prime}{a}\right)^3\\
		&+\left(\frac{353 a^6 M^6}{32\omega_k^{12}}+\frac{1405 a^5 M^5}{128\omega_k^{11}}-\frac{697 a^4 M^4}{512\omega_k^{10}}-\frac{755a^3 M^3}{512\omega_k^9}\right)			\frac{a^\prime}{a}\frac{a^ {\prime\prime}}{a}\frac{a^{\prime\prime\prime}}{a}\\
  		&+\left(-\frac{69 a^4 M^4}{256 \omega_k^{10}}-\frac{69 a^3 M^3}{256 \omega_k^{9}}-\frac{a^2 M^2}{512\omega_k^8}\right)\left(\frac{a^{\prime\prime\prime}}{a}\right)^2+\left(-\frac{113 a^4 M^4}{256\omega_k^{10}}-\frac{7a^3M^3}{16\omega_k^9}+\frac{a^2 M^2}{256\omega_k^8} \right)\frac{a^{\prime\prime\prime\prime}}{a}\frac{a^{\prime\prime}}{a}\\
		&+\left(\frac{1645a^6 M^6}{512\omega_k^{12}}+\frac{205 a^5 M^5}{64\omega_k^{11}}-\frac{349 a^4 M^4}{1024\omega_k^{10}}-\frac{419 a^3 M^3}{1024\omega_k^9}				\right)\frac{a^{\prime\prime\prime\prime}}{a}\left(\frac{a^\prime}{a}\right)^2\,.
	\end{split}
\end{equation}
For $h^{\rm II}_k$, one can make use of the relation $G^{(n)}(M)=F^{(n)}(-M)$.

\section{Appendix: Adiabatic expansion of $\left\langle T_{\mu\nu} \right\rangle$  for spin-1/2 fields}\label{sec:appendixC}

The unrenormalized components of the VEV of the EMT,  $\langle T_{\mu\nu}^{\delta \psi} \rangle$,  for spin-1/2 fermions can be obtained through the adiabatic expansion, which we compute up to $6th$ order\footnote{Details of the corresponding computation for scalar fields were provided in \cite{Moreno-Pulido:2020anb, Moreno-Pulido:2022phq}. A summary of these calculations is presented in Sec.\,\ref{ZPEScalar} of the current work.}. For the $00$ component we have
\begin{align}\label{eq:ExpansionT00}
    \left\langle T_{00}^{\delta\psi} \right\rangle =\left\langle T_{00}^{\delta\psi} \right\rangle^{(0)}+\left\langle T_{00}^{\delta\psi} \right\rangle^{(2)}+\left\langle T_{00} ^{\delta\psi}\right\rangle^{(4)}+\left\langle T_{00}^{\delta\psi} \right\rangle^{(6)}+\cdots
\end{align}
The various terms in the expansion \eqref{eq:ExpansionT00} read as follows:
\begin{align}
\left\langle T_{00}^{\delta\psi} \right\rangle^{(0)}=\frac{1}{2\pi^2 a}\int_0^\infty dk k^2\left[-\frac{2\omega_k}{a}\right]\,,
\end{align}
\begin{align}
\left\langle T_{00}^{\delta\psi} \right\rangle^{(2)}=\frac{1}{2\pi^2 a}\int_0^\infty dk k^2\left[-\frac{a\Delta^2}{\omega_k}-\frac{a^3 M^4}{4\omega_k^5}\left(\frac{a^\prime}{a}\right)^2+\frac{a M^2}{4\omega_k^3}\left(\frac{a^\prime}{a}\right)^2\right]\,,
\end{align}
\begin{align}
\begin{split}
\left\langle T_{00}^{\delta\psi} \right\rangle^{(4)}&=\frac{1}{2\pi^2 a}\int_0^\infty dk k^2\Bigg[\frac{a^3 \Delta^4}{4\omega_k^3}+\left(\frac{5a^5M^4 \Delta^2}{8\omega_k^7}-\frac{7a^3 M^2 \Delta^2}{8 \omega_k^5}+\frac{a\Delta^2}{4\omega_k^3}\right)\left(\frac{a^\prime}{a}\right)^2\\
&+\left(\frac{105 a^7 M^8}{64\omega_k^{11}}-\frac{63a^5 M^6}{32\omega_k^9}+\frac{21 a^3 M^4}{64\omega_k^7}\right)\left(\frac{a^\prime}{a}\right)^4+\left(-\frac{7a^5 M^6}{8\omega_k^9}+\frac{7 a^3 M^4}{8\omega_k^7}\right)\frac{a^{\prime\prime}}{a}\left(\frac{a^\prime}{a}\right)^2 \\
&+\left(-\frac{a^3 M^4}{16\omega_k^7}+\frac{a M^2}{16\omega_k^5}\right)\left(\frac{a^{\prime\prime}}{a}\right)^2+\left(\frac{a^3 M^4}{8\omega_k^7}-\frac{a M^2}{8\omega_k^5}\right)\frac{a^\prime}{a}\frac{a^{\prime\prime\prime}}{a}
\Bigg]\,,
\end{split}
\end{align}
\begin{align*}
\begin{split}
\left\langle T_{00}^{\delta\psi} \right\rangle^{(6)}&=\frac{1}{2\pi^2 a}\int_0^\infty dk k^2\Bigg[-\frac{a^5 \Delta^6}{8\omega_k^5}+\left(\frac{35 a^7  M^4}{32\omega_k^9}+\frac{55 a^5 M^2}{32\omega_k^7}-\frac{5a^3 }{8\omega_k^5}\right)\left(\frac{a^\prime}{a}\right)^2\Delta^4\\
&+\left(-\frac{1155 a^9 M^8}{128\omega_k^{13}}+\frac{987 a^7 M^6}{64\omega_k^{11}}-\frac{903 a^5 M^4}{128\omega_k^9}+\frac{21a^3 M^2}{32\omega_k^7}\right)\left(\frac{a^\prime}{a}\right)^4\Delta^2\\
&+\left(\frac{63 a^7 M^6}{16\omega_k^{11}}-\frac{91 a^5 M^4}{16\omega_k^9}+\frac{7a^3 M^2}{4\omega_k^7}\right)\frac{a^{\prime\prime}}{a}\left(\frac{a^\prime}{a}\right)^2\Delta^2\\
&+\left(\frac{7a^5 M^4 }{32\omega_k^9}-\frac{9 a^3 M^2}{32\omega_k^7}+\frac{a}{16\omega_k^5}\right)\left(\frac{a^{\prime\prime}}{a}\right)^2\Delta^2+\left(-\frac{7a^5 M^4 }{16\omega_k^9}+\frac{9a^3M^2}{16\omega_k^7}-\frac{a}{8\omega_k^5}\right)\frac{a^\prime}{a}\frac{a^{\prime\prime\prime}}{a}\Delta^2\\
&+\bigg(-\frac{25025 a^{11} M^{12}}{512\omega_k^{17}}+\frac{39039 a^9 M^{10}}{512 \omega_k^{15}}-\frac{14883 a^7 M^8}{512\omega_k^{13}}+\frac{869 a^5 M^6}{512\omega_k^{11}}\bigg)\left(\frac{a^\prime}{a}\right)^6\\
&+\left(\frac{3003 a^9 M^{10}}{64\omega_k^{15}}-\frac{2013 a^7 M^8}{32\omega_k^{13}}+\frac{1023 a^5 M^6}{64\omega_k^{11}}\right)\frac{a^{\prime\prime}}{a}\left(\frac{a^\prime}{a}\right)^4+\left(-\frac{5a^5 M^6}{16\omega_k^{11}}+\frac{5a^3 M^4}{16\omega_k^9} \right)\left(\frac{a^{\prime\prime}}{a}\right)^3\\
&+\left(-\frac{891a^7M^8}{128\omega_k^{13}}+\frac{501a^5 M^6 }{64\omega_k^{11}}-\frac{111 a^3 M^4}{128\omega_k^9}\right)\left(\frac{a^{\prime\prime}}{a} \frac{a^{\prime}}{a}\right)^2\\
\end{split}
\end{align*}
\vspace{0.5cm}
\begin{align} \label{T00(6)}
\begin{split}
&\phantom{aaaaaa}+\left(-\frac{429a^7 M^8}{64\omega_k^{13}}+\frac{249a^5 M^6}{32\omega_k^{11}}-\frac{69a^3 M^4}{64\omega_k^9}\right)\left( \frac{a^\prime}{a} \right)^3\frac{a^{\prime\prime\prime}}{a}+\left(\frac{15 a^5 M^6}{16\omega_k^{11}}-\frac{15 a^3M^4}{16\omega_k^9}\right)\frac{a^\prime}{a}\frac{a^{\prime\prime}}{a}\frac{a^{\prime\prime\prime}}{a}\\
&\phantom{aaaaaa}+\left(-\frac{a^3 M^4}{64\omega_k^9}+\frac{a M^2}{64\omega_k^7}\right)\left(\frac{a^{\prime\prime\prime}}{a}\right)^2+\left(\frac{9a^5 M^6}{16\omega_k^{11}}-\frac{9a^3 M^4}{16 \omega_k^9}\right)\frac{a^{\prime\prime\prime\prime}}{a}\left(\frac{a^\prime}{a}\right)^{2}\\
&\phantom{aaaaaa}+\left(\frac{a^3 M^4}{32\omega_k^9}-\frac{a M^2}{32\omega_k^7}\right)\frac{a^{\prime\prime}}{a}\frac{a^{\prime\prime\prime\prime}}{a}+\left(-\frac{a^3M^4}{32\omega_k^9}+\frac{aM^2}{32\omega_k^7}\right)\frac{a^\prime}{a}\frac{a^{\prime\prime\prime\prime\prime}}{a}\Bigg]\,.
\end{split}
\end{align}
To obtain the component $\langle T_{11}^{\delta\psi}\rangle$ a similar expansion as in \eqref{eq:ExpansionT00} holds.  However, it is possible to use the following relation with the previously computed  $ \langle T_{00}^{\delta \psi}\rangle$ component:
\begin{align}
\left\langle T_{11}^{\delta\psi}\right\rangle=-\frac{1}{3\mathcal{H}}\left(\left\langle T_{00}^{\delta\psi}\right\rangle^\prime+\mathcal{H}\left\langle T_{00}^{\delta\psi}\right\rangle \right)\,.
\end{align}
As a result we find:
\begin{align}
\left\langle T_{11}^{\delta\psi} \right\rangle^{(0)}&=\frac{1}{2\pi^2 a}\int_0^\infty dk k^2\left[\frac{2a M^2}{3 \omega_k}-\frac{2\omega_k}{3a}\right]\,,
\end{align}
\vspace{0.5cm}
\begin{align}
\begin{split}
\left\langle T_{11}^{\delta\psi} \right\rangle^{(2)}&=\frac{1}{2\pi^2 a}\int_0^\infty dk k^2\Bigg[\left(-\frac{a^3 M^2}{3\omega_k^3}+\frac{a}{3\omega_k}\right)\Delta^2+\left(-\frac{5a^5M^6}{12\omega_k^7}+\frac{a^3 M^4}{3\omega_k^5}+\frac{a M^2}{12\omega_k^3}\right)\left(\frac{a^\prime}{a}\right)^2\\
&+\left(\frac{a^3 M^4}{6\omega_k^5}-\frac{a M^2}{6\omega_k^3}\right)\frac{a^{\prime\prime}}{a}\Bigg]\,,
\end{split}
\end{align}
\begin{align}
\begin{split}
\left\langle T_{11}^{\delta\psi} \right\rangle^{(4)}&=\frac{1}{2\pi^2 a}\int_0^\infty dk k^2\Bigg[\left(\frac{a^5 M^2}{4\omega_k^5}-\frac{a^3}{4\omega_k^3}\right)\Delta^4+\left(-\frac{5a^5 M^4}{12\omega_k^7}+\frac{7a^3 M^2}{12\omega_k^5}-\frac{a }{6\omega_k^3}\right)\frac{a^{\prime\prime}}{a}\Delta^2\\
&+\left(\frac{35 a^7 M^6}{24\omega_k^9}-\frac{25 a^5 M^4}{12\omega_k^7}+\frac{13a^3M^2}{24\omega_k^5}+\frac{a}{12\omega_k^3}\right) \left(\frac{a^\prime}{a}\right)^2 \Delta^2\\
&+\left(\frac{385 a^9 M^{10}}{64\omega_k^{13}}-\frac{483a^7 M^8}{64\omega_k^{11}}+\frac{91a^5 M^6}{64\omega_k^9}+\frac{7a^3 M^4}{64\omega_k^7}\right)\left(\frac{a^\prime}{a}\right)^4\\
&+\left(-\frac{77a^7M^8}{16\omega_k^{11}}+\frac{21a^5 M^6}{4\omega_k^9}-\frac{7a^3 M^4}{16\omega_k^7}\right)\frac{a^{\prime\prime}}{a}\left(\frac{a^\prime}{a}\right)^2+\left(\frac{7a^5 M^6}{16\omega_k^9}-\frac{11a^3 M^4}{24\omega_k^7}+\frac{a M^2}{48\omega_k^5}\right)\left(\frac{a^{\prime\prime}}{a}\right)^2\\
&+\left(\frac{7a^5 M^6}{12\omega_k^9}-\frac{13a^3 M^4}{24\omega_k^7}-\frac{a M^2}{24\omega_k^5}\right)\frac{a^\prime}{a}\frac{a^{\prime\prime\prime}}{a}+\left(-\frac{a^3 M^4}{24\omega_k^7}+\frac{a M^2}{24\omega_k^5}\right)\frac{a^{\prime\prime\prime\prime}}{a}\Bigg]\,,
\end{split}
\end{align}
\vspace{0.5cm}
\begin{align}\label{T11(6)}
\begin{split}
\left\langle T_{11}^{\delta\psi} \right\rangle^{(6)} & = \frac{1}{2\pi^2 a}\int_0^\infty dk k^2\Bigg[\left(-\frac{5a^7 M^2}{24\omega_k^7}+\frac{5a^5}{24\omega_k^5}\right)\Delta^6+\left( \frac{35 a^7 M^4}{48\omega_k^9}-\frac{55a^5 M^2}{48\omega_k^7}+\frac{5a^3}{12 \omega_k^5}\right) \frac{a^{\prime\prime}}{a} \Delta^4 \\
&+\left(-\frac{105 a^9 M^6}{32\omega_k^{11}}+\frac{35 a^7 M^4}{6\omega_k^9}-\frac{265 a^5 M^2}{96\omega_k^7}+\frac{5a^3 }{24\omega_k^5}\right)\left(\frac{a^\prime}{a}\right)^2\Delta^4\\
&+\left(-\frac{5005 a^{11}M^{10}}{128\omega_k^{15}}+\frac{9163 a^9 M^8}{128\omega_k^{13}}-\frac{4683 a^7 M^6}{128\omega_k^{11}}+\frac{497 a^5 M^4}{128\omega_k^9}+\frac{7a^3 M^2}{32\omega_k^7}\right)\left(\frac{a^\prime}{a}\right)^4\Delta^2\\
&+\left(\frac{847a^9 M^8}{32\omega_k^{13}}-\frac{343 a^7 M^6}{8\omega_k^{11}}+\frac{553 a^5 M^4}{32\omega_k^9}-\frac{7a^3M^2}{8\omega_k^7}\right)\frac{a^{\prime\prime}}{a}\left(\frac{a^\prime}{a}\right)^2\Delta^2\\
&+\left(-\frac{63 a^7 M^6}{32\omega_k^{11}}+\frac{35a^5M^4}{12\omega_k^9}-\frac{31a^3 M^2}{32\omega_k^7}+\frac{a}{48\omega_k^5}\right)\left(\frac{a^{\prime\prime}}{a}\right)^2\Delta^2\\
&+\left(-\frac{21 a^7 M^6}{8\omega_k^{11}}+\frac{175a^5 M^4}{48\omega_k^9}-\frac{47 a^3 M^2}{48\omega_k^7}-\frac{a}{24\omega_k^5}\right)\frac{a^\prime}{a}\frac{a^{\prime\prime\prime}}{a}\Delta^2 \\
&+\left(\frac{7a^5 M^4}{48\omega_k^9}-\frac{3a^3 M^2}{16\omega_k^7}+\frac{a}{24\omega_k^5}\right)\frac{a^{\prime\prime\prime\prime}}{a}\Delta^2\\
&-\left(\frac{425425a^{13} M^{14}}{1536\omega_k^{19}}-\frac{355355 a^{11}M^{12}}{768\omega_k^{17}}+\frac{25883 a^9 M^{10}}{128\omega_k^{15}}-\frac{12221 a^7 M^8}{768\omega_k^{13}}-\frac{869a^5 M^6}{1536 \omega_k^{11}}\right)\left(\frac{a^\prime}{a}\right)^6\\
&+\left(\frac{85085 a^{11} M^{12}}{256\omega_k^{17}}-\frac{124839 a^9 M^{10}}{256\omega_k^{15}}+\frac{40623 a^7 M^8}{256\omega_k^{13}}-\frac{869 a^5 M^6}{256\omega_k^{11}}\right)\frac{a^{\prime\prime}}{a}\left(\frac{a^\prime}{a}\right)^4\\
&+\left(-\frac{11869 a^9 M^{10}}{128\omega_k^{15}}+\frac{15301 a^7 M^8}{128\omega_k^{13}}-\frac{3395 a^5 M^6}{128\omega_k^{11}}-\frac{37 a^3 M^4}{128\omega_k^{9}}\right)\left(\frac{a^\prime}{a} \frac{a^{\prime\prime}}{a}\right)^2\\
&+\left(\frac{671 a^7 M^8}{192\omega_k^{13}}-\frac{391 a^5 M^6}{96 \omega_k^{11}}+\frac{37 a^3 M^4}{64\omega_k^{9}}\right)\left(\frac{a^{\prime\prime}}{a}\right)^3\\
&+\left(-\frac{715 a^9 M^{10}}{16 \omega_k^{15}}+\frac{3597a^7 M^8}{64\omega_k^{13}}-\frac{357 a^5 M^6}{32\omega_k^{11}}-\frac{23 a^3 M^4}{64\omega_k^9}\right)\frac{a^{\prime\prime\prime}}{a}\left(\frac{a^\prime}{a}\right)^3\\
&+\left(\frac{473a^7 M^8}{32\omega_k^{13}}-\frac{263 a^5 M^6}{16\omega_k^{11}}+\frac{53 a^3 M^4}{32\omega_k^9}\right)\frac{a^{\prime\prime\prime}}{a}\frac{a^{\prime\prime}}{a}\frac{a^\prime}{a}\phantom{............}\\
&+\left(-\frac{23 a^5 M^6}{64\omega_k^{11}}+\frac{17 a^3  M^4}{48\omega_k^9}+\frac{a M^2}{192\omega_k^7}\right)\left(\frac{a^{\prime\prime\prime}}{a}\right)^2\phantom{............}\\
&+\left(\frac{275 a^7 M^8}{64\omega_k^{13}}-\frac{149a^5 M^6}{32\omega_k^{11}}+\frac{23 a^3 M^4}{64\omega_k^9}\right)\left(\frac{a^{\prime}}{a}\right)^2\frac{a^{\prime\prime\prime\prime}}{a}\phantom{............}\\
&+\left(-\frac{19 a^5 M^6}{32\omega_k^{11}}+\frac{29 a^3 M^4}{48\omega_k^9}-\frac{a M^2}{96\omega_k^7}\right)\frac{a^{\prime\prime}}{a}\frac{a^{\prime\prime\prime\prime}}{a}\phantom{............}\\
&+\left(-\frac{9a^5 M^6}{32\omega_k^{11}}+\frac{13a^3 M^4}{48\omega_k^9}+\frac{a M^2}{96\omega_k^7}\right)\frac{a^{\prime}}{a}\frac{a^{\prime\prime\prime\prime\prime}}{a}+\left(\frac{a^3 M^4}{96\omega_k^9}-\frac{a M^2}{96\omega_k^7}\right)\frac{a^{\prime\prime\prime\prime\prime\prime}}{a}\Bigg]\,.\phantom{............}
\end{split}
\end{align}
We refer the reader to the master formula in Appendix A.2 of \cite{Moreno-Pulido:2022phq} for the explicit computation
of the integrals in the above results. We refrain from providing additional details here, as the remaining calculations can be handled straightforwardly with the help of the mentioned formula.

\end{document}